\documentclass[fleqn,usenatbib]{mnras}

\usepackage{newtxtext,newtxmath}

\usepackage[T1]{fontenc}

\DeclareRobustCommand{\VAN}[3]{#2}
\let\VANthebibliography\thebibliography
\def\thebibliography{\DeclareRobustCommand{\VAN}[3]{##3}\VANthebibliography}

\usepackage{graphicx}	
\usepackage{amsmath}	

\usepackage{amssymb}	

\defcitealias{2012MNRAS.426..140D}{DVS12}
\defcitealias{DallaVecchiaSchaye2008}{DVS08}

\newcommand{\kms}{km s$^{-1}$}

\title[A new thermal-kinetic SN feedback model]{A thermal-kinetic subgrid model for supernova feedback in simulations of galaxy formation}

\author[E. Chaikin et al.]{Evgenii Chaikin,$^{1}$\thanks{E-mail: chaikin@strw.leidenuniv.nl},
Joop Schaye,$^{1}$
Matthieu Schaller,$^{2,1}$
Alejandro Ben\'itez-Llambay,$^{3}$ \newauthor
Folkert S. J. Nobels$^{1}$ and
Sylvia Ploeckinger$^{2,4}$
\\
$^{1}$Leiden Observatory, Leiden University, PO Box 9513, 2300 RA Leiden, the Netherlands\\
$^{2}$Lorentz Institute for Theoretical Physics, Leiden University, PO Box 9506, 2300 RA Leiden, the Netherlands\\
$^{3}$Dipartimento di Fisica G. Occhialini, Universit\`a degli Studi di Milano Bicocca, Piazza della Scienza, 3 I-20126 Milano MI, Italy\\
$^{4}$Department of Astrophysics, University of Vienna, T{\"u}rkenschanzstrasse 17, 1180 Vienna, Austria
}

\date{Accepted XXX. Received YYY; in original form ZZZ}

\pubyear{2015}

\begin{document}

\label{firstpage}
\pagerange{\pageref{firstpage}--\pageref{lastpage}}
\maketitle

\begin{abstract}
We present a subgrid model for supernova feedback designed for cosmological simulations of galaxy formation that may include a cold interstellar medium (ISM). The model uses thermal and kinetic channels of energy injection, which are built upon the stochastic kinetic and thermal models for stellar feedback used in the \textsc{owls} and \textsc{eagle} simulations, respectively. In the thermal channel, the energy is distributed statistically isotropically and injected stochastically in large amounts per event, which minimizes spurious radiative energy losses. In the kinetic channel, we inject the energy in small portions by kicking gas particles in pairs in opposite directions. The implementation of kinetic feedback is designed to conserve energy, linear and angular momentum, and is statistically isotropic. To test the model, we run simulations of isolated Milky Way-mass and dwarf galaxies, in which the gas is allowed to cool down to $10$ K. Using the thermal and kinetic channels together, we obtain smooth star formation histories and powerful galactic winds with realistic mass loading factors. Furthermore, the model produces spatially resolved star formation rates (SFRs) and velocity dispersions that are in agreement with observations. We vary the numerical resolution by several orders of magnitude and find excellent convergence of the global SFRs and wind mass loading. We show that large thermal-energy injections generate a hot phase of the ISM and modulate the star formation by ejecting gas from the disc, while the low-energy kicks increase the turbulent velocity dispersion in the neutral ISM, which in turn helps suppress star formation.
\end{abstract}




\section{Introduction}

Prescriptions for star formation and feedback from stars form the backbone of all numerical models of galaxy formation. Shortly after being born in dense, molecular clouds, young stellar populations begin to disrupt their parent clouds through various feedback processes. Among them are radiation pressure, stellar winds, cosmic rays, photoionization and core-collapse supernovae (SNe). Numerical simulations have shown that all of these feedback channels can be important for the structure of the interstellar medium (ISM) \citep[e.g.][]{2012MNRAS.421.3522H,2017ARA&A..55...59N,2019MNRAS.485.3317S}.

In order to produce realistic galaxies, simulations of galaxy formation require detailed modelling of feedback processes from both stars and supermassive black holes. However, since the majority of feedback processes -- including SN feedback -- occur on scales below $\sim 10$ pc, running hydrodynamical simulations of cosmologically representative volumes to redshift $z=0$ that directly model the feedback from individual stars is practically impossible. Instead, subgrid models `mimicking' the effects of individual feedback processes are adopted \citep[see e.g.][for reviews]{2015ARA&A..53...51S,2020NatRP...2...42V}. These models operate below the minimum resolved scale, but aim to produce a galaxy population whose properties are in agreement with observations on resolved scales. For SN feedback this means that whilst the explosion energy from SNe is deposited simultaneously in a partly resolved ISM, the effects of the SN feedback should be such that the simulation produces galaxies with realistic morphologies and histories of stellar mass assembly that follow the observed Kennicutt-Schmidt (KS) star formation law \citep[][]{Kennicutt1998ApJ,2007ApJ...671..333K}, and are able to develop strong galactic-scale
outflows \citep[e.g.][]{2015MNRAS.452.1184M,2015ApJ...808..129L,2017MNRAS.465.1682H}.

In the earliest cosmological simulations including dark matter and baryons, SN feedback in galaxies was implemented as a subgrid model releasing the canonical $10^{51}$ erg of energy into the gas via a direct `thermal dump' \citep{Katz1992}. However, it soon became obvious that this approach is too inefficient at regulating star formation in dense gas because the low resolution of the simulations would greatly enhance radiative energy losses and the injected SN energy would be dissipated too quickly \citep[e.g.][]{Katz1996,2012MNRAS.426..140D}. Various subgrid models for SN feedback in which the energy losses due to the enhanced radiative cooling are strongly reduced have been successfully employed to produce a realistic galaxy population \citep[e.g.][]{2015MNRAS.446..521S,Hopkins2018Fire2,Pillepich2018,2019MNRAS.486.2827D}. 

Generally, subgrid models for SN feedback can be split into three main categories, depending on the form in which the SN energy is injected into the surrounding gas. These are \textit{thermal},
\textit{kinetic}, and \textit{thermal-kinetic} \citep[see e.g.][for comparisons of different SN models]{2010MNRAS.402.1536S,RosdahlSchaye2017,Smith2018,Gentry2020}. In the thermal models, SN energy is added to gas elements by increasing their internal energies; in the kinetic models, gas elements' velocities (or momenta) are modified; and in the thermal-kinetic models, the energy is injected via both thermal and kinetic channels. Depending on the form in which the SN energy is deposited into the ISM, different ways of reducing numerical energy losses due to spurious radiative cooling are implemented. For purely thermal coupling, it is common to employ the so called `delayed cooling' approach where radiative cooling rates of the SN-heated gas elements are temporarily set to zero (or exponentially suppressed) so that the injected SN energy is retained in the ISM for longer \citep[e.g.][]{1997PhDT........19G,Stinson2006,Dubois2015}. As a result, more mechanical work is done by the expanding, hot bubbles on the surrounding, generally colder and denser gas, which makes the SN model overall more efficient at suppressing star formation. For purely kinetic coupling, in order to make SN feedback more efficient, it is common to use the `hydrodynamical decoupling' approach. In this method, the hydrodynamical forces acting on the gas elements that have been directly affected by SN feedback are temporally switched off -- until these gas elements have escaped the star-forming ISM to become part of a galactic-scale outflow \citep[e.g][]{SpringelHernquist2003,OppenheimerDave2006,Vogelsberger2013}. 

The delayed cooling and hydrodynamical decoupling methods have the drawback that they result in artefacts that become more prominent at higher resolution, when the ISM is better resolved. In the kinetic models with hydrodynamical decoupling, these artefacts may include insufficient turbulence of the gas in the ISM and the absence of emerging SN superbubbles -- both of which are direct consequences of the SN energy freely leaving the galaxy in the (decoupled) outflowing gas, without interacting with the ISM where it has been deposited. The thermal models strengthened by delayed cooling produce an excessive amount of dense gas that has short cooling times but which is (by construction) not allowed to cool.

One of the possible ways to strengthen SN feedback without directly suppressing the ability of gas elements to interact and cool immediately after the feedback, is to inject the SN energy stochastically: in larger amounts per SN event but with a lower frequency in time. The energy in this approach can be deposited either in thermal form via heating gas to high temperatures ($T\sim 10^{7.5}$ K) (\citealt{2012MNRAS.426..140D}, henceforth \citetalias{2012MNRAS.426..140D}) or in kinetic form via kicking gas elements with high kick velocities ($\Delta v_{\rm kick} \sim 10^3$ km s$^{-1}$) (e.g. \citealt{DallaVecchiaSchaye2008}, henceforth \citetalias{DallaVecchiaSchaye2008}). By using such temperatures and velocities, which are similarly high as in SN (super)bubbles, excessive thermal losses are avoided and it becomes possible to regulate galaxy star formation and generate galactic winds with wind mass loading factors similar to observations \citep[e.g.][]{2020MNRAS.494.3971M}. 

A disadvantage of the stochastic models with high energies per SN energy injection (corresponding to $\Delta T\gtrsim 10^{7.5}$ K) is that the number of SN energy injections per star particle becomes small, so in galaxies made up of a modest number of star particles SN feedback may be undersampled. Furthermore, with such high energies per SN event, SN feedback will tend to regulate the galaxy star formation rate (SFR) mostly through ejecting gas from the ISM, whereas in reality SNe are also expected to inhibit star formation by increasing turbulence in the ISM \citep[e.g.][]{2006ApJ...653.1266J, 2011ApJ...731...41O}. In other words, the turbulence in the ISM gas generated by the \citetalias{2012MNRAS.426..140D} stochastic feedback might be too weak and/or too local. A solution to these shortcomings within the \citetalias{2012MNRAS.426..140D}-like models' framework can be to extend the original model by combining large and rare thermal energy injections with low and frequent energy input in kinetic form\footnote{Low-energy injections ($\Delta T\lesssim 10^{6}$ K) can only be distributed in kinetic form, as opposed to thermal form, because of the strong radiative cooling at such low $\Delta T$.}.

The ISM turbulence may also be underestimated if the gas in the ISM is assumed to follow an effective equation of state (eEOS), $P(\rho)\propto \rho^{\gamma}$, where $P$ and $\rho$ are, respectively, the gas pressure and density, and $\gamma$ is the polytropic index  \citep[e.g.][]{2008MNRAS.383.1210S}. An eEOS is employed when the simulation is unable to accurately model the multiphase structure of the ISM due to the lack of physics, resolution, and/or because running the simulation becomes too computationally expensive due to the very small time-steps reached in dense gas. For these reasons, nearly all previous cosmological simulations have relied on using an eEOS \citep[e.g.][]{2010MNRAS.402.1536S,2015MNRAS.446..521S,2017MNRAS.465.2936M,Pillepich2018,2019MNRAS.486.2827D}, but recently first attempts have been made to abandon it and probe the multiphase ISM more directly \citep[e.g.][]{2021A&A...651A.109D,2023MNRAS.522.3831F}.

In this work, we present a new stochastic thermal-kinetic model for SN feedback designed for large cosmological simulations, including those that (partly) resolve a cold ISM, without employing an eEOS. In essence, our model is based on the works of \citetalias{DallaVecchiaSchaye2008} and \citetalias{2012MNRAS.426..140D} with a number of significant modifications. Specifically, (i) while the feedback in our model is also done stochastically, we include not only large thermal injections but also low-energy kicks, (ii) the SN energy is distributed statistically isotropically, and (iii) energy, linear momentum, and angular momentum are exactly conserved. We describe our SN model in Section \ref{sec:methods}. In Section \ref{sec: numerical simulations} we introduce the numerical simulations that we use to test and validate our SN model. In Section \ref{sec: results} we present the results of the simulations, and we discuss them in Section \ref{sec: discussion}. Finally, in Section \ref{sec: conclusions} we summarise our conclusions.

\section{SN feedback model}
\label{sec:methods}

Modern cosmological simulations including hydrodynamics cannot yet resolve individual stars and instead use star particles as the smallest building blocks of stellar mass. Each star particle represents a coeval stellar population determined by the assumed initial mass function (IMF) $\Phi(m)$. For the tests presented in this work, we adopt the \citet{Chabrier2003} IMF. In order to compute the number of SNe per star particle, we integrate the IMF between $m_{\rm min}=8 \, \rm M_\odot$ and $m_{\rm max}=100 \, \rm M_\odot$. More precisely, given a star particle of initial mass $m_*$ and age $t$, we compute how much SN energy it will release in a time-step\footnote{The energy $\Delta E_{\rm SN}$ is computed every time-step since a star particle is born. This differs from the approach taken by \citetalias{DallaVecchiaSchaye2008} and \citetalias{2012MNRAS.426..140D}, where a star particle's SN energy would only become eligible for stochastic injection after a delay of 30 Myr since its birth.} $\Delta t$,
\begin{equation}
     \Delta E_{\rm SN}(t,\Delta t, m_*, f_{\rm E}) = 10^{51} \, \mathrm{erg} \, f_{\rm E} \, \, m_{\rm *}  \, \int_{m_{\rm d}(t+\Delta t)}^{m_{\rm d}(t)}  \, \Phi(m) \, \mathrm{d}m \, ,
    \label{eq: energy_per_dt}
\end{equation}
where the parameter $f_{\rm E}$ gives the energy of a single SN, in units of $10^{51}$ erg, and $m_{\rm min} \leq m_{\rm d}(t) < m_{\rm max}$ is the zero-age main sequence mass of the stars that die at age $t$, which in this work is computed using the metallicity-dependent stellar-lifetime tables from \citet{Portinari1998}. In our model, the energy $\Delta E_{\rm SN}$ is deposited in the surrounding gas in both kinetic and thermal forms. We use a free parameter, $f_{\rm kin}$, to split the SN energy between the two channels: $f_{\rm kin} \, \Delta E_{\rm SN}$ is released in kinetic form and the remainder, $(1-f_{\rm kin}) \, \Delta E_{\rm SN}$, is injected thermally. 

The following discussion assumes that the model has been implemented in a smoothed particle hydrodynamics (SPH) code, but the scheme could easily be adapted to other types of hydro solvers. We further assume that star particles follow the same algorithm of gas-neighbour finding as gas particles, and therefore have the same expected number of gas neighbours inside the kernel as is the case for gas particles.

\subsection{Thermal channel}
\label{paragraph: thermal model}

The prescription for the thermal channel represents an isotropic version of the \citetalias{2012MNRAS.426..140D} stochastic model and is described in detail in \citet{Chaikin2022}. 

Based on a probability, every time-step $\Delta t$ a star particle may inject $\Delta E_{\rm heat}$ erg of energy into one or several gas neighbours. The value of $\Delta E_{\rm heat}$ is determined by specifying the desired temperature increase $\Delta T$ of the heated gas particles of mass $m_{\rm gas}$ using the expression

\begin{equation}
    \Delta E_{\rm heat}(m_{\rm gas},\Delta T) = \frac{k_{\rm B}\Delta T}{(\gamma-1)}\frac{m_{\rm gas}}{\mu_{\rm ionized} m_{\rm p}} \, ,
    \label{eq: heating_energy}
\end{equation}
in which $\gamma = 5/3$ is the ratio of specific heats for an ideal monatomic gas, $k_{\rm B}$ is the Boltzmann constant, $m_{\rm p}$ is the proton mass, and $\mu_{\rm ionized}=0.6$ is the mean molecular weight of a fully ionized gas. For a star particle that can release $(1-f_{\rm kin}) \, \Delta E_{\rm SN}$ of thermal energy in the time-step [$t,t+\Delta t$), the probability of heating a gas neighbour by the temperature $\Delta T$ is given by

\begin{equation}
    p_{\rm heat}(f_{\rm E}, f_{\rm kin}, \Delta T, m_{*}, m_{\rm ngb}, t, \Delta t) = (1-f_{\rm kin}) \, \frac{\Delta E_{\mathrm{SN}}(t, \Delta t, m_*, f_{\rm E})}{\Delta E_{\rm heat}(m_{\rm ngb},\Delta T)} \, ,
    \label{eq:prob_heat}
\end{equation}
where $m_{\rm ngb}$ is the total mass of the gas elements in the star particle's kernel. The probability $p_{\rm heat}$ cannot be greater than unity: if $p_{\rm heat}>1$, then we adjust (i.e. increase) the heating temperature $\Delta T$ until $p_{\rm heat}=1$. Note that as long as gas and stellar particles have comparable masses, the condition $p_{\rm heat}>1$ can only be triggered if the heating temperature $\Delta T\ll 10^7$ K.

\subsubsection{The number of heating events}
\label{sec:num_of_heating_events}

To calculate the number of thermal energy injections for a star particle of age $t$, initial mass $m_*$, time-step $\Delta t$ and gas mass inside the star particle kernel $m_{\rm ngb}$, we compute the heating probability $p_{\rm heat}$ (equation \ref{eq:prob_heat}) at the beginning of the time-step and initialise the particle's number of energy-injection events $N_{\rm heat}$ with zero. Next, for each gas neighbour, we draw a random number $r$ from a uniform distribution $0\leq r<1$. Every time we find $r<p_{\rm heat}$, the value of $N_{\rm heat}$ is incremented by one. For the commonly used heating temperature of $\Delta T\sim 10^{7.5}$ K \citep[e.g.][]{2015MNRAS.446..521S}, the average number of heating events per time-step is $N_{\rm heat} \ll 1$, and over the star particle's lifetime the total number of heating events is
\begin{align}
\langle N_{\rm heat, tot} \rangle  &= \frac{(1-f_{\rm kin}) \, E_{\rm SN, tot}(m_*, f_{\rm E})  \, }{\Delta E_{\rm heat}(m_{\rm gas},\Delta T)} \nonumber \, , \\
&= 0.91 \, (1-f_{\rm kin}) \, f_{\rm E} \,  \, \left(\frac{m_{\rm *}}{m_{\rm gas}}\right)\, \left(\frac{\Delta T}{\mathrm{10^{7.5} \, K}}\right)^{-1} \, ,
\label{eq:N_heat}
\end{align}
where for simplicity we assumed the gas particle mass $m_{\rm gas}$ to be the same for all gas neighbours and $E_{\rm SN, tot}$ is the total SN energy budget of the star particle, 
\begin{equation}
     E_{\rm SN,tot} = 10^{51} \, \mathrm{erg} \, f_{\rm E} \, \, m_{\rm *}  \, \int_{m_{\rm min}}^{m_{\rm max}}  \, \Phi(m) \, \mathrm{d}m \, .
    \label{eq: total_sn_energy} 
\end{equation}

\subsubsection{Distributing the heating events among gas neighbours}
\label{sec: distribute_heating_events}

To select the gas neighbours that will receive the $N_{\rm heat}$ thermal energy-injection events in the time-step $\Delta t$, we adopt the isotropic algorithm from \citet{Chaikin2022}. The algorithm works as follows:

\begin{enumerate}
    \item We create $N_{\rm rays}$ randomly directed rays originating from the position of the star particle.

    \item  For each ray $j$, we compute great-circle distances with each gas neighbour $i$ on a unit sphere using the haversine formula\footnote{The haversine formula uses the latitude and longitude coordinates of the gas neighbour $i$ and ray $j$ in the reference frame positioned at the star particle.}
\begin{equation}
 \Omega_{ij} = 2\arcsin \sqrt{\sin^2\left(\frac{\theta_{j}-\theta_{ i}}{2}\right) + \cos(\theta_{i})\cos(\theta_{ j})\sin^2\left(\frac{\varphi_{j}-\varphi_{ i}}{2}\right)} \, ,
 \label{eq:arclength}
\end{equation}
and find the gas particle $i$ that minimizes the value of $\Omega_{ij}$\footnote{This implies that although each ray always points to a single gas neighbour, a gas neighbour may be pointed to by more than one ray (which may result in an increase of its temperature by more than $\Delta T$).}.

\item  If $N_{\rm heat}\leq N_{\rm rays}$, we randomly pick $N_{\rm heat}$ rays out of $N_{\rm rays}$ rays and inject the energy into the gas neighbours `attached to' these rays. If $N_{\rm heat}> N_{\rm rays}$, we increase the heating temperature $\Delta T$ by $N_{\rm heat}/N_{\rm rays}$ and inject the energy defined by the new value of $\Delta T$ into the gas particles corresponding to all $N_{\rm rays}$ rays. 
\end{enumerate}
We note that in the tests presented in this work, the chance of the second scenario, $N_{\rm heat}> N_{\rm rays}$, occurring is negligibly small, and here we only describe it for completeness. Given a heating temperature $\Delta T$, to obtain an estimate on the minimum number of rays required to avoid the $N_{\rm heat}> N_{\rm rays}$ case, one needs to estimate how many heating events a star particle is expected to distribute during its longest possible time-step. Equation (\ref{eq:N_heat}) determines the expected number of heating events accumulated over the star particle's lifetime, so it can be used as an upper bound on the expected number of heating events per time-step.

\subsection{Kinetic channel}
\label{paragraph:stochastic_kinetic_feedback}

The kinetic channel uses a modified version of the \citetalias{DallaVecchiaSchaye2008} kinetic stochastic model.

To parametrize the amount of energy released in one stochastic kinetic event, instead of specifying the heating temperature $\Delta T$, we specify the desired kick velocity $\Delta v_{\rm kick}$. Analogously to equation (\ref{eq:prob_heat}), in a simulation time-step from $t$ to $t+\Delta t$ during which the star particle releases the kinetic energy $f_{\rm kin} \Delta E_{\rm SN}$, the probability that the star particle will kick a gas neighbour in this time-step can be written as
\begin{align}
    p_{\rm kick}(f_{\rm E}, f_{\rm kin}, \Delta v_{\rm kick}, m_{*}, m_{\rm ngb}, t, \Delta t) &= f_{\rm kin} \frac{\Delta E_{\rm SN}(t, \Delta t, m_*, f_{\rm E}) }{\Delta E_{\rm kick}(m_{\rm ngb}, \Delta v_{\rm kick})} \nonumber \, , \\
    &= f_{\rm kin}\frac{2\,  \Delta E_{\rm SN}(t, \Delta t, m_*, f_{\rm E})}{m_{\rm ngb} \, \Delta v_{\rm kick}^2} \, ,
    \label{eq: prob_kick}
\end{align}
where $\Delta E_{\rm kick}(m_{\rm ngb}, \Delta v_{\rm kick}) = m_{\rm ngb} \, \Delta v_{\rm kick}^2 / 2$ is the energy required to make the mass $m_{\rm ngb}$ (that is initially at rest in the reference frame of the star particle) move with velocity $\Delta v_{\rm kick}$.

\subsubsection{Enforcing conservation of energy and linear momentum}

Owing to the symmetries of the Lagrangian from which the equations of motion in SPH are derived, SPH schemes naturally conserve energy, linear and angular momentum on a global scale \citep[e.g.][]{price2012}. It is therefore desirable to respect these global conservation laws for any SN feedback model that is implemented in an SPH code. Indeed, violating conservation laws during SN feedback might lead to undesired behaviour in galaxy simulations \citep[e.g.][]{hopkinsfeedback2018}. 

In the model of \citetalias{DallaVecchiaSchaye2008}, individual particles are kicked in random angular directions with a fixed velocity, which in general violates the conservation of linear momentum and energy (although one might argue that on average the errors approximately cancel out, after a sufficiently large number of kicks). We upgrade the \citetalias{DallaVecchiaSchaye2008} kinetic model by ensuring that the linear momentum and energy are both conserved to the floating-point precision. To achieve this, we introduce an algorithm that accounts for the relative motion between stars and the local gas. Our algorithm is somewhat similar to that presented in \citet{hopkinsfeedback2018} but is designed for stochastic, \citetalias{DallaVecchiaSchaye2008}-like models where the number of particle kicks may be small and vary from time-step to time-step.

\subsubsection{Kicking an arbitrary number of neighbours inside the kernel}
Consider a system comprising a star particle and $N_{\rm ngb}$ gas neighbours at a certain time-step $\Delta t$ where the star particle releases some energy $\Delta E_{\rm kin}$. The star injects this energy in kinetic form by kicking its gas neighbours (i.e. by modifying their (peculiar) velocities recorded at the beginning of the time-step). In the reference frame where the star particle is at rest, the total kinetic energy of the system immediately before the feedback event, $E_{\rm kin, tot}$, is

\begin{equation}
 \sum_j\frac{m_j|\boldsymbol{v}^\prime_j|^2}{2} = E_{\rm kin, tot}  \, ,
   \label{eq: E_conservation_before_kick}
\end{equation}
where $\boldsymbol{v}^\prime_j$ is the velocity relative to the star particle of the $j^{\rm th}$ gas neighbour prior to the kick, $m_j$ is the mass of particle $j$, and the sum is computed from $j=1$ to $N_{\rm ngb}$. Immediately after the kinetic feedback, the energy of the system should have increased by $\Delta E_{\rm kin}$ and equation (\ref{eq: E_conservation_before_kick}) becomes

\begin{equation}
\sum_j\frac{m_j|\boldsymbol{v}^\prime_j+ \Delta \boldsymbol{ v}_j|^2}{2} = E_{\rm kin,tot} + \Delta E_{\rm kin} \, ,
    \label{eq: E_conservation_after_kick}
\end{equation}
where $\Delta \boldsymbol{v}_j$ is the change in $j^{\rm th}$ gas neighbour's velocity due to the kick. We can subtract equation (\ref{eq: E_conservation_before_kick}) from equation (\ref{eq: E_conservation_after_kick}) and rewrite the result in the `lab frame' where the star is moving with velocity $\boldsymbol{v}_*$\footnote{Here we assumed that the particle mass $m_j$ is constant or, in other words, that the energy is injected into gas particles separately from the transfer of mass of the SN ejecta. We opted to `decouple' these two processes because our kinetic energy feedback is probabilistic, whereas the injection of mass is done continuously and deterministically \citep[e.g.][]{2009MNRAS.399..574W} and results in only small changes since the mass loss is divided over $N_{\rm ngb}$ particles and spread over many time-steps. A more general form of equation (\ref{eq: E_is_conserved}) including the change in the stellar and gas masses can be found in Appendix E of \citet{hopkinsfeedback2018}. 
},

\begin{equation}
   \sum_j\frac{m_j|\Delta \boldsymbol{v}_j|^2}{2} + \sum_j\, m_j\Delta\boldsymbol{v}_j\cdot(\boldsymbol{v}_j- \boldsymbol{v}_{*}) = \Delta E_{\rm kin} \, .
\label{eq: E_is_conserved}
\end{equation}
In the equation above, the velocity of the $j^{\rm th}$ gas particle in the lab frame prior to the kick, $\boldsymbol{v}_j$, is related to that in the frame comoving with the star particle via $\boldsymbol{v}_j = \boldsymbol{v}^{\prime}_j + \boldsymbol{v}_*$. 

Mathematically, injecting energy in kinetic form while accounting for the motion of gas neighbours relative to the star and conserving total SN energy means that the kick velocity $\Delta \boldsymbol{v}_j$ has to be a solution of equation (\ref{eq: E_is_conserved}). Another important constraint that can be put on the velocity $\Delta \boldsymbol{v}_j$ is the requirement that the total linear momentum is conserved, 

\begin{equation}
    \sum_j\, m_j\, \Delta \boldsymbol{v}_j= 0 \, .
    \label{eq: p_is_conserved}
\end{equation}
In the kinetic channel of our model, we solve both equations (\ref{eq: p_is_conserved}) and (\ref{eq: E_is_conserved}).

\subsubsection{Kicking pairs}
\label{parapgraph:kicks_in_pairs}

\begin{figure*}
  \includegraphics[width=0.23\textwidth]{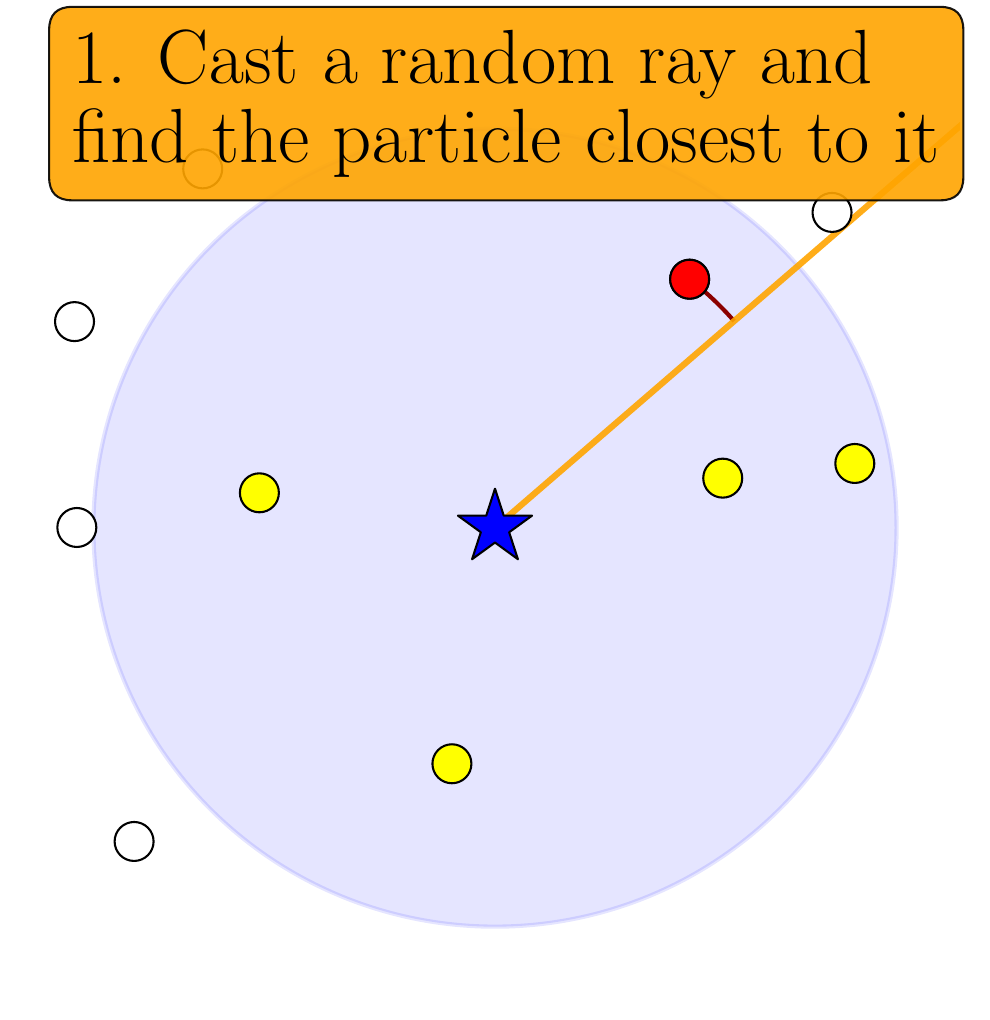}
  \includegraphics[width=0.23\textwidth]{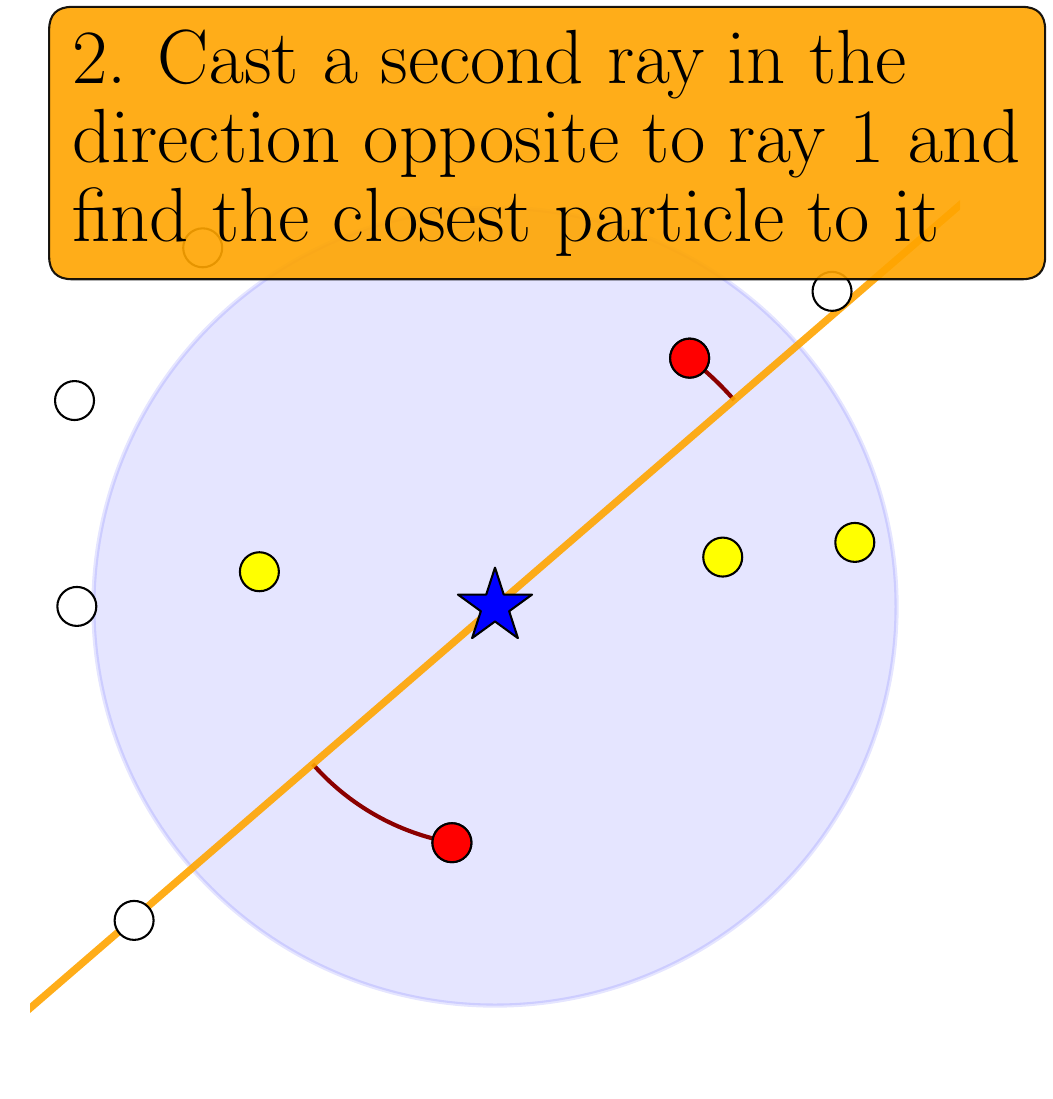}
  \includegraphics[width=0.23\textwidth]{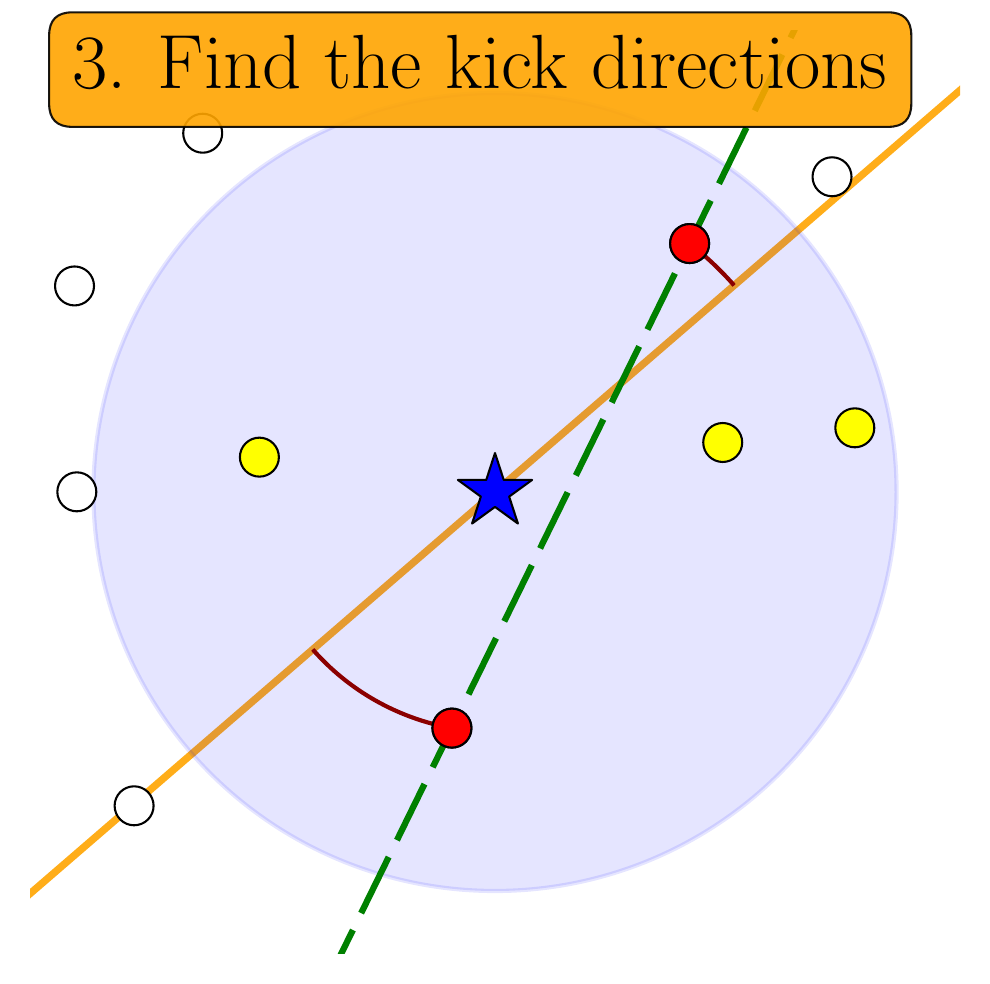}
  \includegraphics[width=0.23\textwidth]{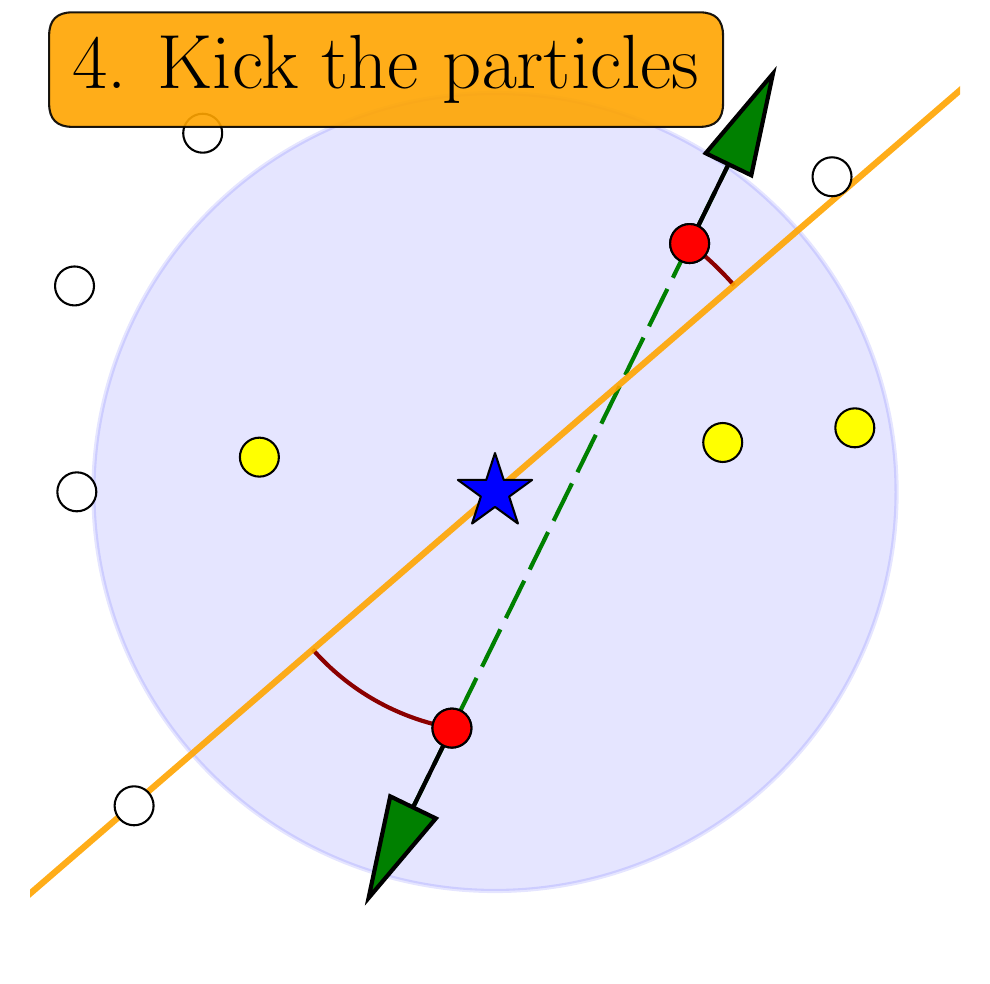}
  \caption{Illustration of the algorithm used for the kinetic channel of the subgrid model for SN feedback. The star particle doing feedback is shown as a blue star in the centre of each panel, and its kernel is given by the shaded circular region. Gas particles that are outside (inside) the kernel are shown in white (yellow). \textit{Step 1.} In a given time-step, cast a ray (orange line) in a random angular direction from the position of the star particle and find the closest gas neighbour to the ray (red circle) by minimizing the arc length on a unit sphere between the gas neighbours and the ray; \textit{Step 2.} Cast a second ray in the direction opposite to the first ray and repeat step 1 for the second ray. \textit{Step 3.} Find the line connecting the selected gas neighbours (green dashed line). \textit{Step 4.} Kick the two gas neighbours along the line connecting them in opposite directions (away from each other). }
  \label{fig: kin_feedback_illustration}
\end{figure*}

In order to conserve linear momentum, at least two gas particles need to be kicked simultaneously (otherwise the only solution to equation (\ref{eq: p_is_conserved}) is $\Delta \boldsymbol{v}_j = 0$). For a pair of two gas neighbours equation (\ref{eq: E_is_conserved}) can be written in the form
\begin{equation}
     \sum_{j=+,-} \frac{m_j|\Delta \boldsymbol{v}_j|^2}{2} + \sum_{j=+,-} m_j\Delta\boldsymbol{v}_j\cdot(\boldsymbol{v}_j- \boldsymbol{v}_{*}) = \Delta E_{\rm pair} \, ,
\label{eq_pair_general}
\end{equation}
where the sum is computed over the first and second particles in the pair, which are denoted by the indices `+' and `-'. 

Without loss of generality, we may write that the first and second particle in the pair are kicked with velocities of magnitudes $|\Delta \boldsymbol{v}_{+}| = w_+ \Delta v_{\rm pair}$ and $|\Delta \boldsymbol{v}_{-}| = w_- \Delta v_{\rm pair}$, where the expected kick velocity, $\Delta v_{\rm pair}$, is related to the kinetic energy injected into the pair, $\Delta E_{\rm pair}$, via $\Delta v_{\rm pair} = \sqrt{2\, \Delta E_{\rm pair} / (m_+ + m_-) }$, and the weights $w_{+,-}$ are yet to be found. The two kicks need to be executed in opposite directions, which can be defined by unit vectors $\boldsymbol{n}_\pm = \pm \boldsymbol{n}$. The linear momentum conservation equation (\ref{eq: p_is_conserved}) then obtains the form,
\begin{equation}
   m_{+} \,  w_{+} \,  \Delta v_{\rm pair}\boldsymbol{n} - m_{-}\, w_{-} \, \Delta v_{\rm pair} \boldsymbol{n} = 0 \, .
    \label{eq: pair_momentum}
\end{equation}
Making the ansatz,
\begin{equation}
   w_{\pm} =  \beta\frac{\sqrt{m_+ m_-}}{m_{\pm}}\, ,
\end{equation}
and plugging it into the energy conservation equation (\ref{eq_pair_general}), we can write down the equation for $\beta$,
\begin{equation}
  \beta^2 + 2\, \beta \,\frac{\sqrt{m_+ m_-}}{m_+ + m_-}\sum_{j=+,-} \, \frac{|\boldsymbol{v}_j-\boldsymbol{v}_{*}|}{\Delta v_{\rm pair}} \cos \theta_j= 1\, ,
  \label{eq:eq_beta_pair}
\end{equation}
which has a simple solution

\begin{equation}
    \beta =  \sqrt{\alpha^2 + 1} -\alpha \, , 
    \label{eq:beta_alpha}
\end{equation}
with 
\begin{equation}
\alpha = \frac{\sqrt{m_+ m_-}}{m_+ + m_-}\sum_{j=+,-} \, \frac{|\boldsymbol{v}_j-\boldsymbol{v}_{*}|}{\Delta v_{\rm pair}} \cos \theta_j, 
\label{eq:alpha}
\end{equation}
and
\begin{equation}
    \cos \theta_{j} =  \frac{(\boldsymbol{v}_{j}-\boldsymbol{v}_{*})\cdot\boldsymbol{n}_j }{|\boldsymbol{v}_{j}-\boldsymbol{v}_{*}|}\, .
    \label{eq: sign_in_cos}
\end{equation}
Inspection of the above equations shows that the two particles forming a pair are kicked with actual\footnote{Here and in the following we use the word `actual' to refer to the kick velocities that are \textit{in fact} applied to the gas neighbours, as opposed to the desired kick velocity $\Delta v_{\rm kick}$, which is used to set the energy of one kick event in the rest-frame of the star particle.} velocities 
\begin{equation}
    \Delta \boldsymbol{v}_{\pm} \equiv w_{\pm} \, \Delta v_{\rm pair} \, \boldsymbol{n}_\pm = \pm \, \beta\frac{\sqrt{m_+ m_-}}{m_{\pm}}\Delta v_{\rm pair} \, \boldsymbol{n} \, ,
    \label{eq:v_kick_pair_beta}
\end{equation}
which are different from the expected kick velocity in the pair $\Delta v_{\rm pair}$ if (1) there is relative motion between the star and the surrounding gas ($\alpha \neq 0$), and/or the kicked particles have different masses ($m_+ \neq m_-$). If the gas particles have different masses, then the more massive particle receives a smaller kick velocity.

\subsubsection{The direction of kicks}

To complete the prescription for the kinetic channel, it is necessary to specify the normal vector $\boldsymbol{n}$ in equation (\ref{eq:v_kick_pair_beta}). Requiring that the angular momentum with respect to the star particle is conserved gives the only possible solution, which is to kick the two gas particles away from each other along the line connecting the two particles. If the two particles have coordinates $(x_+,y_+,z_+)$ and $(x_-,y_-,z_-)$ (in any reference frame), the normal vector can be computed as

\begin{align}
 \boldsymbol{n}  &= \frac{1}{\sqrt{(x_+ - x_-)^2 + (y_+ - y_-)^2 + (z_+ - z_-)^2}}\begin{bmatrix}
       x_+ - x_- \\
       y_+ - y_- \\
       z_+ - z_-
     \end{bmatrix} \, .
\label{eq: kick_direction}
\end{align}
As a consequence, the direction of the imparted linear momentum may not precisely stem from the star particle.

\subsubsection{The number of kick events}
\label{sec:num_of_kick_events}

We define a `kick event' as a kinetic energy injection event in which \textit{two} particles are kicked, as described in $\S$\ref{parapgraph:kicks_in_pairs}.

Given a star particle of initial mass $m_*$, age $t$, and the number of gas neighbours in the kernel $N_{\rm ngb}$ with the total mass $m_{\rm ngb}$, to obtain the number of kick events in a time-step [$t, t+\Delta t$), we first compute the kick probability $p_{\rm kick}$ using equation (\ref{eq: prob_kick}) and initialise the number of kick events $N_{\rm kick}$ with zero. Then, for each gas neighbour of the star particle, we draw a random number $r$ from a uniform distribution $0\leq r<1$, and if $r<p_{\rm kick, pair}$, where $p_{\rm kick, pair} = 0.5 \, p_{\rm kick}$\footnote{$p_{\rm kick,pair}$ is the probability of kicking a \textit{pair} of gas neighbours. It is two times smaller than the probability of kicking one gas neighbour, $p_{\rm kick}$, since for a fixed $\Delta v_{\rm kick}$, kicking two particles requires twice as much energy.}, we increase the value of $N_{\rm kick}$ by one. The kinetic energy associated with each kick event is given by

\begin{align}
    \Delta E_{\rm pair}(m_{\rm ngb}/N_{\rm ngb}, \Delta v_{\rm kick}) &= 2 \, \Delta E_{\rm kick}(m_{\rm ngb}/N_{\rm ngb}, \Delta v_{\rm kick}) \nonumber \, , \\
    &= \frac{m_{\rm ngb}}{N_{\rm ngb}} \Delta v_{\rm kick}^2 \, ,
    \label{eq: E_pair}
\end{align}
where $\Delta E_{\rm kick}(m_{\rm ngb}/N_{\rm ngb}, \Delta v_{\rm kick}) = 0.5 \, m_{\rm ngb} \Delta v_{\rm kick}^2/ N_{\rm ngb}$ is the energy that is expected to be needed to kick a gas neighbour with velocity $\Delta v_{\rm kick}$, and the additional factor of 2 is used because two such gas neighbours are kicked in the kick event. Note that we use the mean particle mass in the kernel, $m_{\rm ngb}/N_{\rm ngb}$, instead of the mean particle mass in the pair, $0.5(m_- + m_+)$, because we do not know a priori which two neighbours from $N_{\rm ngb}$ will receive the kick event and, hence, what the values of $m_-$ and $m_+$ will be.

In the limiting case $p_{\rm kick,pair} \geq 1$, we set $N_{\rm kick} = N_{\rm ngb}$ and use $\Delta E_{\rm pair} = f_{\rm kin} \, \Delta E_{\rm SN} / N_{\rm kick}$ instead of equation (\ref{eq: E_pair}) in order to release all available kinetic energy in the time-step\footnote{Note that when $p_{\rm kick,pair} \geq 1$, naively setting $N_{\rm kick} = N_{\rm ngb}/2$ is unlikely to result in kicking all $N_{\rm ngb}$ gas neighbours in the kernel just once because $N_{\rm ngb}$ can be odd and because some gas particles may receive multiple kicks due to the random orientation of rays. This is why we choose $N_{\rm kick} = N_{\rm ngb}$ instead.}. This means that the star will execute $2\, N_{\rm ngb}$ kicks, i.e. many gas neighbours will be kicked multiple times.

Analogously to equation (\ref{eq:N_heat}), the expected number of kick events over a star particle's lifetime is computed by dividing the total energy available for kinetic SN feedback, $f_{\rm kin} \, E_{\rm SN,tot}(m_*, f_{\rm E})$, by the energy of one kick event, $\Delta E_{\rm pair}(m_{\rm gas}, \Delta v_{\rm kick})$,
\begin{align}
    \langle N_{\rm kick,tot}\rangle &= \frac{f_{\rm kin} \, E_{\rm SN,tot}(m_*, f_{\rm E})}{\Delta E_{\rm pair}(m_{\rm gas}, \Delta v_{\rm kick})} \nonumber \\
    &= 23.7 \, f_{\rm E}\,  \left(\frac{f_{\rm kin}}{0.1}\right) \left(\frac{m_{\rm *}}{m_{\rm gas}}\right)\,\left(\frac{\Delta v_{\rm kick}}{50 \, \rm km \, s^{-1}}\right)^{-2} \, ,
    \label{eq:number_of_kicks}
\end{align}
where for simplicity we assumed that all gas particles have the mass $m_{\rm gas}$. For $f_{\rm kin} \gtrsim 0.1$ and relatively small kick velocities ($\Delta v_{\rm kick}\lesssim 10^2$ km s$^{-1}$), the number of kick events is $\gg 1$. In contrast, $f_{\rm kin} \approx 1$ and a desired kick velocity of $\Delta v_{\rm kick}\sim 10^3$ \kms{} yield $\langle N_{\rm kick,tot}\rangle \sim 1$, which energy-wise is similar to heating a gas particle by $\Delta T \sim 10^{7.5}$ K.

\subsubsection{Distributing the kick events among gas neighbours}
\label{paragraph: kin_algorithm}

Consider a star particle that has $N_{\rm ngb}$ gas neighbours inside its kernel and for which the number of kick events in the current time-step, $N_{\rm kick}$, is computed as described in $\S$\ref{sec:num_of_kick_events}. The energy of each kick event is proportional to $\Delta v_{\rm kick}^2$. To distribute the $N_{\rm kick}$ kick events among the neighbours, we use the following algorithm:

\begin{enumerate}
\item We cast $N_{\rm rays}$ rays in randomly chosen directions from the position of the star (i.e. the probabilities of spherical angular coordinates $\varphi$ and $\theta$, which define the direction, are uniform in $\varphi$ and $\cos\theta$).
   
\item For each ray from the previous step, we cast a new ray pointing in the opposite direction. After this step, we will have obtained two sets of $N_{\rm rays}$ rays.
      
\item For each ray $j$, we compute the great-circle distances between the ray and the gas neighbours $i$, $\Omega_{ij}$, and find the gas neighbour with the smallest $\Omega_{ij}$, which is carried out in the same way as in $\S$\ref{sec: distribute_heating_events}.

\item  If $N_{\rm kick}\leq N_{\rm rays}$, we randomly pick $N_{\rm kick}$ rays from the first set. For each of these rays, we select the ray in the second set pointing in the opposite direction. The gas particles attached to the two rays form a pair. We kick the two particles in the pair with the kick velocity given by equation (\ref{eq:v_kick_pair_beta}) along the directions given by equation (\ref{eq: kick_direction}). If, on the other hand, $N_{\rm kick} > N_{\rm rays}$, then we increase the kinetic energy per pair, $\Delta E_{\rm pair}$, by $N_{\rm kick}/N_{\rm rays}$ and kick the gas particles corresponding to all $2\times N_{\rm rays}$ rays using the updated value of $\Delta E_{\rm pair}$.

\end{enumerate}

The above algorithm is illustrated in Fig. \ref{fig: kin_feedback_illustration}. We emphasize that in our algorithm, great circle distances are calculated on a sphere of fixed (unit) radius, which guarantees that two rays pointing in opposite directions cannot correspond to the same gas particle.

\subsubsection{Kick collisions}
\label{paragraph:collisions}

\begin{figure}
    \centering
    \includegraphics[width=0.23\textwidth]{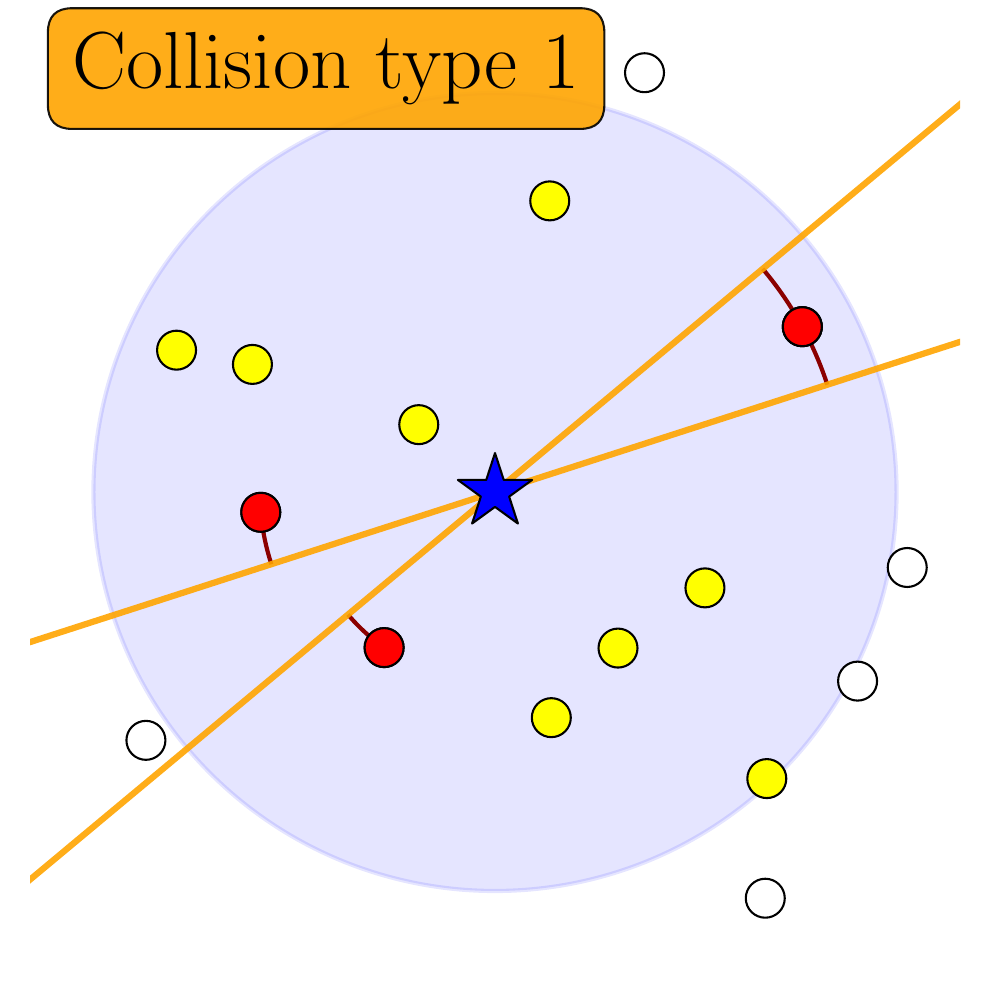}
    \includegraphics[width=0.23\textwidth]{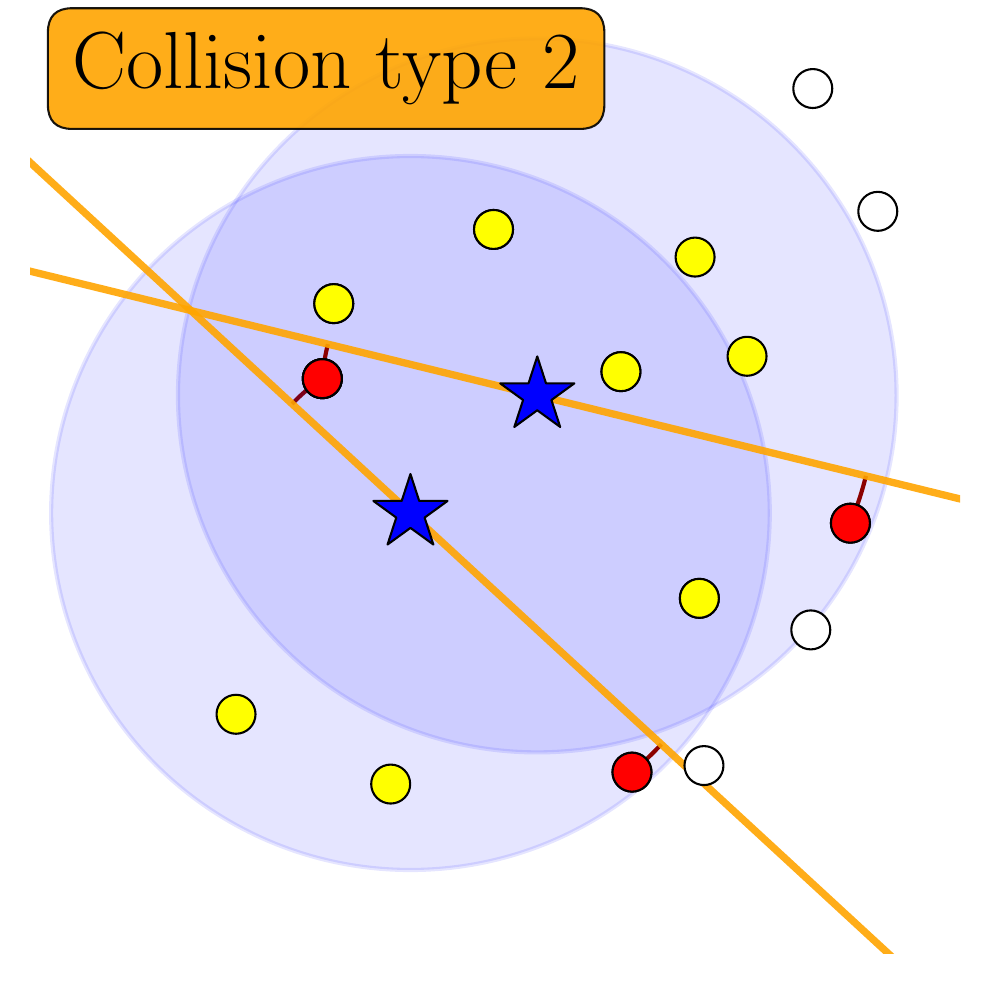}
    \caption{An illustration of two types of ray collisions in the kinetic channel of the prescription for SN feedback. Star particles are shown as blue stars and their kernels as shaded circular regions. Gas particles that are outside (inside) the kernel are shown in white (yellow). \textit{Left:} Collision type 1. In a given time-step, a single star particle has two kick events ($N_{\rm kick} = 2$). For each kick event, the star has two rays (orange lines) pointing in opposite directions from the star; the star kicks the two gas particles that are closest to these rays (red circles). In this example, rays from two independent kick events happen to share the same closest gas particle (the red circle in the top right part of the left-hand panel) resulting in a collision. \textit{Right:} Collision type 2. Two star particles with overlapping kernels want to kick the same gas particle. }
    \label{fig:collisions}
\end{figure}

We forbid any gas particle from being kicked more than once in a single time-step. This requirement is essential because otherwise our algorithm from $\S$\ref{parapgraph:kicks_in_pairs} may give wrong values to the actual kick velocities defined in equation (\ref{eq:v_kick_pair_beta}).

More precisely, all coefficients in $\S$\ref{parapgraph:kicks_in_pairs} are calculated based on the gas particle velocities at the beginning of a time-step; if a gas particle is kicked multiple times in the same time-step, then for the second and subsequent kicks, the particle velocity prior to the kick will be incorrect because the velocity has been updated due to the preceding kick(s). As a consequence, incorrectly calculated coefficients in $\S$\ref{parapgraph:kicks_in_pairs} will lead to a violation of energy conservation. Even worse, in extreme cases, gas particles might be accelerated to unrealistically high velocities. In Appendix \ref{sec:rel_motion_multi_kick} we discuss in detail the consequences of allowing gas particles to be kicked more than once in one time-step.

In our model, there are two possible types of collisions: \textit{type 1}, where a single star particle attempts to kick the same gas neighbour via more than one ray, and \textit{type 2}, where two or more star particles with overlapping kernels have chosen to kick the same gas particle. We illustrate both collision types in Fig. \ref{fig:collisions}. We avoid collisions as follows: 

\begin{enumerate}
    \item Each star particle carries a counter of its past kick events that have not been distributed due to a kick collision. Additionally, it carries an energy reservoir in which the kinetic energy from its undistributed kick events is (temporarily) stored.
    \item The energy in the reservoir, $E_{\rm reservoir}$, and the counter of the undistributed kick events, $N_{\rm kick,failed}$, are both initialised with zeros when a star particle is born.
    \item When a star particle tries to kick a pair of gas neighbours with the energy per kick event $\Delta E_{\rm pair}$ but fails to do so because of a collision, the energy $\Delta E_{\rm pair}$ is added to the reservoir and the counter $N_{\rm kick,failed}$ is incremented by one.
    \item Star particles attempt to distribute their previously undistributed kick events every time-step. More precisely, in a given time-step, a star particle first computes the number of kick events, $N_{\rm kick}$, as described in $\S$\ref{sec:num_of_kick_events}, each of which has the energy $\Delta E_{\rm pair}$. To account for the undistributed kick events from the past, we redefine $\Delta E_{\rm pair}$ as 
    $\Delta E_{\rm pair}^\prime \equiv (\Delta E_{\rm pair} N_{\rm kick} + E_{\rm reservoir})/N_{\rm kick}^\prime$ where $N_{\rm kick}^\prime \equiv N_{\rm kick} + N_{\rm kick,failed}$ if $p_{\rm kick,pair} < 1$, and $N_{\rm kick}^\prime \equiv N_{\rm kick}$ otherwise. After these redefinitions, we set $N_{\rm kick,failed}$ and $E_{\rm reservoir}$ back to zero, while $\Delta E_{\rm pair}^\prime$ and $N_{\rm kick}^\prime$ are then used in all steps in $\S$\ref{paragraph: kin_algorithm} instead of $\Delta E_{\rm pair}$ and $N_{\rm kick}$, respectively.
\end{enumerate}

In the cases of `colliding' kick events, to decide which events are successful and which are not, we give kick events different priorities:

\begin{itemize}
    \item In the case of a collision of type $2$, the kick event of the star particle with the largest $\Delta E_{\rm pair}$ is given priority\footnote{Two star particles may have different energies per kick event, $\Delta E_{\rm pair}$, if, for example, one star particle originally had the probability of kicking a pair of gas neighbours $p_{\rm kick,pair}>1$, so $\Delta E_{\rm pair}$ was increased to release all available energy in this time-step.}. If the star particles have equal  $\Delta E_{\rm pair}$, then the star particle with the larger internal id in the simulation has priority. 
    \item In order to avoid collisions of type $1$, we use the rays' internal indices: each pair of two anti-parallel rays (constructed from the two sets of rays) is labelled with an index taking values from $0$ to $N_{\rm rays}-1$. The pair of two gas neighbours corresponding to the pair of two anti-parallel rays with the ray's lowest internal index has priority.
\end{itemize}

We stress that since kicks can only happen in pairs, if either particle in the pair is not allowed to be kicked, then neither of the two particles in the pair is kicked.

\section{Numerical simulations}
\label{sec: numerical simulations}

\subsection{Code and setup}

We implemented our new SN feedback model in the SPH code \textsc{swift}\footnote{\textsc{swift} is publicly available at \url{http://www.swiftsim.com}.} \citep{2016pasc.conf....2S,2018ascl.soft05020S}. For the hydrodynamics solver, we use the energy-density SPH scheme \textsc{Sphenix} \citep{2022MNRAS.511.2367B}, which has been designed for next-generation cosmological simulations with \textsc{eagle}-like subgrid physics. We use the same parameters of the SPH scheme as in the original paper, including the quartic spline for the SPH kernel and the Courant–Friedrichs–Lewy (CFL) parameter $C_{\rm CFL}=0.2$, which limits time-steps of gas particles. Furthermore, we do not allow the ratio between time-steps of any two neighbouring gas particles to be greater than $4$ and use the \citet{2012MNRAS.419..465D} time-step limiter. Finally, every time a gas particle is kicked in SN kinetic feedback, we update its SPH signal velocity via
\begin{equation}
    v_{\mathrm{ sig, \, new}, i} = \max(2 \, c_{\mathrm{s},i},  v_{\mathrm{ sig, \, old}, i} + \beta_{\rm V} \, \Delta v) \, ,
\end{equation}
where $v_{\mathrm{ sig, \, old}, i}$ and $v_{\mathrm{ sig, \, new}, i}$ are the particle's signal velocity immediately before and after the kick, respectively, $c_{\mathrm{s},i}$ is the particle's speed of sound, $\Delta v$ is (the absolute value of) the actual kick velocity defined in equation (\ref{eq:v_kick_pair_beta}), and $\beta_{\rm V}$ is a dimensionless constant which in  \citet{2022MNRAS.511.2367B} is equal to 3 and we adopt the same value here.

The target particle smoothing length in our simulations is set to $1.2348$ times the local inter-particle separation, which for the quartic spline gives the expected number of gas neighbours in the kernel, $\langle N_{\rm ngb}\rangle \approx 65$.  

\subsection{Initial conditions}

We run simulations of a Milky Way-mass galaxy (H12) and a dwarf galaxy (H10); the initial conditions for both cases were generated  using the \textsc{makenewdisk} code \citep{springel2005} with the modifications introduced by Nobels et al. (in preparation). Our model for the H12 galaxy consists of a dark matter halo with an external \citet{hernquist1990} potential, a total mass $M_{200} = 1.37 \times 10^{12} \, \rm  M_{\odot}$, concentration $c=9$ (defined for a \citealt{1996ApJ...462..563N} equivalent halo), and spin parameter $\lambda=0.033$, where the dark-matter potential is analytic. Our model for the H10 galaxy uses the same functional form, but the total mass of the halo is $M_{200} = 1.37 \times 10^{10} \, \rm  M_{\odot}$ and the concentration is $c=14$. In both cases, the halo contains an exponential disc of stars and gas with total mass $M_{\rm disc} = 0.04 \, M_{200}$, and the initial gas fraction in the disc is set to $30$ per cent. For the H12 galaxy, the gas initially has solar metallicity, $Z_\odot=0.0134$ \citep{2009ARA&A..47..481A}, while for H10 it is 10 per cent of $Z_\odot$. The stellar scale height of the H12 galaxy is $\approx 0.43$ kpc, while for H10 it is $\approx 0.072$ kpc. In both galaxies, the scale height of the stellar disc is $10$ times smaller than the disc's radial scale length. The scale height of the gaseous component is set such that the gas stays in vertical hydrostatic equilibrium at the temperature of $T=10^4$ K.

In order to suppress the initial burst of star formation in the first $\approx 0.1$ Gyr of the simulations, which is the time that the galaxy needs to reach a quasi-equilibrium between stellar feedback and star formation, we assume that a fraction of star particles in the initial conditions (ICs) was formed within the last 100 Myr before the start of the simulation\footnote{The remaining star particles are assumed to have formed an infinitely long time ago, so they do not do any stellar feedback throughout the simulation.}. The stellar ages of these particles are sampled from a uniform distribution. We assume a constant SFR of $10\, \rm M_{\rm \odot}$ yr$^{-1}$ for the H12 galaxy and $0.01\, \rm M_{\rm \odot}$ yr$^{-1}$ for H10, which determines the total number of star particles with assigned stellar ages. The total mass of star particles with assigned stellar ages is approximately $2.6$ ($0.26$) per cent of the mass of all star particles in the ICs for the H12 (H10) galaxy. For H12, we assign a stellar age only to those star particles whose cylindrical distance (in the disc plane) is smaller than $10$ kpc from the galactic centre and whose height is within $1$ kpc from the disc midplane. For H10, we scale the values of $10$ kpc and $1$ kpc in proportion to the ratio of the virial radii of the two haloes. The particles that are assigned a stellar age are selected randomly from all stellar particles in the ICs satisfying the spatial criteria.

\subsection{Subgrid model for galaxy evolution}

\subsubsection{Radiative cooling and heating}
\label{sec:cooling}

The gas radiative cooling and heating rates are computed using the tables from \cite{2020MNRAS.497.4857P}\footnote{We use their fiducial version of the tables, \textsf{UVB\_dust1\_CR1\_G1\_shield1} (for the naming convention and more details we refer the reader to Table 5 in \citealt{2020MNRAS.497.4857P}). }, which were generated with the photoionization code \textsc{cloudy} \citep{Cloudy17}.  In the \cite{2020MNRAS.497.4857P} fiducial model the gas remains in ionization equilibrium in the presence of a modified version of the redshift-dependent, ultraviolet and X-ray background of \cite{2020MNRAS.493.1614F}, cosmic rays, and the local interstellar radiation field. The intensity of the latter two components is assumed to scale with the local Jeans column density as the star formation surface density in the KS law. The self-shielding column density is also assumed to scale like the local Jeans column density. We compute the cooling rates by interpolating the cooling tables over gas density and temperature at redshift $z=0$ and metallicity $Z=Z_\odot$ (Milky Way-mass galaxy) or $Z=0.1 \, Z_\odot$ (dwarf galaxy). The gas is allowed to cool down to 10 K and we do not use an effective pressure floor to model the ISM.

\begin{table*}
\caption{Numerical simulations used in this work. Column (1): the names of the simulations; column (2): $M_{200}$ is the total mass of the halo; column (3): $m_{\rm gas}$ is the gas particle mass; column (4): $\varepsilon_{\rm soft,gas}$ is the Plummer-equivalent gravitational softening length (for baryons); column (5): $f_{\rm kin}$ is the fraction of the SN energy injected in kinetic form (the remaining fraction, $1-f_{\rm kin}$, is injected in thermal form); column (6): $\Delta v_{\rm kick}$ is the desired kick velocity in SN kinetic feedback; column (7): other changes in the subgrid model relative to the fiducial set-up (see text for details). The fiducial simulations of the Milky Way-mass and dwarf galaxies are highlighted with a thicker font.}
	\centering
	\begin{tabular}{llrlrrl} 
	     & $M_{200}$ & $m_{\rm gas}$ & $\varepsilon_{\rm soft,gas}$ & $f_{\rm kin}$ & $\Delta v_{\rm kick}$ & Other variation(s) in subgrid \\
	     Simulation name & [$\rm M_\odot $] & [$\rm M_\odot $] & [kpc] &  & [km s$^{-1}$] & model (relative to fiducial case) \\
	 
	    \hline
	    \hline
&  & &  &  & &  \\	
\multicolumn{2}{l}{Variations in the fraction of SN energy released in kinetic form} & &  &  &  &  \\	  
	  \hline  
\textsf{H12\_M5\_fkin0p0} & $1.37 \times 10^{12}$ & $10^5$ & $0.2$  & $0.0$ & -- & -- \\
\textbf{\textsf{H12\_M5\_fkin0p1\_v0050}} & $\boldsymbol{1.37 \times 10^{12}}$ & $\boldsymbol{10^5}$ & $\boldsymbol{0.2}$  & $\boldsymbol{0.1}$ & $\boldsymbol{50}$ & \textbf{--} \\
\textsf{H12\_M5\_fkin0p3\_v0050} & $1.37 \times 10^{12}$ & $10^5$ & $0.2$  & $0.3$ & $50$ & -- \\
\textsf{H12\_M5\_fkin1p0\_v0050} & $1.37 \times 10^{12}$ & $10^5$ & $0.2$  & $1.0$ & $50$ & -- \\
\hline
\textsf{H10\_M3\_fkin0p0}        & $1.37 \times 10^{10}$ & $1.56 \times 10^3$ & $0.05$  & $0.0$ & -- & -- \\
\textbf{\textsf{H10\_M3\_fkin0p1\_v0050}} & $\boldsymbol{1.37 \times 10^{10}}$ & $\boldsymbol{1.56 \times 10^3}$ & $\boldsymbol{0.05}$  & $\boldsymbol{0.1}$ & $\boldsymbol{50}$ & \textbf{--} \\
\textsf{H10\_M3\_fkin0p3\_v0050} & $1.37 \times 10^{10}$ & $1.56 \times 10^3$ & $0.05$  & $0.3$ & $50$ & -- \\
\textsf{H10\_M3\_fkin1p0\_v0050} & $1.37 \times 10^{10}$ & $1.56 \times 10^3$ & $0.05$  & $1.0$ & $50$ & -- \\
&  & &  &  &  &  \\	
Variations in the desired kick velocity &  &  &  &  & &  \\	 
\hline
\textsf{H12\_M5\_fkin1p0\_v0010} & $1.37 \times 10^{12}$ & $10^5$ & $0.2$  & $1.0$ & $10$ & -- \\
\textsf{H12\_M5\_fkin1p0\_v0200} & $1.37 \times 10^{12}$  & $10^5$ & $0.2$  & $1.0$ & $200$ & -- \\
\textsf{H12\_M5\_fkin1p0\_v0600} & $1.37 \times 10^{12}$  & $10^5$ & $0.2$  & $1.0$ & $600$ & -- \\
\textsf{H12\_M5\_fkin1p0\_v1000} & $1.37 \times 10^{12}$  & $10^5$ & $0.2$  & $1.0$ & $1000$ & -- \\
&  & &  &  &  &  \\	
Variations in numerical resolution &  & &  &  &  &  \\	 
\hline
\textsf{H12\_M4\_fkin0p1\_v0050} & $1.37 \times 10^{12}$ & $1.25 \times 10^4$ & $0.1$  & $0.1$ & $50$ & -- \\
\textsf{H12\_M6\_fkin0p1\_v0050} & $1.37 \times 10^{12}$ & $0.80 \times 10^6$ & $0.4$  & $0.1$ & $50$ & -- \\
\textsf{H12\_M7\_fkin0p1\_v0050} & $1.37 \times 10^{12}$ & $0.64 \times 10^7$ & $0.8$  & $0.1$ & $50$ & -- \\
\textsf{H10\_M2\_fkin0p1\_v0050} & $1.37 \times 10^{10}$ & $1.95 \times 10^2$ & $0.025$  & $0.1$ & $50$ & -- \\
\textsf{H10\_M4\_fkin0p1\_v0050} & $1.37 \times 10^{10}$ & $1.25 \times 10^4$ & $0.1$  & $0.1$ & $50$ & -- \\
\textsf{H10\_M5\_fkin0p1\_v0050} & $1.37 \times 10^{10}$ & $10^5$ & $0.2$  & $0.1$ & $50$ & -- \\

&  & &  &  &  &  \\	
Other variations in the subgrid model &  &  &  &  & &  \\
\hline
\textsf{H12\_M5\_fkin0p0\_TempDens} & $1.37 \times 10^{12}$ & $10^5$ & $0.2$  & $0.0$ & $50$ & Temperature-density criterion\\
\textsf{H12\_M5\_fkin0p1\_v0050\_TempDens} & $1.37 \times 10^{12}$ & $10^5$ & $0.2$  & $0.1$ & $50$ & Temperature-density criterion \\
\textsf{H12\_M5\_fkin0p3\_v0050\_TempDens} & $1.37 \times 10^{12}$ & $10^5$ & $0.2$  & $0.3$ & $50$ & Temperature-density criterion \\
\textsf{H12\_M5\_fkin1p0\_v0050\_TempDens} & $1.37 \times 10^{12}$ & $10^5$ & $0.2$  & $1.0$ & $50$ & Temperature-density criterion \\
\hline

\textsf{H12\_M5\_fkin0p1\_v0050\_NoRelMotion} & $1.37 \times 10^{12}$  & $10^5$ & $0.2$  & $0.1$ & $50$ & No relative motion \\
\textsf{H12\_M5\_fkin0p1\_v0050\_NoRelMotion\_MulKicks} & $1.37 \times 10^{12}$ & $10^5$ & $0.2$  & $0.1$ & $50$ & No relative motion + multiple kicks\\
\textsf{H12\_M5\_fkin1p0\_v0050\_NoRelMotion} & $1.37 \times 10^{12}$ &  $10^5$ & $0.2$  & $1.0$ & $50$ & No relative motion \\
\textsf{H12\_M5\_fkin1p0\_v0050\_NoRelMotion\_MulKicks} & $1.37 \times 10^{12}$ & $10^5$ & $0.2$  & $1.0$ & $50$ & No relative motion + multiple kicks\\
\hline

\end{tabular}
\label{tab:runs}
\end{table*}

\subsubsection{Star formation}
\label{sec: star_formation}

To decide whether a gas particle is star-forming or not, we use a gravitational instability criterion (Nobels et al., in preparation). Briefly, the gas is allowed to form stars when it is locally unstable against gravitational collapse. Mathematically, this condition can be expressed by requiring that the kinetic energy of a gas element due to its thermal motion and turbulent motion is smaller than its (absolute) gravitational binding energy, 
\begin{equation}
     \alpha \equiv \frac{\sigma_{\rm 3D,turb}^2 + \sigma_{\rm th}^2}{G \rho^{1/3} \langle m_{\rm ngb}\rangle ^{2/3}} < 1 \, ,
     \label{eq:virial_crit}
\end{equation}
where $G$ is the gravitational constant, $\rho$ is the mass density of the gas element, $\langle m_{\rm ngb}\rangle$ is the average mass in the kernel of the gas element, $\sigma_{\rm 3D,turb}$ is the three-dimensional turbulent velocity dispersion, and $\sigma_{\rm th}$ is the gas thermal velocity dispersion. We compute $\langle m_{\rm ngb}\rangle $ as $\langle m_{\rm ngb}\rangle = \langle N_{\rm ngb} \rangle \, m_{\rm gas}$ where $m_{\rm gas}$ is the mass of the gas element to which the star formation criterion is applied and $\langle N_{\rm ngb} \rangle \approx 65$. The thermal velocity dispersion $\sigma_{\rm th}$ is computed as

\begin{equation}
    \sigma_{\rm th} = \sqrt{\frac{3P}{\rho}}\, ,
\end{equation}
where $P$ is the pressure of the gas element. Finally, the turbulent velocity dispersion $\sigma_{\rm 3D,turb}$ of the gas element (which for clarity we label below with an index $i$ to distinguish it from its neighbours $j$) is given by
\begin{equation}
    \sigma_{\mathrm{3D,turb},i}^2 = \frac{1}{\rho_i} \sum_j \, m_{j} |\boldsymbol{v}_i - \boldsymbol{v}_j|^2 \, W(r_{ij}, h_i) \, ,
\label{eq:v_sigma}
\end{equation}
where the sum is computed over the gas neighbours of gas element $i$, $\boldsymbol{v}_j$ are their peculiar velocities, $r_{ij}$ is the distance between gas elements $i$ and $j$, and $W$ is the SPH kernel function centred on gas particle $i$, which has a smoothing length $h_i$.

If a gas element satisfies the gravitational instability criterion ($\alpha < 1$), it is star-forming. The process of star formation occurs stochastically: we first compute the SFR of the gas element $\dot m_{\rm sf}$ following the \citet{1959ApJ...129..243S} law 
\begin{equation}
\dot m_{\rm sf} = \varepsilon \frac{m_{\rm gas}}{t_{\rm ff}}\, ,
\end{equation}
where $t_{\rm ff} = [3\uppi/(32 G \rho)]^{1/2}$ is the free-fall time-scale and $\varepsilon = 0.01$ is the star formation efficiency on this time-scale. We then compute the probability that this gas element will become a star particle, $p_{\rm sf}$, which is  realised by multiplying $\dot m_{\rm sf}$ by the element's current time-step and dividing by its current mass $m_{\rm gas}$. 

In addition to the default, gravitational instability criterion, as one of the variations in our subgrid model we consider a temperature-density criterion for star formation in which a gas element is star-forming if its hydrogen number density is higher than $10^2$ cm$^{-3}$ or its temperature is lower than $10^3$ K. The remaining steps, including the computation of the gas element's SFR, are the same as in the fiducial scenario with the gravitational instability criterion. In Appendix \ref{sec:sf_crit} we show that our conclusions are insensitive to the choice of star formation criterion.

\subsubsection{Early stellar feedback}

We model several early stellar feedback processes, all of which are subdominant to the feedback from SNe. Our stellar-evolution model for early feedback processes uses the Binary Population and
Spectral Synthesis (\textsc{bpass})
tables \citep{BPASS2017,BPASS2018} version 2.2.1 with a \citet{Chabrier2003} IMF whose minimum and maximum stellar masses are $0.1 \, \rm M_\odot$ and $100 \, \rm M_\odot$, respectively. The early feedback processes we include are stellar winds, radiation pressure and H\textsc{ii} regions. 

The implementation and effects of these three early feedback processes will be described in detail in Ploeckinger et al. (in preparation). Briefly, in the stellar-wind feedback, star particles stochastically kick their gas neighbours with a fixed\footnote{Here we do not apply our energy-conserving algorithm developed for SN kinetic feedback in \S\ref{paragraph:stochastic_kinetic_feedback} because we expect the stellar early feedback to be momentum-driven. That is, the stellar early feedback does not drive sufficiently fast winds to include an energy-driven phase.} kick velocity of $50$ \kms{} using the cumulative momentum provided by the \textsc{bpass} tables. The feedback from radiation pressure is based on the photon energy spectrum, which also comes from the \textsc{bpass} tables. To compute the photon momentum exerted onto the gas, we use the optical depth from \citet{2020MNRAS.497.4857P}, which is derived from the local Jeans column density. The radiation pressure feedback is also stochastic and uses a fixed kick velocity of $50$ \kms. Finally, young star particles stochastically ionize and heat the surrounding gas to a temperature $T = 10^{4}$ K, following a Str{\"o}mgren sphere approximation. The probability of becoming an H\textsc{ii} region is a function of the gas density and \textsc{bpass} ionizing photon flux. A gas particle becomes an H\textsc{ii} region and during this time it is not allowed to be star-forming even if it satisfies the star formation criterion. A new set of gas particles is selected as H\textsc{ii} regions every 2 Myr, as long as the conditions for the Str{\"o}mgren sphere are fulfilled.

\subsubsection{Supernova feedback}

The feedback from core-collapse SNe uses the stochastic model introduced and described in $\S$\ref{sec:methods}. We adopt an SN energy in units of  $10^{51}$ erg, $f_{\rm E}=2$, and a heating temperature of $\Delta  T = 10^{7.5}$ K. The values of these two parameters are chosen such that the thermal feedback remains efficient at most gas densities reached in our simulations and that stellar particles have on average at least one thermal injection event in their lifetime (variations in $\Delta  T$ and $f_{\rm E}$ have been investigated in e.g. \citetalias{2012MNRAS.426..140D} and \citealt{Crain2015}). The other two parameters, the fraction of the energy released in kinetic form, $f_{\rm kin}$, and the desired kick velocity, $\Delta v_{\rm kick}$, take a range of values in our simulations, with their fiducial values set to $f_{\rm kin}=0.1$ and $\Delta v_{\rm kick}=50$ \kms. 

Unless stated otherwise, the kinetic channel follows the algorithm detailed in $\S$\ref{paragraph:stochastic_kinetic_feedback} including the corrections introduced in $\S$\ref{paragraph:collisions} to avoid collisions of kick events. For both the kinetic and thermal channels, we set the number of rays per star particle $N_{\rm rays}=8$. This means that the maximum number of gas neighbours a star particle can kick in a single time-step is equal to $2 \, N_{\rm rays}= 16$; while the maximum number of heated particles per time-step is equal to $8$. 

We explore two variations in the kinetic channel of the SN feedback model that differ from the fiducial algorithm described in $\S$\ref{paragraph:stochastic_kinetic_feedback}. In the first variation, we neglect the correction due to the relative motion of gas around stars, which is done by setting $\beta=1$ in equation (\ref{eq:v_kick_pair_beta}). In the second variation, we additionally allow gas particles to be kicked more than once in a single time-step, which is done by not applying the corrections introduced in $\S$\ref{paragraph:collisions} to the steps in $\S$\ref{paragraph: kin_algorithm}. The impact of these two variations is studied in Appendix \ref{sec:rel_motion_multi_kick}.

In addition to core-collapse SN feedback, we include type-Ia SN feedback. Because, energy-wise, it is subdominant to core-collapse SN feedback, for simplicity we implement type-Ia SN feedback as a purely thermal ($f_{\rm kin}=0$) isotropic stochastic feedback following the algorithm from  $\S$\ref{paragraph: thermal model}. As for core-collapse SN feedback, the heating temperature in the type-Ia SN feedback is $\Delta  T = 10^{7.5}$ K and the maximum number of rays $N_{\rm rays}=8$. To evaluate the type-Ia energy budget per star particle, we use a delay time distribution (DTD),
\begin{equation}
    \mathrm{DTD}(t) = \frac{\nu}{\tau} \exp\left(-\frac{t-t_{\rm delay}}{\tau}\right) \Theta(t-t_{\rm delay}) \, , 
    \label{eq:dtd}
\end{equation}
where $\nu = 2 \times 10^{-3} \, \rm M^{-1}_{\odot}$ is the SNIa efficiency, $\tau = 2$ Gyr is the SNIa time-scale and $t_{\rm delay} = 40$ Myr is the time (delay) between the birth of a star particle and when type-Ia SNe are first allowed to go off. To compute the energy from all individual type-Ia SNe in a time-step $[t,t+\Delta t)$, we integrate equation (\ref{eq:dtd}) from $t$ to $t+\Delta t$ and use an energy per type-Ia SN of $10^{51}$ erg. This total SNIa energy is then used in equation (\ref{eq:prob_heat}) to calculate the heating probability in the type-Ia SN feedback. We emphasize that type-Ia SN feedback is always subdominant to core-collapse SN feedback, that the model for type-Ia SN feedback is not varied in our simulations, and that all further discussion refers entirely to core-collapse SN feedback.

Lastly, we note that for simplicity we do not include metal enrichment from stars and our galaxies do not contain supermassive black holes. 

\subsection{Runs}

The names of the simulations with the $M_{\rm 200} = 1.37 \times 10^{12} \, \rm M_\odot$ halo begin with \textsf{H12} and the names of the runs with the $M_{\rm 200} = 1.37 \times 10^{10} \, \rm M_\odot$ halo start with \textsf{H10}. The simulation resolutions with the gas particle mass of $m_{\rm gas} = 1.95 \times 10^2\, \rm M_\odot$\footnote{Note that at our highest resolution, $m_{\rm gas} = 1.95 \times 10^2\, \rm M_\odot$, representing a stellar population by simply integrating the IMF is not entirely correct because $\sim 100 \, \rm M_\odot$-mass stars are nearly as massive as stellar particles themselves. However, this does not affect our conclusions because we explore this resolution only as part of the convergence test (see $\S$\ref{sec: resolution}).}, $1.56 \times 10^3  \, \rm M_\odot$, $1.25 \times 10^{4} \, \rm M_\odot$, $10^{5} \, \rm M_\odot$, $0.80 \times 10^{6} \, \rm M_\odot$, and  $0.64 \times 10^{7} \, \rm M_\odot$ are denoted \textsf{M2}, \textsf{M3}, \textsf{M4}, \textsf{M5}, \textsf{M6}, and \textsf{M7}, respectively. Our fiducial resolution for the Milky Way-mass (dwarf) galaxy is M5 (M3). Additionally, in the simulation names we use key words \textsf{fkinXpX} and \textsf{vXXXX} (where `X' is a 0-9 digit) to indicate the fraction of SN energy that is injected in kinetic form $f_{\rm kin}$ and the desired kick velocity $\Delta v_{\rm kick}$, respectively. Names of runs with purely thermal SN feedback do not contain the \textsf{vXXXX} suffix. 

The names of the runs that do not account for the relative star-gas motion have the suffix \textsf{NoRelMotion}. If gas particles are allowed to be kicked more than once in a single time-step in the kinetic feedback, then the name has the suffix \textsf{MulKicks}. The runs using the temperature-density criterion for star formation have the suffix \textsf{TempDens}; if this suffix is not present, the run uses the default, gravitational instability criterion. All simulations in this work were run for 1 Gyr and are summarised in Table \ref{tab:runs}.
 
\section{Results}
\label{sec: results}

\begin{figure}
    \centering
      \includegraphics[width=0.47\textwidth]{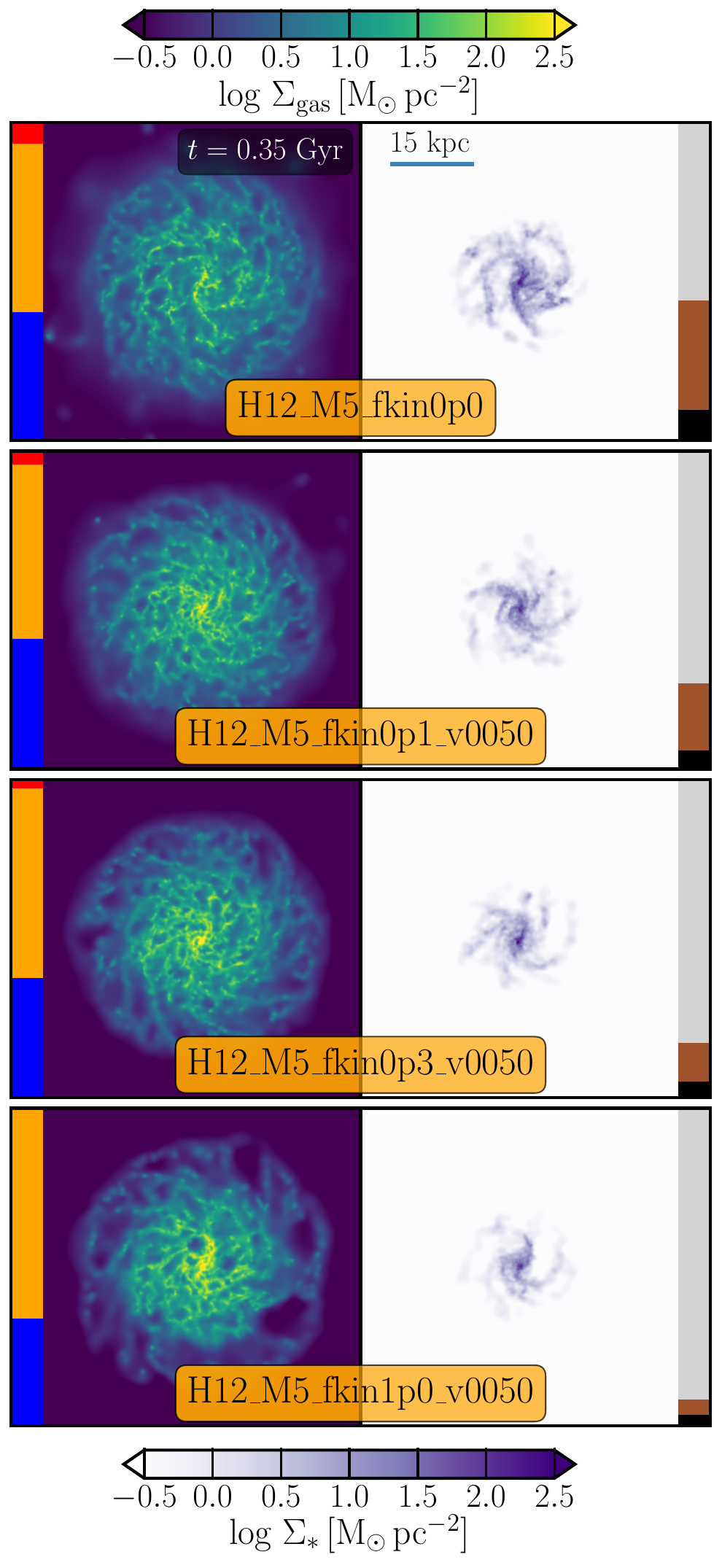}
    \caption{The H12 galaxy with M5 resolution at time $t=0.35$ Gyr shown face-on. The colour scale indicates the mass surface density of the gas (\textit{left}) and of the newly born stars (\textit{right}). The fraction of SN energy injected in kinetic form increases from top to bottom: $f_{\rm kin} = 0, 0.1, 0.3, 1$. The desired kick velocity is set to $\Delta v_{\rm kick} = 50$ \kms{} in the cases with $f_{\rm kin}>0$.  In each row, the left-hand colour bar shows the mass fractions of the hot ($T>10^5$ K, red), warm ($10^3<T\leq10^5$ K, orange), and cold ($T\leq10^3$ K, blue) gas whose height is $< 5 \, h_z$ where $h_z\approx 0.43$ kpc is the initial stellar scale height in the H12 galaxy. For visualisation purposes, in the colour bar the mass of the hot gas has been increased by a factor of 10. The right-hand colour bar shows the mass fractions of all gas at heights  $< 5 \, h_z$ (light-grey), $\geq 5 \, h_z$ (brown), and the gas that has turned into stars by $t=0.35$ Gyr (black). Higher values of $f_{\rm kin}$ lead to higher surface densities in the centre of the galaxy, less gas in the hot phase, less gas outside the stellar disc, and overall fewer stars formed by $t=0.35$ Gyr.}
    \label{fig:morphology_M12_main}
\end{figure}

 In section \ref{sec: vary_fkin}, we vary the fraction of SN energy injected in kinetic form, $f_{\rm kin}$, between $0$ and $1$ while keeping the other parameters fixed to their fiducial values, to see how the galaxy responds to the two different types of energy injection. In our fiducial model for SN feedback, the energy released in one thermal injection event ($\Delta T = 10^{7.5}$ K) is more than two orders of magnitude greater than the energy in one kick event ($\Delta v_{\rm kick} = 50$ \kms). Next, in Section \ref{sec: vary_vkick}, we vary the desired kick velocity, $\Delta v_{\rm kick}$, between $10$ and $10^3$ \kms{} to study and quantify the differences between kinetic SN feedback using low- and high-energy kicks. In order to maximize the differences in galaxy properties between the models with different $\Delta v_{\rm kick}$, we consider solely fully kinetic models ($f_{\rm kin}=1$). Finally, in Section \ref{sec: resolution}, we investigate numerical convergence for our fiducial model.

\subsection{Variations in the fraction of kinetic energy}
\label{sec: vary_fkin}

The figures in this section are shown at time $t=0.35$ Gyr or use a time interval centred around $t=0.35$ Gyr (from $0.2$ to $0.5$ Gyr), which is the moment in time when the models with different $f_{\rm kin}$ have comparable total SFRs (see $\S$\ref{sec: sfh}). This is done in order to ensure that the differences in the galaxy properties are due to the different $f_{\rm kin}$, and are not a mere consequence of different total SFRs.

\subsubsection{Morphology}

Fig. \ref{fig:morphology_M12_main} displays the mass surface density of the gas (left column) and stars (right column) in the simulations of the H12 halo with M5 resolution at time $t=0.35$ Gyr. The galaxies are viewed face-on. We only include the star particles that were born during the simulation (i.e. the star particles that are not part of the ICs). We take the fiducial model, \textsf{H12\_M5\_fkin0p1\_v0050}, and study the effects of varying the fraction of SN energy  injected in kinetic form, $f_{\rm kin}$. The other parameters, including the desired kick velocity, $\Delta v_{\rm kick} = 50$ \kms{}, are kept fixed. Each row in the figure corresponds to a different value of $f_{\rm kin}$ (from top to bottom, $f_{\rm kin}=0, 0.1, 0.3, 1$). 

The left-hand colour bar in each row shows the mass fractions of the hot ($T>10^5$ K, red), warm ($10^3<T\leq10^5$ K, orange), and cold ($T\leq10^3$ K, blue) phases computed for the gas whose scale height is less than $5 \, h_z$, where $h_z\approx 0.43$ kpc is the initial scale height of the stellar disc in the H12 galaxy. For visualisation purposes, the mass fraction of the hot gas in the colour bar has been increased by a factor of 10. The colour bar on the right shows the mass fractions of all gas located at heights $<5 \, h_z$ (light-grey), located at heights $\geq 5 \, h_z$  (brown), and the gas that has turned into stars during $0<t<0.35$ Gyr (black).

There are three points that can be taken from Fig. \ref{fig:morphology_M12_main}. First, the gas distribution is less centrally concentrated for lower values of $f_{\rm kin}$, which is especially evident in the simulation with the purely thermal SN feedback (\textsf{H12\_M5\_fkin0p0}, $f_{\rm kin} = 0$). The reason for this behaviour is that unlike the kinetic feedback with low-energy kicks, large thermal-energy injections in the thermal feedback are capable of blowing superbubbles and responsible for the launching of vigorous galactic winds that evacuate the gas from the ISM. These superbubbles constitute the hot ISM and are visible in the figure as low-density regions in an otherwise high gas surface density. The mass (fraction) of this hot gas decreases with $f_{\rm kin}$, and for $f_{\rm kin}=1$ disappears nearly completely, as indicated by the left-hand colour bars. The production of superbubbles is further enhanced in the models with less energy available for the kinetic channel because if the SN kicks occur more rarely (due to lower $f_{\rm kin}$), then the impact of young stars on their local ISM is delayed, making it more likely for stars to form in clusters, which in turn increases the probability that bubbles from individual SN events will overlap and form superbubbles. 

Second, in the purely kinetic model (\textsf{H12\_M5\_fkin1p0\_v0050}, $f_{\rm kin}=1$), the gas is overall more dense and compact -- especially close to the galactic centre -- compared to the models including thermal feedback. The purely kinetic model clearly suffers from not being able to eject the gas from the galactic disc to large distances. Instead, a large fraction of the gas accumulates around the galactic centre resulting in high surface densities ($\Sigma_{\rm gas} \gtrsim 10^{2.5} \, \rm M_\odot \, pc^{-2}$). The cases with $f_{\rm kin}=0.1$ (\textsf{H12\_M5\_fkin0p1\_v0050}) and $f_{\rm kin}=0.3$ (\textsf{H12\_M5\_fkin0p3\_v0050}) exhibit properties that are intermediate between those found in the purely thermal and kinetic models.

Third, the distributions of the stellar component show a somewhat different trend: in the purely thermal model, by the time $t=0.35$ Gyr the galaxy has formed notably more stars than in the model with $f_{\rm kin}=1$. The lower the value of $f_{\rm kin}$, the more stars the galaxy has formed. In addition, for lower values of $f_{\rm kin}$, the star formation extends to larger radii in the galactic plane. However, these variations in stellar mass are not the main cause of the differences in the gas surface densities seen in the left column of the figure. Instead, the gas surface densities are highly sensitive to the ability of the models to push the gas out of the ISM through a strong galactic wind. This can be inferred from the right-hand colour bars, which show that the gas mass fraction at heights $\geq 5 \, h_z$ strongly decreases with $f_{\rm kin}$, while the changes in the stellar mass remain subdominant. We will present a quantitative analysis of the mass loading of galactic winds in $\S$\ref{sec:wind_mass_loading}.

\begin{figure}
    \centering
    \includegraphics[width=0.47\textwidth]{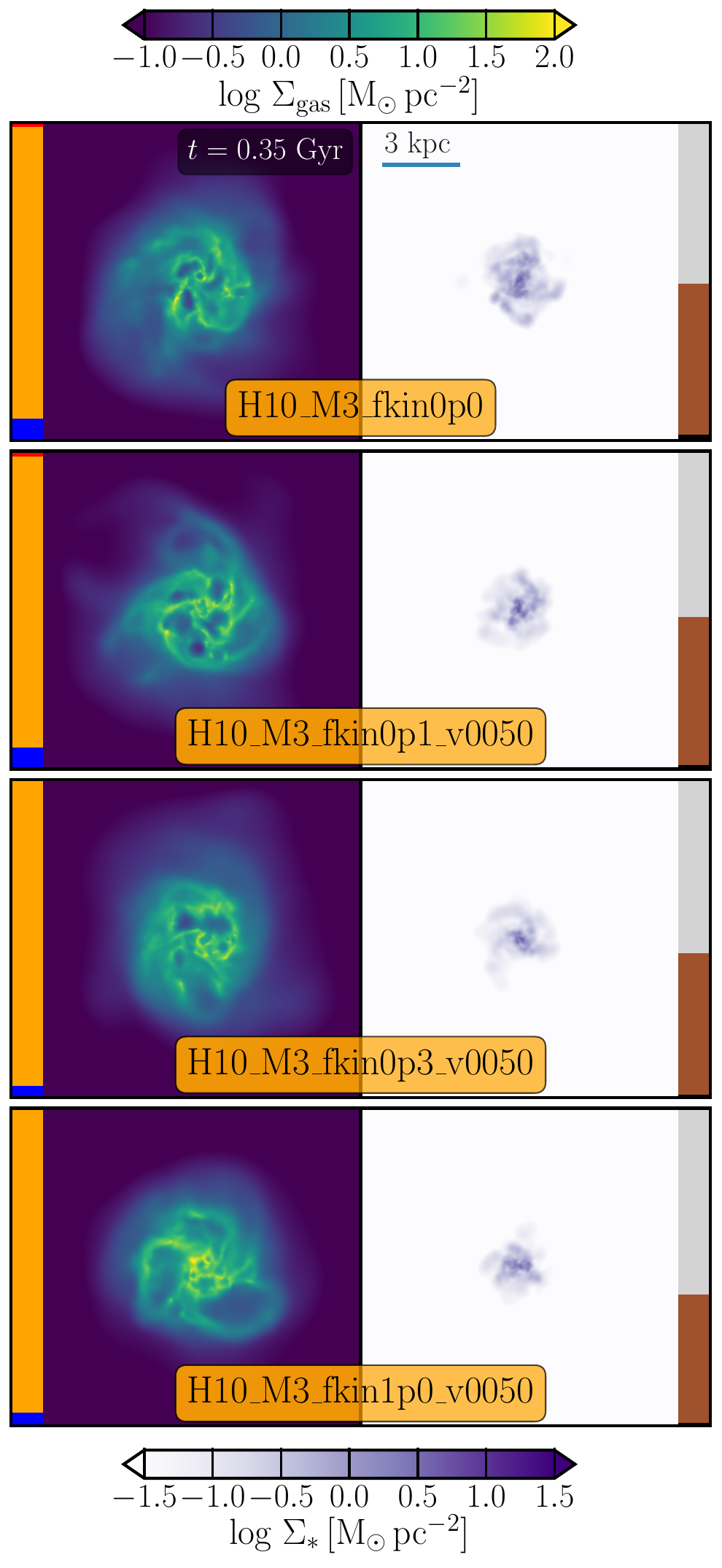}
    \caption{As Fig. \ref{fig:morphology_M12_main} but for the H10 galaxy with M3 resolution. Note the different spatial and colour scales. The distributions of gas (\textit{left}) and stellar (\textit{right}) mass surface density show the same trends with $f_{\rm kin}$ as in Fig. \ref{fig:morphology_M12_main}, albeit to a much smaller degree. }
    \label{fig:morphology_M10_main}
\end{figure}

Fig. \ref{fig:morphology_M10_main} shows the models with the same variations in $f_{\rm kin}$ as in Fig. \ref{fig:morphology_M12_main} but for the H10 galaxy with M3 resolution. In this case, the mass surface densities of both gas and stars vary much less with $f_{\rm kin}$ than for the Milky Way-mass object. None the less, the trends seen in Fig. \ref{fig:morphology_M12_main} remain: smaller values of $f_{\rm kin}$ lead to higher stellar surface densities and lower gas surface densities. One substantial difference between the H10 and H12 galaxies is that the mass fraction of the cold gas is significantly lower in H10. This is because the ISM of the dwarf galaxy has $10$-times lower metallicity and possesses less dense gas. 

\subsubsection{Gas surface density profiles}

\begin{figure}
    \centering
    \includegraphics[width=0.45\textwidth]{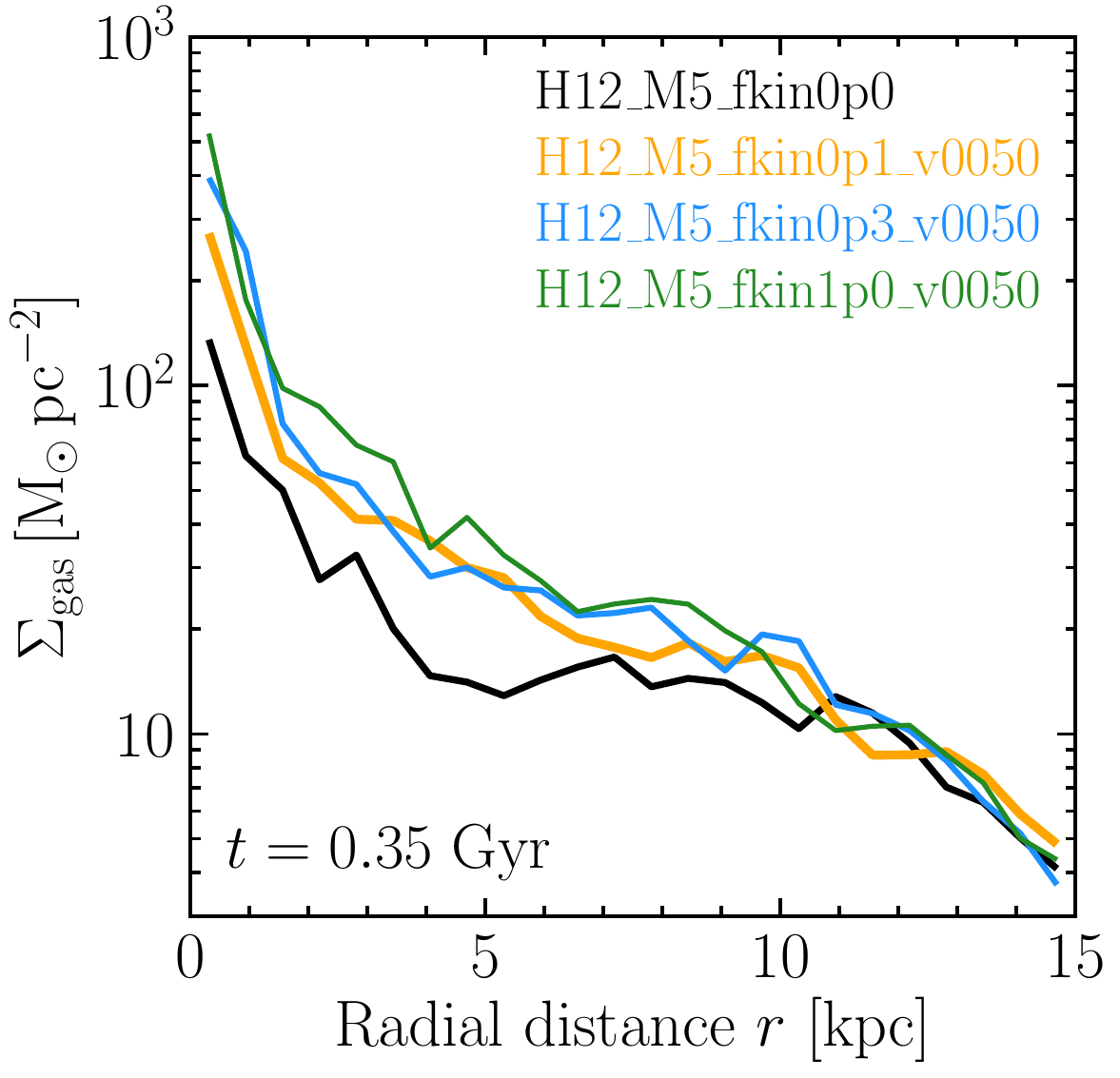}
    \caption{Radial profiles of the gas surface density in the galactic plane, shown for the H12 galaxy with M5 resolution at time $t=0.35$ Gyr for four values of the fraction of kinetic energy in SN feedback, $f_{\rm kin} = 0$ (black), $0.1$ (orange), $0.3$ (blue), and $1$ (green). The desired kick velocity is $\Delta v_{\rm kick} = 50$ \kms{} in the runs with kinetic feedback. On average, increasing $f_{\rm kin}$ leads to a higher gas surface density and a steeper radial profile.}
    \label{fig:surface_density_M12_main}
\end{figure}

Fig. \ref{fig:surface_density_M12_main} displays radial profiles of the gas surface density at time $t=0.35$ Gyr in the simulations with different $f_{\rm kin}$, for the H12 halo with M5 resolution. The galaxy is viewed face-on and the profiles are computed in radial bins of equal size. We show four runs that use $f_{\rm kin}$: $0$ (black), $0.1$ (orange), $0.3$ (blue), and $1$ (green). The desired kick velocity is set to $\Delta v_{\rm kick} = 50$ \kms{} in all cases where $f_{\rm kin}>0$.

The galaxies on average have higher gas surface densities in the models with higher $f_{\rm kin}$ and the profiles become steeper as we increase $f_{\rm kin}$. Near the galactic centre, the differences in the gas surface density between the models with $f_{\rm kin}=1$ and $f_{\rm kin}=0$ reach approximately $0.5$ dex. At large radii ($r \gtrsim 11$ kpc), the gas surface densities in all models converge. The gas mass density profiles in the H10 galaxies exhibit similar behaviour with $f_{\rm kin}$ as for the H12 case, but the variations are much smaller than for H12 (not shown).

We emphasize that by varying $f_{\rm kin}$ we change the energy in both the thermal and kinetic channels of our SN feedback model. By running additional tests, we verified that the large differences between the models with $f_{\rm kin}=0$ and $0.1$ in Fig. \ref{fig:surface_density_M12_main} are due to the presence of SN kinetic feedback in the latter case, while the reduction of the energy in the thermal channel by 10 per cent has little effect.

\subsubsection{Star formation history}
\label{sec: sfh}

\begin{figure}
    \centering
    \includegraphics[width=0.46\textwidth]{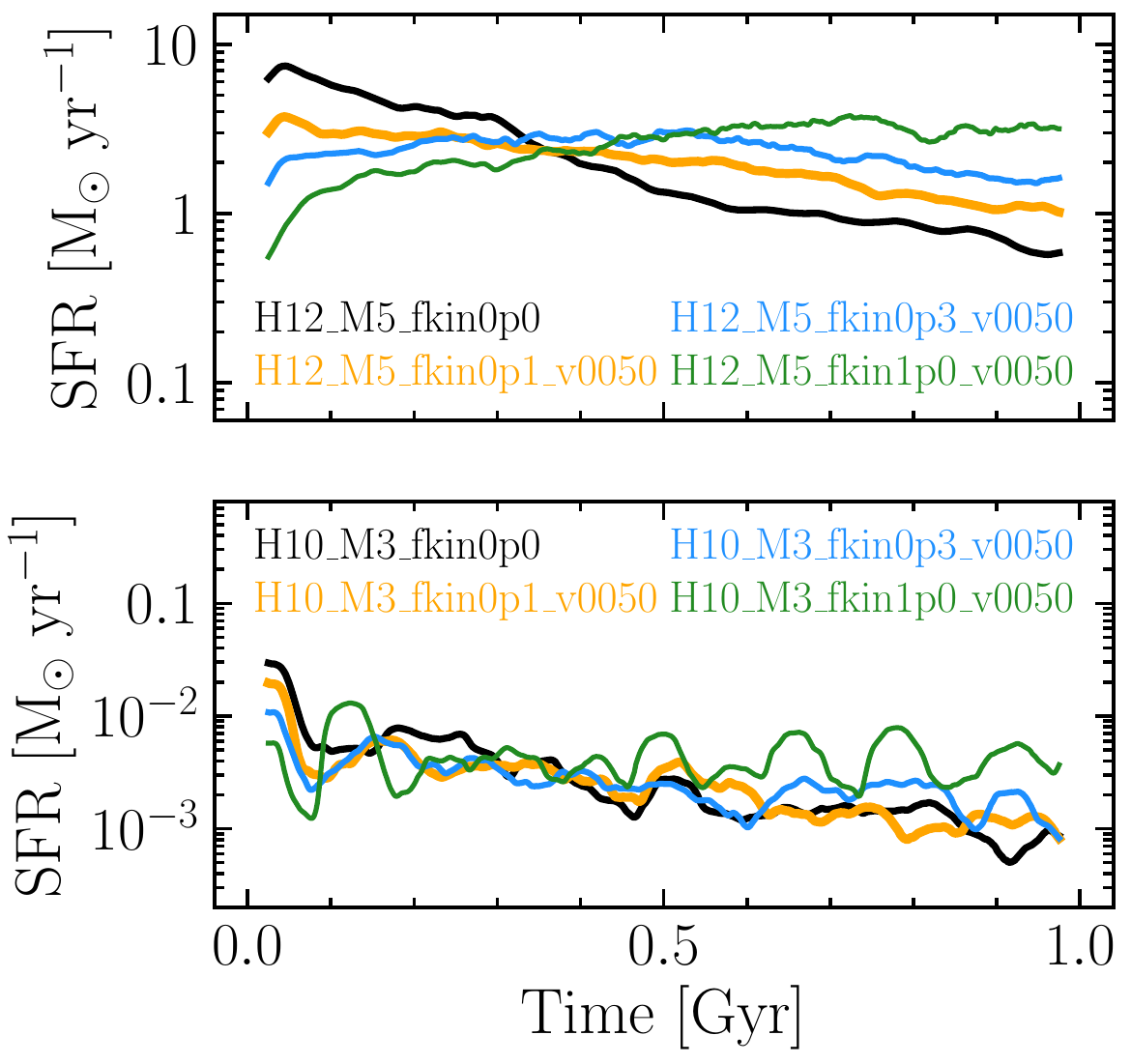}
      \caption{Star formation rates versus time in the H12 galaxy with M5 resolution (\textit{top panel}) and the H10 galaxy with M3 resolution (\textit{bottom panel}) for four values of $f_{\rm kin}$ (colour-coded). The desired kick velocity is set to $\Delta v_{\rm kick} = 50$ \kms{} in the simulations with $f_{\rm kin}>0$. Higher $f_{\rm kin}$ suppresses the initial burst of star formation more efficiently but leads to higher SFRs at late times. The H12 galaxy is more sensitive to the variations in $f_{\rm kin}$ than the H10 galaxy is.}
    \label{fig:SFH_main}
\end{figure}

In Fig. \ref{fig:SFH_main}, we compare star formation histories in the models with different $f_{\rm kin}$ for the H12 halo with M5 resolution (top) and the H10 halo with M3 resolution (bottom). As in the previous figures, we show the results for $f_{\rm kin} = 0$, $0.1$, $0.3$, and $1$, and the desired kick velocity is set to $\Delta v_{\rm kick} = 50$ \kms{} in all cases where $f_{\rm kin}>0$. Every time-step our simulations record the galaxy's total SFR by summing up the contributions of all star-forming gas particles. These time-steps can become small ($\lesssim 10$ kyr), which introduces a noise component into the time evolution of the galaxy's SFR. To reduce this noise, in the figure we show the moving average over 50 Myr.

In the H12 halo, increasing the fraction of kinetic energy suppresses the initial burst of star formation and leads to a more steady SFR over the remaining time, with a higher SFR reached by the end of the simulation. By the time $t = 1$ Gyr, the SFR in the galaxy with purely kinetic feedback ($f_{\rm kin}=1$) is a factor of $5$ higher than in the galaxy with purely thermal feedback ($f_{\rm kin}=0$). This is mainly because in the models with higher $f_{\rm kin}$, the galaxies are able to retain more gas in their ISM at later times (Figs. \ref{fig:morphology_M12_main} and \ref{fig:surface_density_M12_main}), resulting in higher SFRs. More precisely, at $t=1$ Gyr, the galaxy with purely thermal feedback has approximately $36$ per cent of the initial gas mass remaining at heights $< 5 h_z$, while for the purely kinetic run this number is twice as large. As expected, the SFRs in the models with intermediate values of $f_{\rm kin}$ are bracketed by the SFRs in the purely thermal and kinetic models.

The SFRs in the dwarf galaxy with different SN models stay mostly within the range $10^{-3} - 10^{-2} \, \rm M_\odot \, yr^{-1}$. These values are more than two orders of magnitudes lower than those in the H12 halo. The same effect of changing $f_{\rm kin}$ can also be noticed in the H10 halo, though it is strongly suppressed compared to the H12 halo, and is obscured by the large fluctuations in the SFRs. 

\subsubsection{Velocity dispersion and its impact on star formation}
\label{sec:vel_disp_and_sfr}

In this section, we investigate the connection between low-energy kicks in the SN kinetic feedback and the gas turbulent velocity dispersion in the ISM. Because we use the gravitational instability criterion as our default criterion for star formation, the velocity dispersion of the gas plays an important role in shaping the conditions required for star formation.

\begin{figure}
    \centering
    \includegraphics[width=0.45\textwidth]{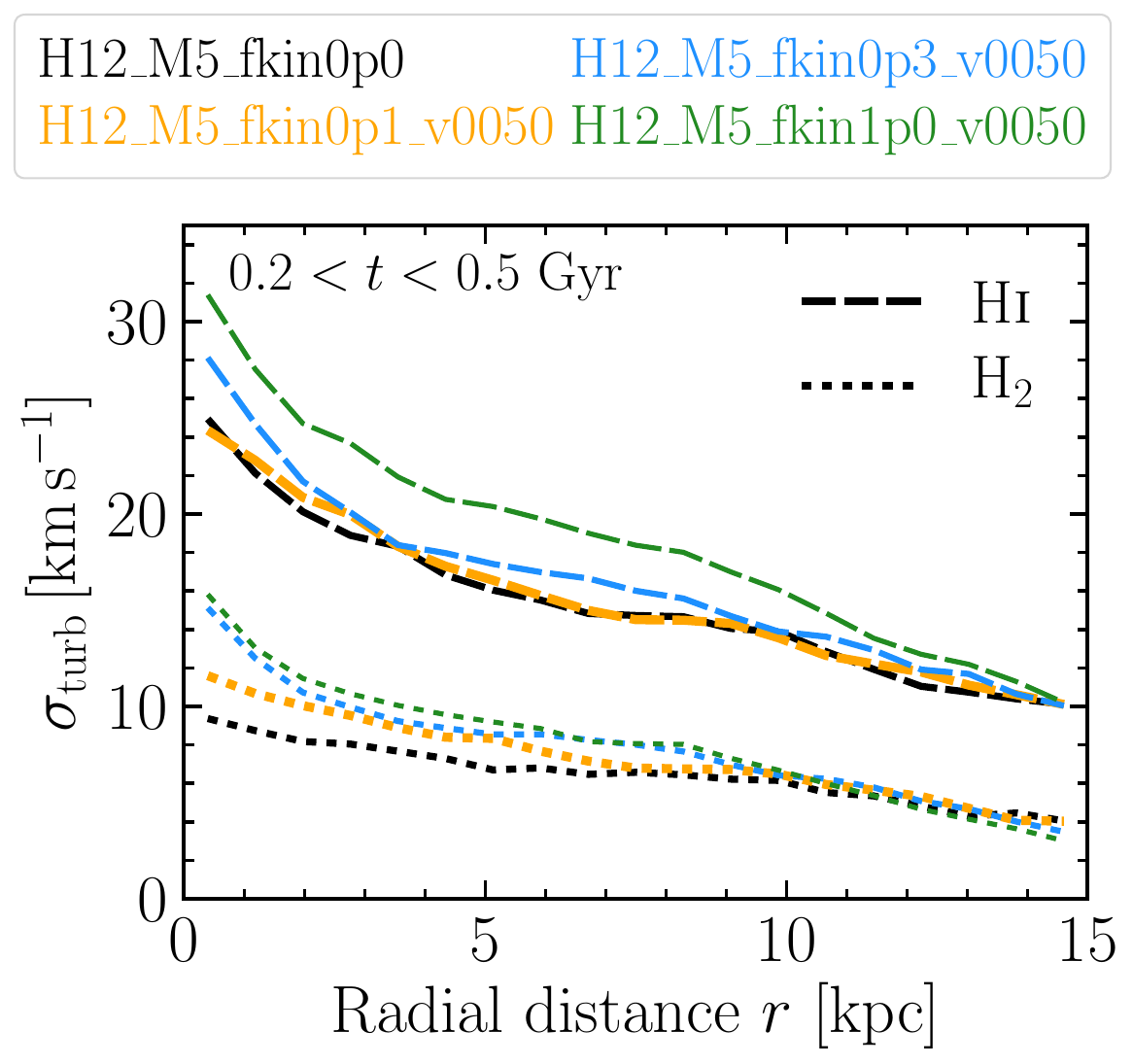}
    \caption{The turbulent velocity dispersion in the atomic (long-dashed) and molecular (short-dashed) gas at height $< 250$ pc displayed versus distance from the galactic centre, averaged over time $0.2<t<0.5$ Gyr. The plot is shown for the H12 galaxy with the desired kick velocity $\Delta v_{\rm kick} = 50$ \kms. Different colours correspond to different values of $f_{\rm kin}$. On average, higher values of $f_{\rm kin}$ yield higher velocity dispersion in both the H\textsc{i} and H$_2$ gas. }
    \label{fig:velocity_dispersion}
\end{figure}

Fig. \ref{fig:velocity_dispersion} shows radial profiles of the one-dimensional mass-weighted turbulent velocity dispersion in the H$_2$ and H\textsc{i} gas for the different values of $f_{\rm kin}$. The molecular and atomic gas mass fractions are taken directly from the \cite{2020MNRAS.497.4857P} tables (see $\S$\ref{sec:cooling}). For a given radial bin, the gas turbulent velocity dispersion, $\sigma_{\rm turb}$, is computed as  

\begin{equation}
    \sigma_{\rm turb}^2(\mathrm{phase}) = \frac{1}{3}\frac{\sum_i \, m_{\mathrm{gas}, i}  f_{\mathrm{phase}, i} \, \sigma_{\mathrm{3D,turb},i}^2}{\sum_i \, m_{\mathrm{gas}, i}  f_{\mathrm{phase}, i}} \, ,
\label{eq: sigma_sim}
\end{equation}
where $m_{\mathrm{gas}, i}$ is the mass of gas particle $i$, $f_{\mathrm{phase}, i}$ is the fraction of hydrogen mass in the phase the velocity dispersion is computed for (molecular, atomic, or neutral), and $\sigma_{\mathrm{3D,turb},i}$ is the gas three-dimensional velocity dispersion estimated within the kernel of particle $i$ (see equation \ref{eq:v_sigma} for details). In each radial bin, we show the average value of the velocity dispersion during times $0.2<t<0.5$ Gyr\footnote{For a given radial-distance bin, the average of the velocity dispersion over time $0.2<t<0.5$ Gyr is calculated by first computing the velocity dispersion in this bin separately for the data from all snapshots with times $0.2<t<0.5$ Gyr, and then taking the average value. Our simulations output snapshots every 5 Myr so we have 60 snapshots between $0.2$ and $0.5$ Gyr.}. We are interested in the gas turbulence within the galactic disc, so when computing $\sigma_{\rm turb}$ we select only those gas particles whose height with respect to the galactic plane is no greater than 250 pc. Without such a cut there would be large contributions to $\sigma_{\rm turb}$ from the outflowing and inflowing gas below and above the disc.

The runs with higher $f_{\rm kin}$ consistently have higher velocity dispersion for both H\textsc{i} and H$_2$ -- except at very large radii where the models converge. The velocity dispersion increases as we approach the galactic centre: at radial distance $r\approx 0$ kpc, the velocity dispersion in the model with $f_{\rm kin}=0$ is approximately 24 \kms{} for H\textsc{i} and 10 \kms{} for H$_2$, while in the purely kinetic model it is 32 \kms{} and 16 \kms, respectively. At distances $r \gtrsim 13$ kpc, the differences in $\sigma_{\rm turb}$ between different models become negligible because at such large radii star formation (and subsequent feedback from SNe) is rare, which leads to the gas properties being similar in all four models. 
 
\begin{figure}
    \centering
    \includegraphics[width=0.45\textwidth]{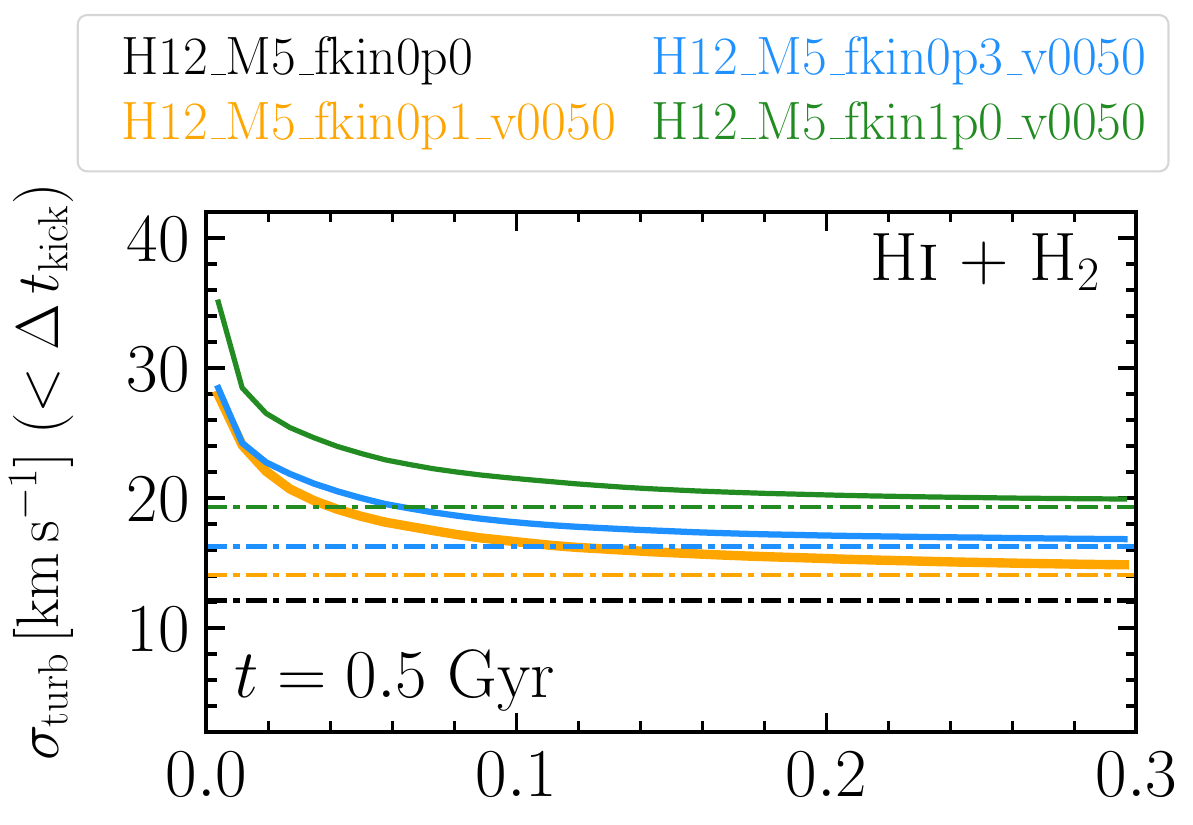} \\
    \includegraphics[width=0.45\textwidth]{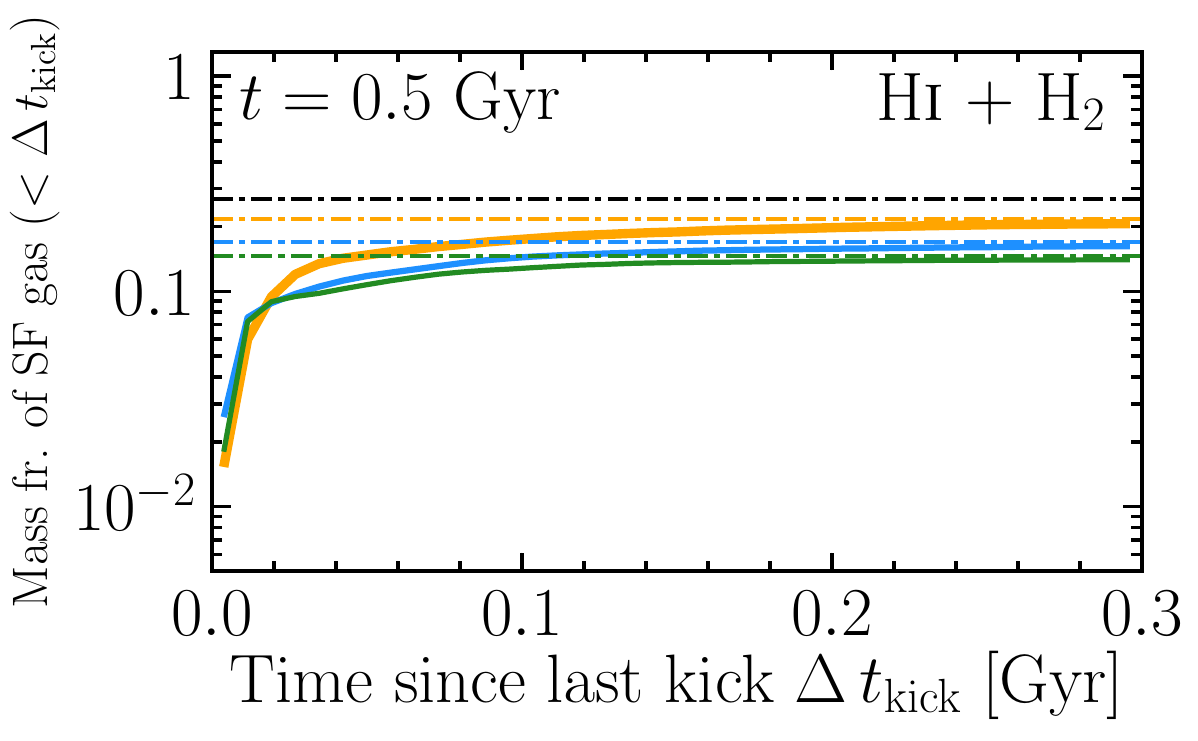}
    \caption{The turbulent velocity dispersion (\textit{top}) and the mass fraction of the gas that is star-forming (\textit{bottom}) at time $t=0.5$ Gyr, both computed in the neutral ISM that has been directly affected by SN kinetic feedback and displayed versus the time since the gas was last kicked. The horizontal dash-dotted lines indicate the average values computed for all neutral ISM. In all cases, we only consider gas particles at height $< 250$ pc. The plots are shown for the H12 galaxy with the desired kick velocity $\Delta v_{\rm kick} = 50$ \kms. Different colours correspond to different values of $f_{\rm kin}$. Low-energy kicks in SN kinetic feedback increase the turbulent velocity dispersion in the neutral ISM, which in turn decreases the fraction of SF gas in the neutral ISM. The smaller the time since the last kick event, the larger the effects. }
    \label{fig:velocity_dispersion_vs_time}
\end{figure}

To establish a more direct connection between the kinetic feedback from SNe and the gas turbulent velocity dispersion, in the top panel of Fig. \ref{fig:velocity_dispersion_vs_time} we plot the turbulent velocity dispersion in the neutral ISM at time $t=0.5$ Gyr as a function of time since the gas last underwent a kick event, $\Delta t_{\rm kick}$. To compute the velocity dispersion for a given $\Delta t_{\rm kick}$, we only select the gas particles that have heights $h < 250$ pc, and have been kicked by SNe in the time interval $[0.5 - \Delta t_{\rm kick}, 0.5]$ Gyr at least once. The results are presented for the different values of $f_{\rm kin}$ in the top panel and are depicted by the solid curves. Additionally, we show the average velocity dispersion for \textit{all} gas in the neutral ISM at height $h<250$ pc (horizontal dash-dotted lines)\footnote{Note that in Fig. \ref{fig:velocity_dispersion_vs_time}, for the model with $f_{\rm kin}=0$ we cannot show the black solid curve because the model has purely thermal feedback.}. As expected, we find that the velocity dispersion computed using only the kicked gas particles is higher than the overall velocity dispersion. Moreover, the smaller the time since the last kick event, the higher the velocity dispersion. At a fixed $\Delta t_{\rm kick}$, the velocity dispersion is higher for higher $f_{\rm kin}$, with the differences in $\sigma_{\rm turb}$ between the models with $f_{\rm kin}=1$ and $f_{\rm kin}=0$ changing from a factor of $\approx 1.6$ at $\Delta t_{\rm kick}\approx 0.2$ Gyr to a factor of $\approx 3$ at $\Delta t_{\rm kick}\approx 0.0$ Gyr.

Finally, we estimate the impact of the kicks on the amount of star-forming (SF) gas. In the bottom panel of Fig. \ref{fig:velocity_dispersion_vs_time}, we re-create the top panel but replace the velocity dispersion in the neutral ISM with the mass fraction in the neutral ISM that is SF. There we find that the fraction of SF gas among the gas particles that have just been kicked ($\Delta t_{\rm kick}\approx 0$ Gyr) is around 2 per cent and asymptotes to $\approx 20$ per cent for $\Delta t_{\rm kick}$ greater than $0.1$ Gyr. The asymptotic value decreases slightly with $f_{\rm kin}$: from $\approx 20$ per cent for $f_{\rm kin}=0.1$ to $\approx 15$ per cent for $f_{\rm kin}=1$. The run without kinetic feedback has a noticeably larger fraction of SF gas, $\approx 30$ per cent (black dash-dotted line), indicating that the star formation is sensitive to even a tiny amount of SN energy injected in kinetic form via low-energy kicks, however long time ago.

\begin{figure}
    \centering
    \includegraphics[width=0.45\textwidth]{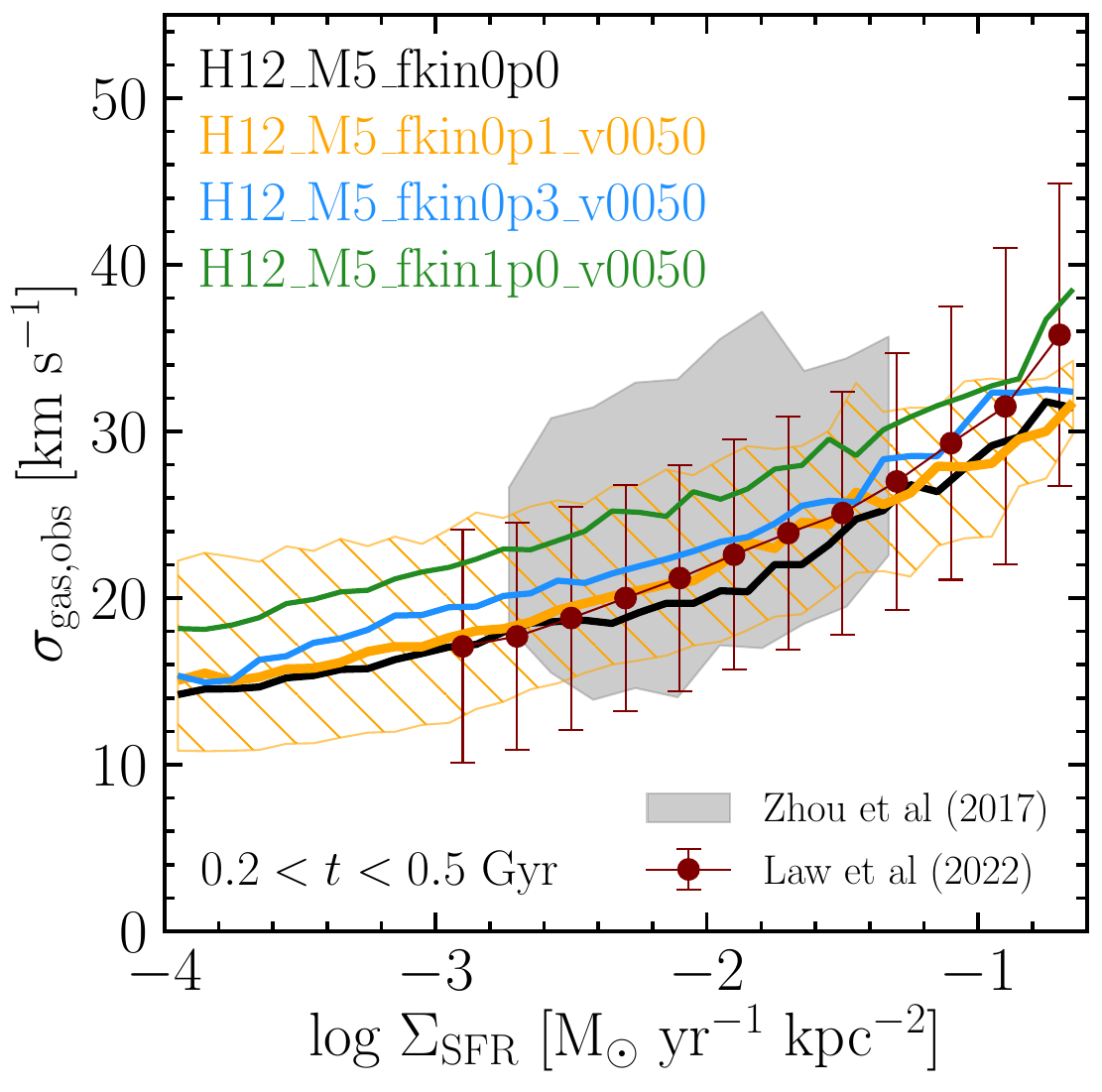}
      \caption{The mock observed H\textsc{i} turbulent velocity, $\sigma_{\rm gas, obs}$, versus SFR surface density, $\Sigma_{\rm SFR}$, computed in pixels of size ($1$ kpc)$^2$, in the H12 galaxy averaged over time $0.2 < t < 0.5$ Gyr, for four different values of $f_{\rm kin}$ (colours). The desired kick velocity is 50 \kms and the galaxy is viewed face-on (see equation \ref{eq: sigma_obs} for the definition of $\sigma_{\rm gas, obs}$). The solid curves show the median values and the hatched orange region marks the $16^{\rm th}$ to $84^{\rm th}$ percentiles in the \textsf{H12\_M5\_fkin01\_v0050} run. For comparison, we show the H$\alpha$-based, spatially resolved observational data from \citet{2017MNRAS.470.4573Z} and \citet{2022ApJ...928...58L}. In all models, $\sigma_{\rm gas, obs}$ is an increasing function of $\Sigma_{\rm SFR}$, and all models show a reasonable match to the observational data, with the best agreement for $f_{\rm kin}= 0.1 - 0.3$.}
    \label{fig:dispersion_SFR}
\end{figure}

In summary, Figs. \ref{fig:velocity_dispersion} and \ref{fig:velocity_dispersion_vs_time} demonstrate that injecting a non-zero fraction of SN energy in kinetic form via low-energy kicks can significantly increase the gas turbulent velocity dispersion in the neutral ISM. As a consequence, star formation generally proceeds on a longer time-scale, SNe are less clustered, and more gas is able to remain in the ISM (Fig. \ref{fig:morphology_M12_main}), leading to higher gas surface densities (Fig. \ref{fig:surface_density_M12_main}).

\subsubsection{Velocity dispersion versus star formation rate surface density}

We can assess the reliability of our predictions for the gas velocity dispersion -- shown in Figs. \ref{fig:velocity_dispersion} and \ref{fig:velocity_dispersion_vs_time} -- by comparing them to spatially resolved H$\alpha$ measurements. In order to perform such a comparison, rather than computing the velocity dispersion on a particle-by-particle basis, as given by equation (\ref{eq: sigma_sim}), we adopt a different approach: (i) we look at the galaxy face-on and bin the galaxy image in pixels of size (1 kpc)$^2$; (ii) in a given pixel $j$, we calculate the mock observed velocity dispersion as

\begin{equation}
    \sigma_{\mathrm{gas, obs}, j}^2 = \frac{\sum_i \, m_{\mathrm{gas}, i} f_{\mathrm{HI}, i} \, (v_{z,i} - \langle v_{z} \rangle_j)^2}{\sum_i \, m_{\mathrm{gas}, i} f_{\mathrm{HI}, i}} \, + \sigma_{\rm th}^2 ,
\label{eq: sigma_obs}
\end{equation}
where the sum is computed over \textit{all} gas particles that are inside pixel $j$ (i.e. including the inflowing and outflowing gas), $v_{z,i}$ is the velocity of the $i^{\rm th}$ gas particle along the $z$ axis, which is perpendicular to the galactic plane, and $\langle v_{z}\rangle_j$ is the average, H\textsc{i}-mass weighted velocity along the $z$ direction in pixel $j$. Additionally, because in the sum above we weigh the velocity dispersion by the H\textsc{i} mass but aim to compare it with H$\alpha$ measurements, which are likely dominated by the gas in H\textsc{ii}  regions, we add in quadrature a thermal velocity dispersion component of $\sigma_{\rm th} = 9$ \kms, corresponding to the gas thermal motion at temperature $10^4$ K \citep[e.g.][]{2005A&A...431..235R}. We compare with observations from the SAMI \citep{2017MNRAS.470.4573Z} and MaNGA \citep{2022ApJ...928...58L} surveys, both of which target galaxies in the nearby Universe, have $\sim$ kpc spatial resolution, and use H$\alpha$ measurements to obtain the SFRs and turbulent velocity dispersions.

Fig. \ref{fig:dispersion_SFR} shows the results of the comparison. We plot the turbulent velocity dispersion as a function of the SFR surface density. As before, we consider the H12 galaxy with M5 resolution for $f_{\rm kin}=0, 0.1, 0.3$ and $1$ (colour-coded) and $\Delta v_{\rm kick} = 50$ \kms{}. In each SFR surface density bin, we calculate the median velocity dispersion using the pixels from all simulation snapshots with times $0.2<t<0.5$ Gyr, and we only consider those pixels that contain at least 20 gas particles. These median values are shown by the solid curves. Similarly to Fig. \ref{fig:velocity_dispersion}, we find that the velocity dispersion increases with $f_{\rm kin}$ and spans a range of values from $\approx 14$ \kms{} to $\approx 38$ \kms.  Regardless of the value of $f_{\rm kin}$, all models are in reasonable agreement with the observational data reported by \citet{2017MNRAS.470.4573Z} and \citet{2022ApJ...928...58L}, which indicates that our numerical set-up as a whole is realistic. Importantly, we recover the trend of $\sigma_{\mathrm{gas, obs}}$ increasing with the SFR surface density that is present in both observational datasets, and this holds for all values of $f_{\rm kin}$. The agreement improves slightly for $f_{\rm kin} = 0.1 - 0.3$ compared to $f_{\rm kin} = 0$, while for $f_{\rm kin} = 1$ the velocity dispersion becomes a little too large. We note that for our fiducial resolution of $m_{\rm gas}=10^{5} \, \rm M_\odot$, our predictions for the velocity dispersion at SFR surfaces densities below $10^{-3} \, \rm M_\odot \, yr^{-1} \, kpc^{-2}$  may not be robust (because this corresponds to only one star particle formed per pixel in 0.1 Gyr).

\subsubsection{Wind mass loading factors}
\label{sec:wind_mass_loading}

\begin{figure}
    \centering
    \includegraphics[width=0.49\textwidth]{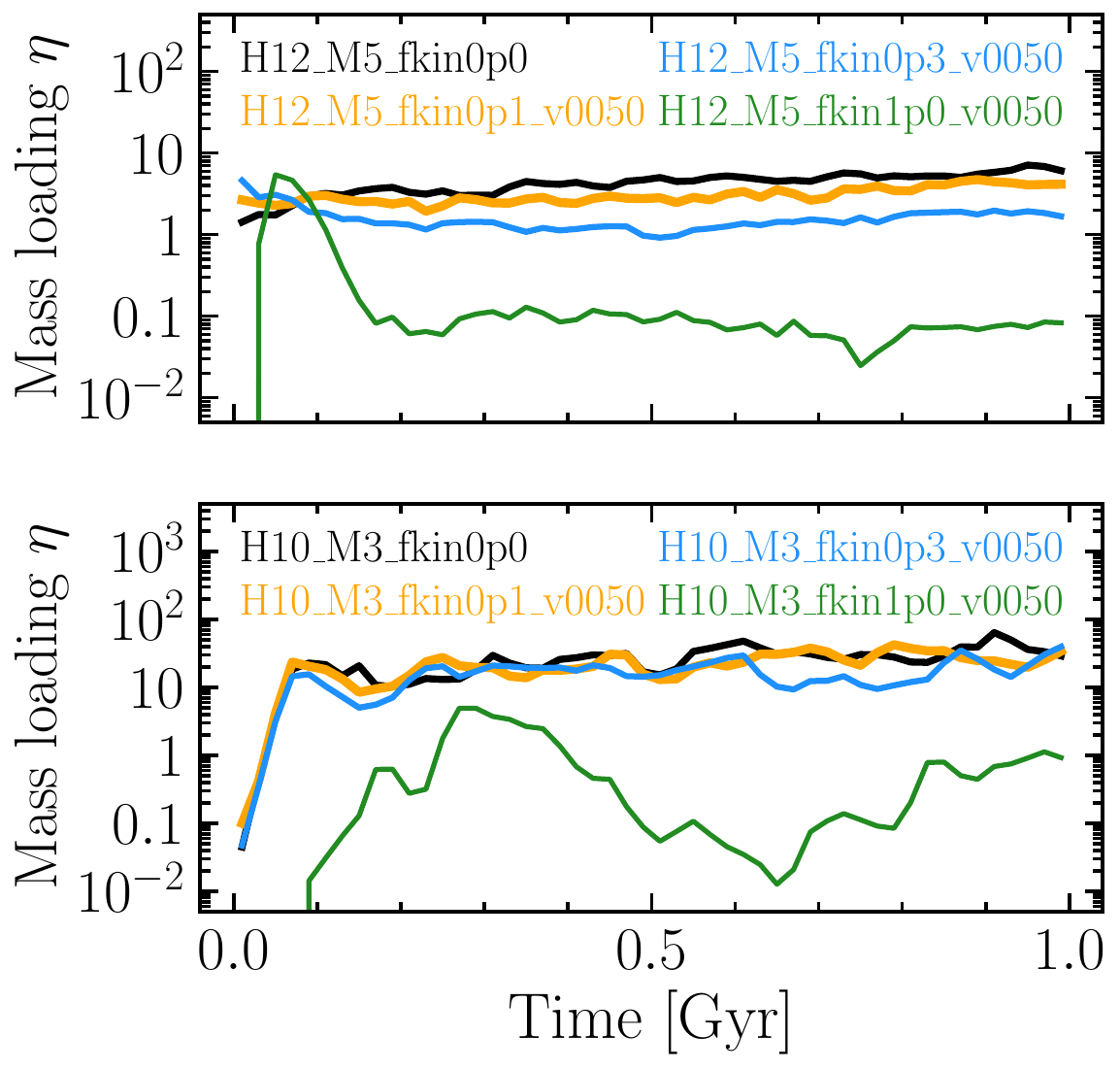}
      \caption{The wind mass loading factor measured at height $d=10 \pm 0.5$ kpc from the galactic plane, in the H12 galaxy with M5 resolution (\textit{top panel}) and H10 galaxy with M3 resolution (\textit{bottom panel}), for different values of $f_{\rm kin}$: $0$ (black), $0.1$ (orange), $0.3$ (blue), and $1$ (green). The desired kick velocity is $\Delta v_{\rm kick} = 50$ \kms{} in all cases where kinetic feedback is present. The mass loading in the H10 galaxy is an order of magnitude higher than in H12. For the chosen desired kick velocity and heating temperature, a large fraction of the SN energy needs to be injected thermally to drive stable galactic outflows.}
    \label{fig:Mass_loading}
\end{figure}

In this section, we investigate how the strength and structure of galactic winds generated in our simulations depend on the manner in which the SN energy is deposited: high-energy injections in the thermal channel versus low-energy kicks in the kinetic channel. To characterise the power of galactic winds, we define the wind mass loading factor $\eta$ at time $t$ and at (absolute) height $d$ from the disc plane using the expression
\begin{equation}
    \eta(t, d, \Delta d) = \frac{1}{\dot{m}_{\rm sf,gal}} \sum_{|z_i\pm d|<\Delta d/2} \, \frac{m_{\mathrm{gas},i} \, |v_{z,i}|}{\Delta d}\, , 
\end{equation}
where $\dot{m}_{\rm sf,gal}$ is the galaxy SFR at time $t$, $v_{z,i}$ is the velocity $z$ component of particle $i$, $m_{\mathrm{gas},i}$ is the mass of particle $i$, $z_i$ is the height of particle $i$ relative to the galactic disc, and to compute the sum we consider all gas particles that (i) are vertically moving away from the disc and (ii) have heights within $d\pm \Delta d/2$ from the disc.

Fig. \ref{fig:Mass_loading} shows wind mass loading factors at distance $d=10\pm 0.5$ kpc in the H10 and H12 galaxies for the different values of $f_{\rm kin}$. Since in these models the kinetic channel uses a low desired kick velocity, $\Delta v_{\rm kick} = 50$ \kms, while the thermal channel has a relatively high heating temperature $\Delta T = 10^{7.5}$ K, a non-zero fraction of the SN energy injected in thermal form is required to drive a sustained and strong galactic wind, as we can observe in the figure. The lower the value of $f_{\rm kin}$, the stronger the galactic wind. For the H12 galaxy, the runs with $f_{\rm kin}\leq 0.3$ have $\eta \sim 1 - 10$; and for the H10 galaxy, these models have $\eta \sim 10 - 10^2$. These values roughly agree, e.g., with the scaling from \cite{2015MNRAS.454.2691M} derived in the \textsc{fire} zoom simulations \citep{2014MNRAS.445..581H}. In contrast, in the purely kinetic model, the H12 galaxy has a mass loading of $\eta \sim 10^{-1}$ and for H10 $\eta \sim 1$.

Since galactic winds with $\eta \gtrsim 1$ are commonly observed \citep[e.g.][]{2005ARA&A..43..769V,2015ApJ...804...83S,2019MNRAS.490.4368S}, we conclude that as long as the desired kick velocity is low, $\Delta v_{\rm kick}\lesssim 10^2$ \kms{}, large thermal energy injections ($f_{\rm kin} \lesssim 0.3$, $\Delta T \sim 10^{7.5}$ K) are a necessary ingredient to make our model agree with observations and with simulations carried out by other research groups.

\subsubsection{Kennicutt-Schmidt star formation law}

\begin{figure}
    \centering
    \includegraphics[width=0.45\textwidth]{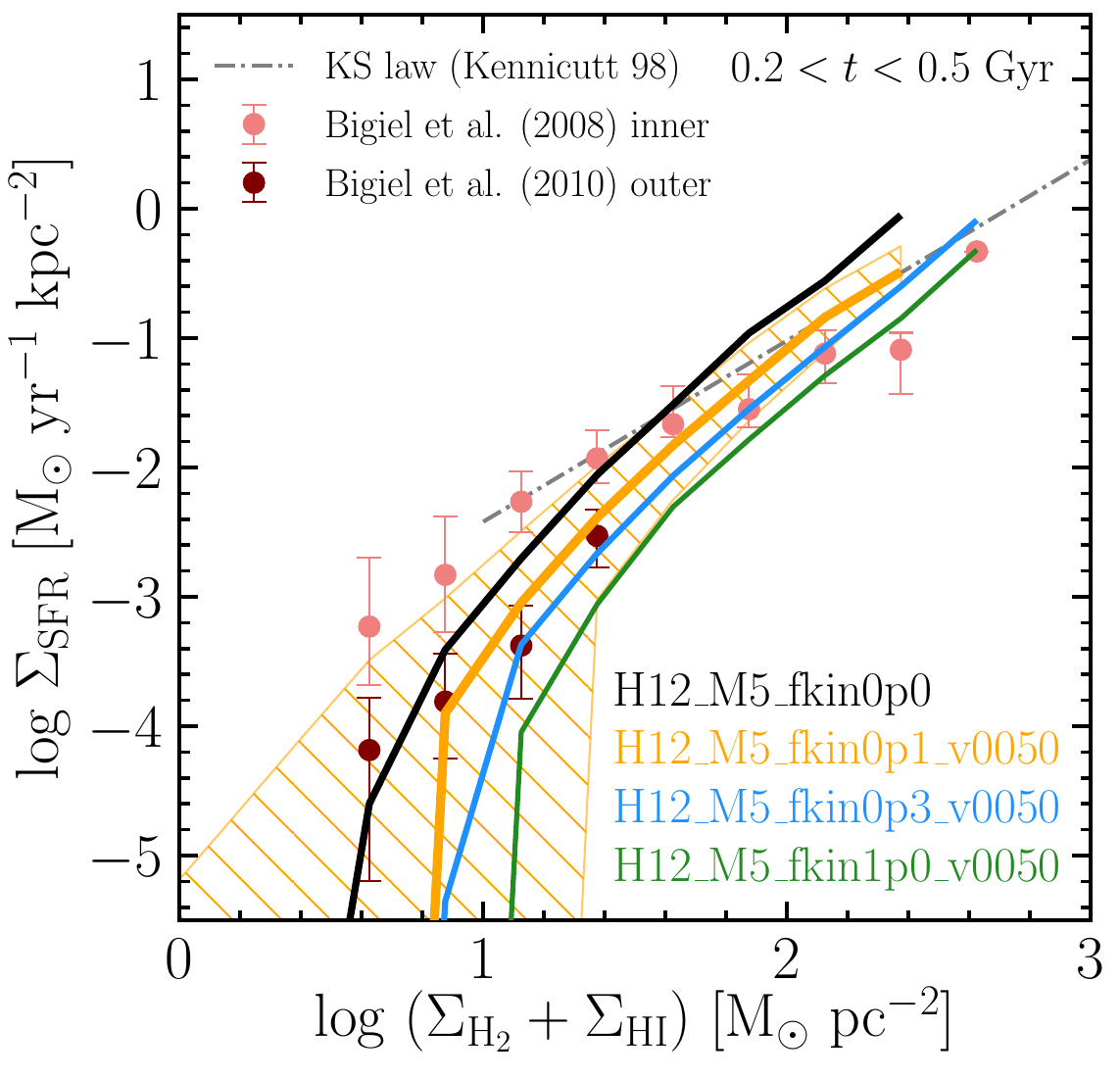}
    \caption{Median star formation rate surface density versus neutral gas surface density in the H12 galaxy averaged over time $0.2 < t < 0.5$ Gyr, for four different values of $f_{\rm kin}$ (colours). The desired kick velocity is set to 50 \kms{} in the runs including kinetic feedback. The galaxy is viewed face-on and the relation is computed in bins of size ($0.75$ kpc)$^2$. The hatched orange region marks the 16$^{\rm th}$ to 84$^{\rm th}$ percentiles in the \textsf{H12\_M5\_fkin01\_v0050} run. For comparison, we show the observational data from \citet{Bigiel2008AJ} and \citet{Bigiel2010AJ} as well as the KS law with a slope of $n=1.4$ \citep[][]{Kennicutt1998ApJ}. The KS relation in the run with $f_{\rm kin}=0.1$ is closest to the observations, while in the run with purely thermal feedback it is too steep.}
    \label{fig:KS_law_main}
\end{figure}

Fig. \ref{fig:KS_law_main} shows SFR surface density, $\Sigma_{\rm SFR}$, as a function of neutral gas surface density, $\Sigma_{\rm H_2} + \Sigma_{\rm HI}$, for the H12 galaxy with M5 resolution. The relation is calculated using square spatial bins of size ($0.75$ kpc)$^2$ and the galaxy is viewed face-on. We again display cases with four different values of $f_{\rm kin}$ (colour-coded). The solid curves show the median values of $\Sigma_{\rm SFR}$. In a given $\Sigma_{\rm H_2} + \Sigma_{\rm HI}$ bin, the median $\Sigma_{\rm SFR}$ is computed among pixels from all snapshots with times $0.2 < t < 0.5$ Gyr. For comparison, the figure additionally shows the observational data from \citet{Bigiel2008AJ} and \citet{Bigiel2010AJ} for the inner and outer parts of discs of nearby spiral galaxies, respectively; as well as the KS law with a slope of $n=1.4$ \citep[][]{Kennicutt1998ApJ}. 

We find that by increasing $f_{\rm kin}$, we decrease $\Sigma_{\rm SFR}$ at a fixed surface density of neutral gas. For example, at log $[(\rm \Sigma_{\rm H_2} + \Sigma_{\rm HI})/M_\odot \, pc^{-2}] = 1.4$, changing the feedback from purely thermal ($f_{\rm kin}=0$) to purely kinetic ($f_{\rm kin} = 1$) reduces $\Sigma_{\rm SFR}$ from $\approx 10^{-2}$ to $\approx 10^{-3}\, \rm M_\odot \, yr^{-1} \, kpc^{-2}$. However, although at a fixed gas surface density the SFR surface density is a decreasing function of $f_{\rm kin}$, the galaxy's total SFR is not necessarily lower for higher $f_{\rm kin}$. In fact, at times $t \gtrsim 0.4$ Gyr the opposite is the case (Fig. \ref{fig:SFH_main}). The reason is that models with higher $f_{\rm kin}$ generally possess more (high-density) gas at these times (Fig. \ref{fig:surface_density_M12_main})

The SFR surface density in the purely thermal model rises too steeply relative to the observed KS law, while in the model with purely kinetic feedback $\Sigma_{\rm SFR}$ cuts off at a too-high gas surface density ($\Sigma_{\rm H_2} + \Sigma_{\rm HI} \approx 12 \, \rm M_\odot \, pc^{-2}$) and the relation undershoots the data. The models with $f_{\rm kin} = 0.1$ and $f_{\rm kin} = 0.3$ produce relations that are within the acceptable range of values, with the $f_{\rm kin} = 0.1$ model showing a slightly better agreement with the observational data.

\subsection{Variations in the desired kick velocity}
\label{sec: vary_vkick}

\begin{figure}
    \centering
    \includegraphics[width=0.45\textwidth]{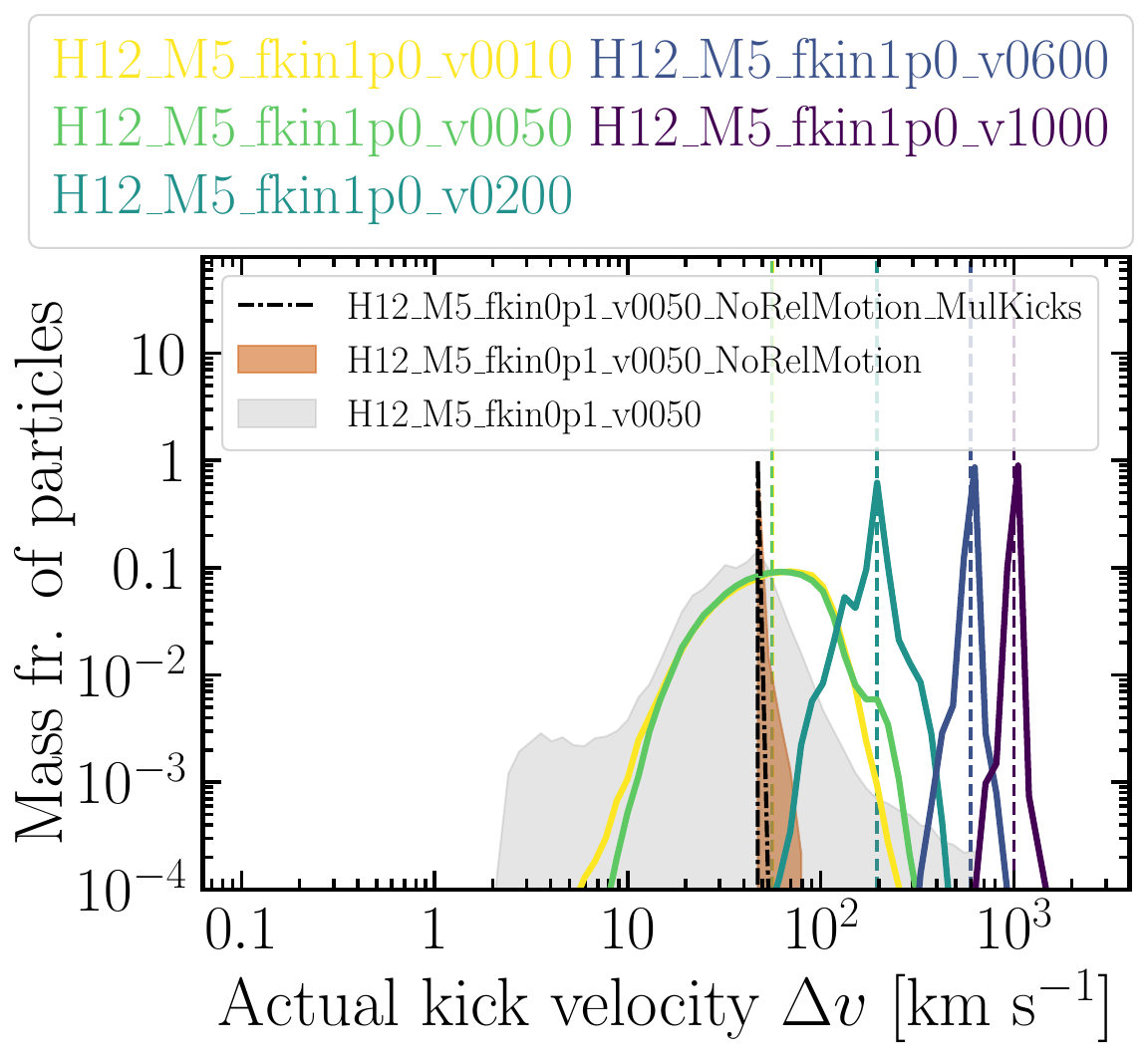}
      \caption{Distribution of actual kick velocities in SN kinetic feedback, in the H12 galaxy with M5 resolution at time $t=1$ Gyr. The solid curves show the runs with the fraction of kinetic energy $f_{\rm kin}=1$. The colour indicates the value of the desired kick velocity in the run, $\Delta v_{\rm kick} = 10$, $50$, $200$, $600$, and $1000$ \kms. The dashed vertical lines indicate the median kick velocity in the distribution. For reference, we also show the distributions of actual kick velocities in the fiducial run with $f_{\rm kin}=0.1$ and $\Delta v_{\rm kick} = 50$ \kms{}(\textsf{H12\_M5\_fkin0p1\_v0050}, grey shaded region) and in its two variations where we do not account for the star-gas relative motion (brown shaded region), and where we additionally do not prevent kick collisions (black dash-dotted curve). Although the desired kick velocity is fixed and single-valued, the actual kick velocities exhibit a large spread, particularly for low desired kick velocities. The spread is caused mainly by the relative star-gas motion corrections.}
    \label{fig:distribution_v_kick}
\end{figure}

Thus far we have exclusively discussed SN kinetic models with the desired kick velocity $\Delta v_{\rm kick} = 50$ \kms. In this section, we explore how galaxy properties depend on $\Delta v_{\rm kick}$. To ease the interpretation and maximize the differences, we will focus on purely kinetic models.

\begin{figure}
    \centering
  \includegraphics[width=0.42\textwidth]{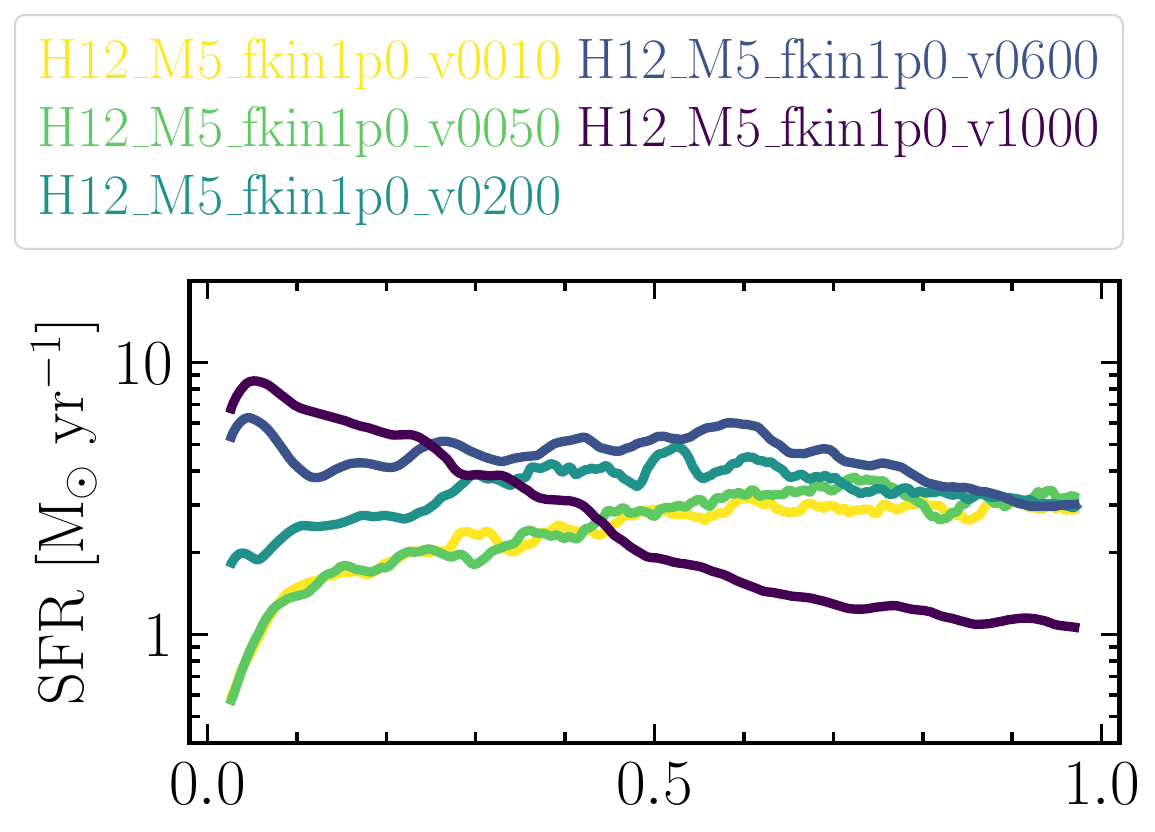} \\ 
  \includegraphics[width=0.43\textwidth]{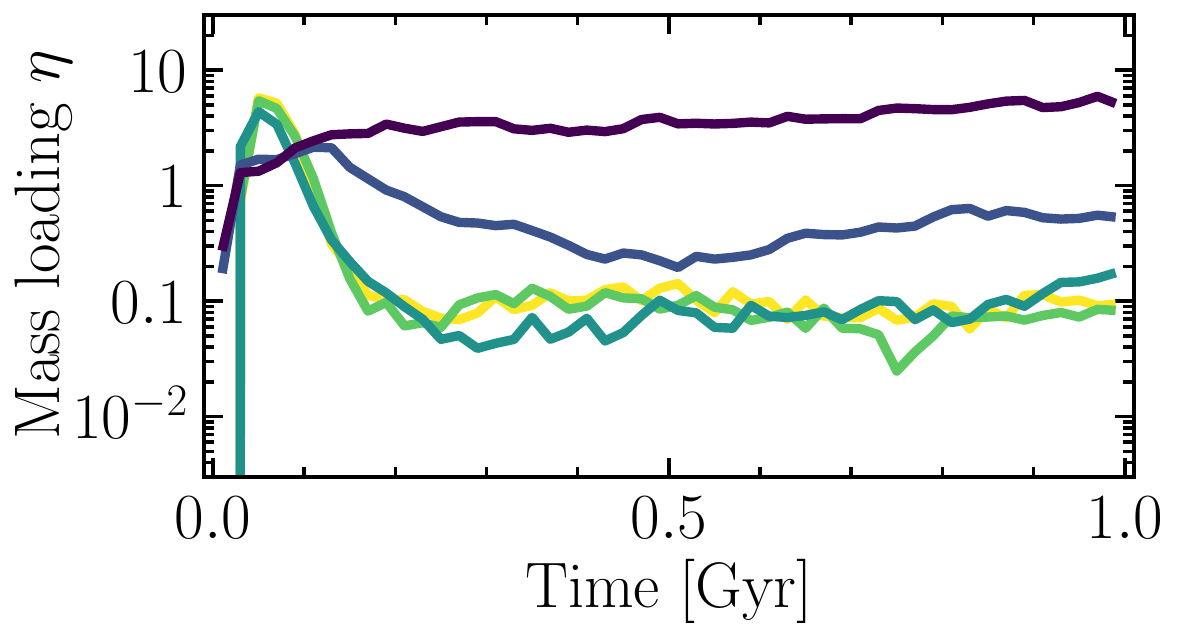} 
\caption{The effect of varying the desired kick velocity, $\Delta v_{\rm kick}$, in the purely kinetic models ($f_{\rm kin}=1$). \textit{Top:} Star formation rate versus time. \textit{Bottom:} Wind mass loading factor at height $d=10\pm 0.5$ kpc versus time. The plots are for the H12 galaxy with M5 resolution. The desired kick velocity $\Delta v_{\rm kick} = 10$, $50$, $200$, $600$, and $1000$ \kms{} (colours). Both SFR and wind mass loading depend strongly on $\Delta v_{\rm kick}$ for $10^2 \lesssim \Delta v_{\rm kick}\lesssim 10^3$ \kms{} and converge for $\Delta v_{\rm kick}\lesssim 10^2$ \kms{}.}
    \label{fig:vary_vkick_compilation_of_plots}
\end{figure}

\subsubsection{Distribution of actual kick velocities}

Fig. \ref{fig:distribution_v_kick} shows the distributions of actual kick velocities, $\Delta v$, in SN kinetic feedback  (defined in equation \ref{eq:v_kick_pair_beta}) at time $t=1$ Gyr in the H12 galaxy with M5 resolution, for five purely kinetic models with desired kick velocities of $\Delta v_{\rm kick} = 10$, $50$, $200$, $600$, and $1000$ \kms{} (differently coloured solid curves). The dashed vertical lines indicate the median kick velocity in the sample. For reference, we also show the kick velocities in our fiducial model (\textsf{H12\_M5\_fkin0p1\_v0050}, grey shaded region), which has $f_{\rm kin}=0.1$ and $\Delta v_{\rm kick} = 50$ \kms, and in the two variations of the fiducial model where we do not account for the star-gas relative motion (\textsf{H12\_M5\_fkin0p1\_v0050\_NoRelMotion}, brown shaded region), and where we additionally do not prevent gas particles from being kicked more than once in a single time-step (\textsf{H12\_M5\_fkin0p1\_v0050\_NoRelMotion\_MulKicks}, black dash-dotted curve). To construct the velocity distributions, we let each gas particle record the velocity it was kicked with in its last SN kinetic-feedback event; and these are the velocities that appear in the plot. 

We first compare the purely kinetic models (solid curves). Although the desired kick velocity specifies a fixed, single value, the actual kick velocities vary, particularly for low desired kick velocities. There are two reasons for these variations:

\begin{itemize}
    \item The actual kick velocities differ from the desired kick velocity because we correct for the relative motion between gas and stars in order to conserve energy (see $\S$\ref{paragraph:stochastic_kinetic_feedback}). The lower the desired kick velocity, the more significant the impact of the relative star-gas motion. Among the five purely kinetic models, for $\Delta v_{\rm kick} = 10$ and $50$ \kms{} the distribution of the actual kick velocities can be described by a lognormal distribution with a width of about one dex. For the two models with the highest desired kick velocities, $\Delta v_{\rm kick} = 600$ and $1000$ \kms{}, the distributions are much narrower, shrinking to below 0.2 dex in $\Delta \log v$. Moreover, the distributions for high $\Delta v_{\rm kick}$ become strongly peaked at $\Delta v=\Delta v_{\rm kick}$. This is a consequence of the desired kick velocity being (much) higher than the average relative velocity between the stars and their gas neighbours, so that the correction due to their relative motion is small.
    
    \item The other cause of the spread in the actual kick velocities is an insufficient number of rays and/or gas neighbours in the stellar kernel to accommodate all kick events. This is the reason why the kick-velocity distributions in the purely kinetic models with $\Delta v_{\rm kick} = 10$ and $50$ \kms{} look remarkably similar, with the median values of the distributions in these two models being $\approx 55$ \kms. Namely, when $\Delta v_{\rm kick} = 10$ \kms{}, the maximum number of rays per star particle, $N_{\rm rays} = 8$, and the expected number of neighbours in the kernel, $\langle N_{\rm ngb}\rangle \approx 65$, are both much smaller than the number of kicks a young particle would try to distribute. If a young star particle releases all its SN energy budget when its age $3 < t_{\rm age} < 43$ Myr and has an average time-step of $\approx 1$ Myr, then according to equation (\ref{eq:number_of_kicks}), for $f_{\rm kin} = 1$, $f_{\rm E} = 2$, and $\Delta v_{\rm kick}=10$ \kms, the number of kick events in one time-step will be $N_{\rm kick} \approx  \langle N_{\rm kick,tot}\rangle / 40 \approx 300$. Our algorithm will first try to distribute the available kinetic energy in $N_{\rm kick}$ kick events, but because it will find (on average) only $N_{\rm ngb} \approx 65$ gas neighbours, this will lead to an increase in the desired kick velocity by $ \sqrt{N_{\rm kick}/N_{\rm ngb}} \, \approx 2$ (see $\S$\ref{sec:num_of_kick_events}). Next, because the maximum number of rays $N_{\rm rays} = 8$, the desired kick velocity will be further increased by $\sqrt{N_{\rm ngb}/N_{\rm rays}} \approx 3$ (see $\S$\ref{paragraph: kin_algorithm}). Therefore, the desired kick velocity after the two corrections will be $\approx 2 \times 3 \times  10 \, \rm km \, s^{-1} \approx 60 \, km \, s^{-1}$, which is close to what we find in Fig. \ref{fig:distribution_v_kick}.

\end{itemize}

\begin{figure}
    \centering
    \includegraphics[width=0.45\textwidth]{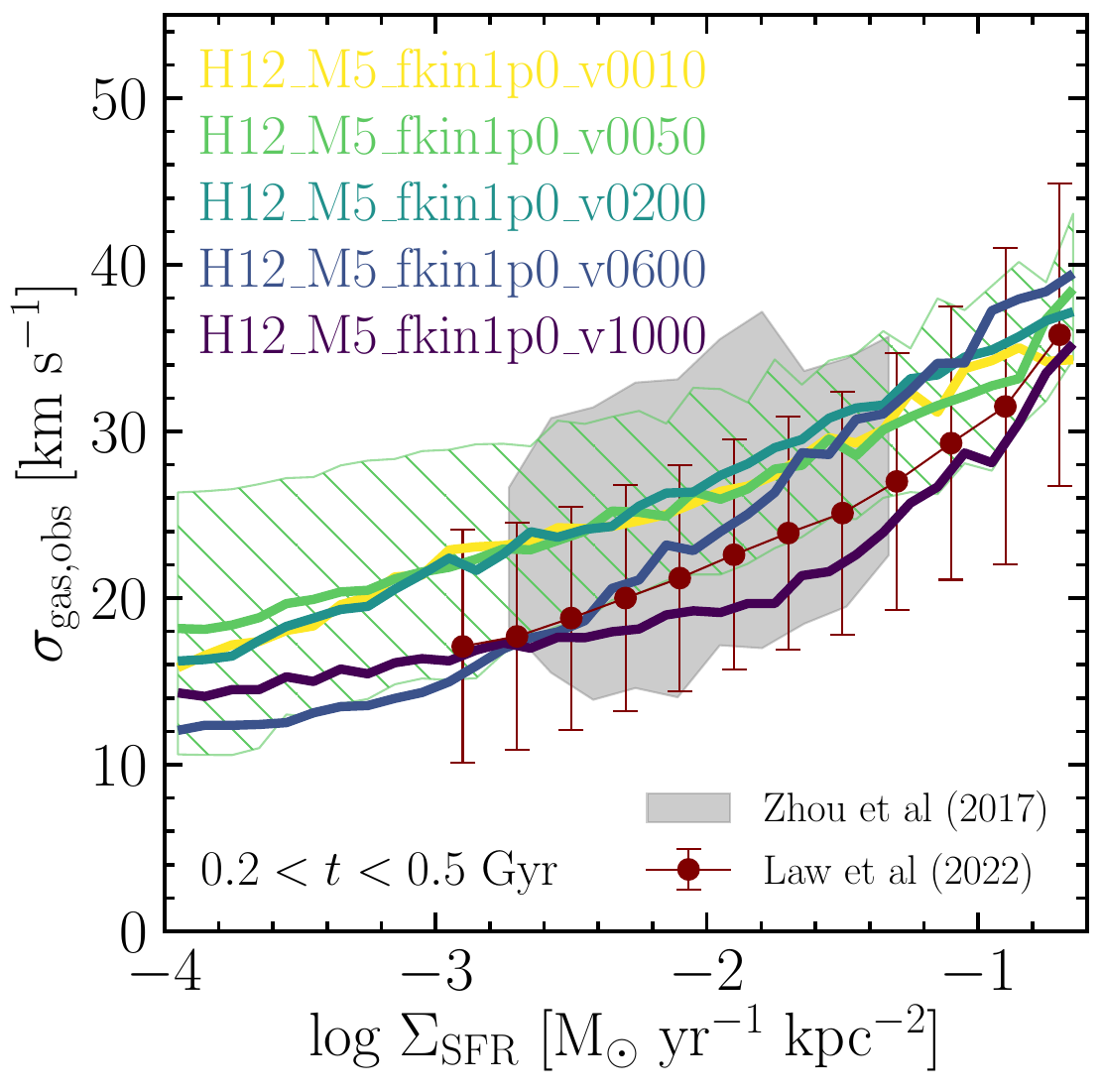}
    \caption{As Fig. \ref{fig:dispersion_SFR}, but varying the desired kick velocity $\Delta v_{\rm kick}$ ($10$, $50$, $200$, $600$, and $1000$ \kms{}, colours) while $f_{\rm kin}$ is equal to 1 in all cases. At most SFR surface densities, kicks with lower $\Delta v_{\rm kick}$ give rise to a higher velocity dispersion if $10^2 \lesssim \Delta v_{\rm kick}\lesssim 10^3$ \kms{}, while for $\Delta v_{\rm kick}\lesssim 10^2$ \kms{} the results converge, which holds for all $\Sigma_{\rm SFR}$.}
    \label{fig:dispersion_SFR_kick}
\end{figure}

The distribution of the kick velocities in the fiducial run (grey shaded region in Fig. \ref{fig:distribution_v_kick}), which in addition to the kinetic feedback uses SN thermal feedback, looks somewhat different. Unlike the purely kinetic run with the same $\Delta v_{\rm kick} = 50$ \kms{}, the distribution in the fiducial model peaks exactly at the desired kick velocity. This is because for $f_{\rm kin} = 0.1$, $f_{\rm E} = 2$, and $\Delta v_{\rm kick}=50$ \kms, the number of kicks per 1-Myr time-step is $\sim 1$, so there are more than enough rays and neighbours. Furthermore, unlike all purely kinetic models, the distribution of the kick velocities in the fiducial model has more extended high- and low-velocity wings, ranging from $\approx 2$ to $\approx 800$ \kms{}. These wings arise because a fraction of gas particles is accelerated to high velocities due to the strong SN thermal feedback, which makes the correction due to relative star-gas motion more significant and results in a larger scatter in the actual kick velocities.

In contrast, when we do not account for the relative star-gas motion and do not prevent kick collisions, the distribution of actual kick velocities expectedly approaches a delta function centred at the desired kick velocity of $\Delta v_{\rm kick}=50$ \kms{} (black dash-dotted curve). If the relative star-gas motion corrections are neglected but we do not allow gas particles to be kicked more than once in a single time-step, the distribution of actual kick velocities still resembles a delta function around $\Delta v_{\rm kick}=50$ \kms{} but has a velocity tail extending to kick velocities up to $\approx 1.6 \, \Delta v_{\rm kick} = 80$ \kms{} (brown shaded region). In order to understand the origin of the tail, we recall that when a stellar particle cannot distribute certain kick events in a given time-step due to kick collisions, it will store the number and total energy of these undistributed events and will attempt to distribute them again in the following time-step, which will happen together with the new kick events from the following time-step (see $\S$\ref{paragraph:collisions} for more details). This increases the chance that the total number of kick events (undistributed + new) will exceed the maximum number of rays and/or gas neighbours, and the desired kick velocity will thus increase, as we have explained above. Note that the figure is shown in log scale, so the number of particles that were kicked with a significantly increased $\Delta v$ due to kick collisions is small relative to the total number of kicked particles in the simulation.

\begin{figure*}
    \centering
    \includegraphics[width=0.50\textwidth]{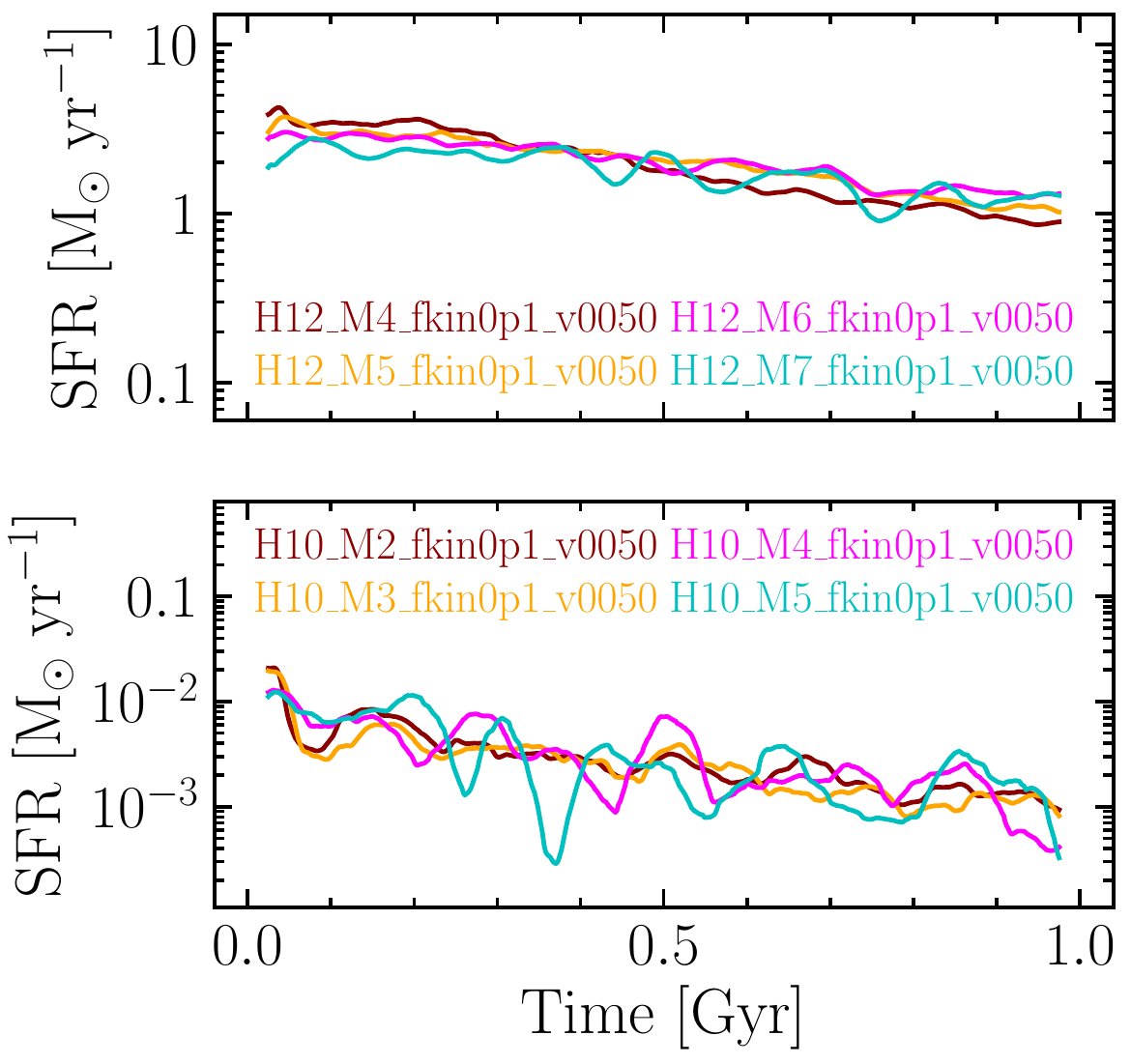}
    \includegraphics[width=0.48\textwidth]{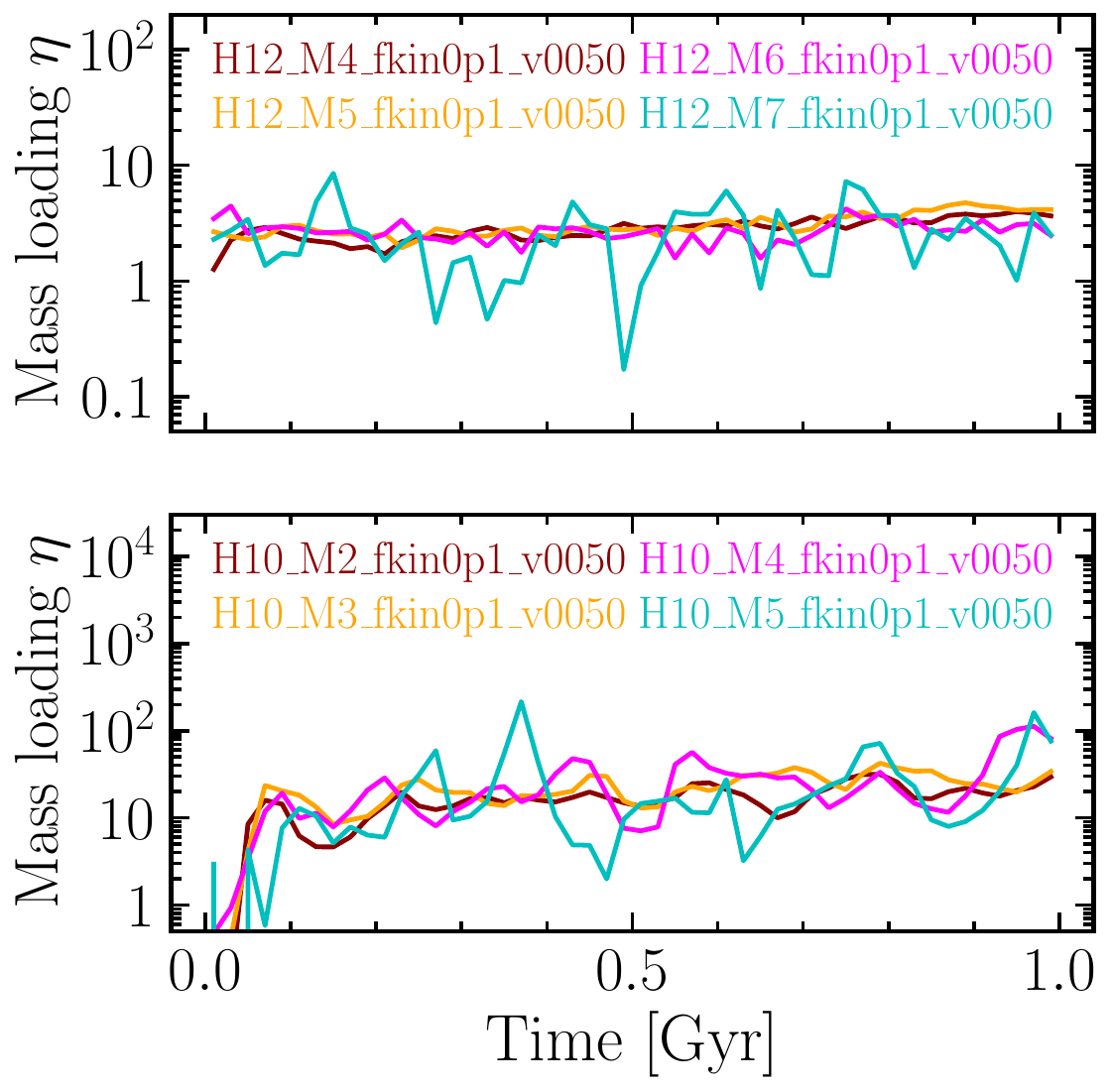}
      \caption{Star formation rate (\textit{left column}) and wind mass loading factor measured at height $d=10 \pm 0.5$ kpc from the galactic plane (\textit{right column}) versus time, for the fiducial SN model ($f_{\rm kin} = 0.1$, $\Delta v_{\rm kick} = 50$ km s$^{-1}$) in the H12 galaxy (\textit{top row}) and the H10 galaxy (\textit{bottom row}) for four different numerical resolutions (different colours). The mass (spatial) resolution is changed by factors of 8 (2) between adjacent resolutions and hence varies by three (one) orders of magnitude for each galaxy. Both the SFRs and wind mass loading factors exhibit excellent convergence, and this is true for both the H12 and H10 galaxies.}
    \label{fig:SFR_and_mass_loading_resolution}
\end{figure*}

\subsubsection{Galaxy properties}

In Fig. \ref{fig:vary_vkick_compilation_of_plots}, we present the temporal evolution of the galaxy SFR (top panel) and of the wind mass loading factor at height $d=10\pm 0.5$ kpc (bottom panel). Both plots are shown for the H12 galaxy with M5 resolution. In each panel, we show the models with $\Delta v_{\rm kick}=10$, $50$, $200$, $600$, and $1000$ \kms{} (colours). All models use purely kinetic SN feedback ($f_{\rm kin} = 1$).

The wind mass loading factor increases with $\Delta v_{\rm kick}$ provided it exceeds 200 \kms. Depending on the value of $\Delta v_{\rm kick}$, the mass loading varies between $\approx 0.1$ and $\approx 5$. Meanwhile, the variations in the star formation history show a more complex behaviour, which can be split into three distinct regimes:
\begin{itemize}
    \item In the models with relatively low desired kick velocities ($\Delta v_{\rm kick}=10$, $50$, and $200$ \kms), after the initial transitory phase ($t \lesssim 0.15$ Gyr), the SFR stays largely within $2-4 \, \rm M_\odot \, yr^{-1}$. In this regime, the regularization of the star formation is achieved by SNe quickly responding to new sites of star formation via relatively frequent, low-energy kicks. The kicks disrupt star-forming clumps of gas and increase the turbulence in the local ISM, which regulates the galaxy's SFR. Note, however, that these low desired kick velocities do not result in significant galactic winds (bottom panel).
    \item  The model with $\Delta v_{\rm kick}=600$ \kms corresponds to an intermediate regime where the number of kicks per time-step is low enough that SNe no longer react to the collapsing gas and prevent stars from forming as quickly and as efficiently as in the first regime. At the same time, the kick velocity is not yet high enough to produce strong galactic winds. Hence, among the five models, $\Delta v_{\rm kick}=600$ \kms{} yields the highest SFRs (if we disregard the first $\approx 0.25$ Gyr of evolution in the model with $\Delta v_{\rm kick}=1000$ \kms).
    \item  The third and final regime of SN kinetic feedback describes the models that efficiently regulate star formation via ejecting gas from the galaxy through strong and sustained galactic winds. The only model that is fully in this regime is that using $\Delta v_{\rm kick}=1000$ \kms. Such a high desired velocity gives rise to a strong, steady galactic wind with a mass loading of $\approx 5$, which is the highest among the considered models, and a smooth, monotonically declining SFR, which becomes the lowest among the considered models after $t \approx 0.45$ Gyr.
\end{itemize}

We note that the SFR and wind mass loading in the purely kinetic run with $\Delta v_{\rm kick}=1000$ \kms{} closely resemble those in the purely thermal run with $\Delta T=10^{7.5}$ K (\textsf{H12\_M5\_fkin0p0}) presented in Figs. \ref{fig:SFH_main} and \ref{fig:Mass_loading}. The resemblance between the thermal and kinetic models, which was also noticed in \citetalias{2012MNRAS.426..140D}, follows from the fact that both models deposit roughly the same amount of energy into the gas in one energy injection event (i.e. the specific energies  $k_{\rm B} \Delta T / [(\gamma-1) \mu_{\rm ionized} m_{\rm p}]$ and $\Delta v_{\rm kick}^2/2$ are similar) and because these energy injection events are powerful enough to create shocks with high Mach numbers. When a gas particle is kicked with $\Delta v_{\rm kick}\sim 1000$ \kms , it will quickly shock-heat its neighbours to high temperatures, so the net effect will be as if the particle was directly heated using a similar amount of energy. However, the models with kinetic and thermal injections of equal energy will only produce comparable total SFRs and wind mass loading for sufficiently high values of the energy per SN event. The thermal feedback with energies per event corresponding to $\Delta T\lesssim 10^{6}$ K will be suppressed by the enhanced radiative energy losses due to numerical overcooling, resulting in the much weaker, momentum-driven winds and more stellar mass formed compared to the kinetic feedback with similar energies per kick event.

Fig. \ref{fig:dispersion_SFR_kick} shows the mock observed velocity dispersion, which is given by equation (\ref{eq: sigma_obs}), versus the SFR surface density, $\Sigma_{\rm SFR}$, for the five models with different $\Delta v_{\rm kick}$. The galaxy is viewed face-on and the SFR surface density and velocity dispersion are computed in the same way as in Fig. \ref{fig:dispersion_SFR}. Also, as in Fig. \ref{fig:dispersion_SFR}, we compare with the observational data from \citet{2017MNRAS.470.4573Z} and \citet{2022ApJ...928...58L}. We find that at nearly all $\Sigma_{\rm SFR}$, the turbulent velocity dispersion decreases with $\Delta v_{\rm kick}$ for $10^2 \lesssim \Delta v_{\rm kick}\lesssim 10^3$ \kms{}, converges for $\Delta v_{\rm kick}\lesssim 10^2$ \kms{}, and -- in all five models -- overlaps with the comparison data. One notable deviation from these trends is the $\Delta v_{\rm kick} = 600$ \kms{} model, which shows the highest velocity dispersion among the five models at $\Sigma_{\rm SFR}>0.1 \, \rm M_\odot \, yr^{-1} \, kpc^{-2}$. The reason is its consistently higher SFR at $0.25<t<0.5$ Gyr (Fig. \ref{fig:vary_vkick_compilation_of_plots}).

\subsection{Variations in numerical resolution}
\label{sec: resolution}

In this section, we explore how our results depend on the numerical resolution of the simulation. This analysis is carried out using the simulations with the fiducial subgrid model for both the Milky Way-mass and the dwarf galaxies, which has $f_{\rm kin}=0.1$ and $\Delta v_{\rm kick}=50$ \kms. Our fiducial resolution for the H12 galaxy is $m_{\rm gas} = 10^5 \, \rm M_\odot$ (M5) and the gravitational softening length is $0.2$ kpc. For the H12 galaxy, we consider three other resolutions, in which the gas particle mass is decreased by a factor of 8 (M4) or increased by factors of 8 (M6) and 64 (M7) with respect to the fiducial resolution (M5). Our fiducial resolution for the H10 galaxy is $m_{\rm gas} = 1.56 \times 10^3  \, \rm M_\odot$ (M3) and like in the H12 case, here we take three variations where the gas particle mass is decreased by a factor of $8$ (M2) or increased by factors of $8$ (M4) and $64$ (M5), relative to $m_{\rm gas} = 1.56 \times 10^3 \, \rm M_\odot$. Together with the change in the gas particle mass, we adjust the gravitational softening length: for each factor-of-eight increase (decrease) in gas particle mass, we increase (decrease) the softening length by a factor of 2.

Fig. \ref{fig:SFR_and_mass_loading_resolution} displays the evolution of galaxy SFRs (left-hand panels) and of the wind mass loading factors at height $d=10\pm 0.5$ kpc (right-hand panels). The top and bottom panels show the H12 and H10 galaxies, respectively. Curves with different colours depict simulations with different resolutions, as indicated in the legends. We find excellent convergence for both the SFR and the wind mass loading, which holds for both the Milky Way-mass galaxy and the dwarf galaxy. We note that we also find good convergence for the KS star-formation law (not shown here, but see Nobels et al., in preparation).

\section{Discussion}
\label{sec: discussion}

\subsection{Comparison with previous work}

\citetalias{DallaVecchiaSchaye2008} ran simulations of isolated Milky Way-mass and dwarf galaxies with initial conditions and resolution similar to ours. They used a purely kinetic stochastic model for SN feedback with kick velocities of $424$, $600$, and $848$ \kms (parametrized by the wind mass loading). The kicks were carried out as single-particle kicks with a fixed velocity in random angular directions. The kicked particles became `wind particles' for a period of time of 15 Myr during which they were not allowed to be kicked again and could not form stars. For the Milky Way-mass galaxy, our purely kinetic models with $\Delta v_{\rm kick} = 600$ and $1000$ \kms{} produce SFRs and wind mass loading factors that evolve qualitatively similarly and differ within roughly a factor of three from those found in \citetalias{DallaVecchiaSchaye2008} for the kick velocities of $424$ \kms{} and $848$ \kms. This quantitative agreement may seem surprising given the large differences between the \citetalias{DallaVecchiaSchaye2008} model and ours, which include the criterion for star formation, the modelling of the ISM (\citetalias{DallaVecchiaSchaye2008} imposed an eEOS), early stellar feedback processes, and the implementation of the SN feedback itself. We attribute this similarity to the fact that (i) for relatively large kick velocities ($\gtrsim 5 \times 10^2$ \kms), the gas radiative energy losses are kept low and the subsequent evolution of the gas particle(s) that received the SN energy is determined mostly by the hydro solver; (ii) both in \citetalias{DallaVecchiaSchaye2008} and our model, gas particles are kicked in random angular directions resulting in a statistically isotropic distribution of kicks. By running additional simulations of the dwarf galaxy with M3 resolution for $\Delta v_{\rm kick} = 600$ and $1000$ \kms{} (not shown in this work), we verified that our SFRs and mass loading factors are comparable to those from \citetalias{DallaVecchiaSchaye2008} not only for the Milky Way-mass galaxy but also for the dwarf galaxy. Finally, we note that \citetalias{DallaVecchiaSchaye2008} found that the galaxy's SFH becomes nearly insensitive to $\Delta v_{\rm kick}$ if $\Delta v_{\rm kick}$ is higher than a certain critical value that produces galactic winds powerful enough to escape the galaxy. Specifically, the SFRs in their Milky Way-mass (dwarf) galaxy are very similar for $\Delta v_{\rm kick} = 600$ and $848$ \kms{} ($\Delta v_{\rm kick} = 424, 600$, and $848$ \kms{}). In our additional tests (not presented in this work), we found slightly larger critical values for $\Delta v_{\rm kick}$: our dwarf galaxy shows convergence in terms of its SFH for $\Delta v_{\rm kick}\gtrsim 500$ km s$^{-1}$ and the Milky Way-mass galaxy for $\Delta v_{\rm kick}\gtrsim 1000$ km s$^{-1}$.

\citetalias{2012MNRAS.426..140D} ran simulations of isolated galaxies with thermal stochastic SN feedback using the same set-up as in \citetalias{DallaVecchiaSchaye2008} with a resolution comparable to ours. They considered SN heating temperatures in the range from $\Delta T = 10^{6.5}$ K to $\Delta T = 10^{8.5}$ K, with $\Delta T = 10^{7.5}$ K being the fiducial value. Their fiducial value for the energy per SN in units of $10^{51}$ erg was set to $f_{\rm E} = 1$. However, \citetalias{2012MNRAS.426..140D} (and also \citetalias{DallaVecchiaSchaye2008}) integrated the stellar IMF from $m_{\rm min} = 6\, \rm M_\odot$, which for the value adopted in this work, $m_{\rm min} = 8 \, \rm M_\odot$,  would require $f_{\rm E} \approx 1.5$ to obtain the same total energy budget. The SFRs and wind mass loading in our runs with purely thermal feedback (\textsf{H12\_M5\_fkin0p0} and \textsf{H10\_M3\_fkin0p0}) resemble those in \citetalias{2012MNRAS.426..140D} for $\Delta T = 10^{7.5}$ K. However, unlike \citetalias{2012MNRAS.426..140D}, we are unable to obtain the KS relation with the correct slope using the models with purely thermal feedback. This difference stems from the fact that in the prescription for star formation, we compute the SFR using the \citet{1959ApJ...129..243S} law, whereas \citetalias{2012MNRAS.426..140D} adopted the pressure law of \cite{2008MNRAS.383.1210S}, which is designed to reproduce the observed KS law for self-gravitating discs. Another noticeable difference between our model and theirs is that in our model the SN energy is distributed isotropically, while in \citetalias{2012MNRAS.426..140D} the energy was distributed in proportion to the gas mass. \citet{Chaikin2022} showed that compared with mass weighting, the isotropic scheme yields SFRs that are a factor of a few smaller.

\subsection{Interpreting the different channels for injecting SN energy}

The models with both kinetic and thermal SN feedback explored in this work (at fixed $f_{\rm E}=2$ and $\Delta T=10^{7.5}$ K) yield SFRs that are stable over time (Fig. \ref{fig:SFH_main}). Some of these results would likely change if the galaxy no longer remains in an isolated environment. In a more realistic, cosmological setting, the galaxy will accrete gas from the halo. For the H12 halo, this means that the purely kinetic model with $\Delta v_{\rm kick} = 50$ \kms{}, which fails to generate steady galactic winds (Fig. \ref{fig:Mass_loading}), will not be able to counteract the cosmic accretion. As a consequence, the gas will keep cooling and precipitating onto the disc at a high rate, eventually making the galaxy overly massive and bulgy (with respect to the expected stellar mass and morphology for its halo mass). For galaxies in a cosmological simulation to look realistic, a powerful mode of SN feedback -- either thermal or kinetic -- is a requirement. Indeed, the \textsc{eagle} simulations \citep{2015MNRAS.446..521S} used the  \citetalias{2012MNRAS.426..140D} thermal model with $\Delta T=10^{7.5}$ K, while the \textsc{owls} simulations  \citep{2010MNRAS.402.1536S} opted for the \citetalias{DallaVecchiaSchaye2008} kinetic model with $\Delta v_{\rm kick} = 600$ \kms.  The need for high $\Delta T$ or $\Delta v_{\rm kick}$ may be partly alleviated if the model for SN feedback makes use of hydrodynamical decoupling or delayed cooling of wind particles. However, even then high energy injections and high wind velocities are preferred. For example, in the \textsc{Illustris} \citep{2014MNRAS.444.1518V} and \textsc{Simba} \citep{2019MNRAS.486.2827D} simulations, whose SN feedback includes decoupled, star-formation driven winds, wind particles in a Milky Way-mass halo at redshift $z=0$ are launched with velocities $\approx 400$ \kms. The decoupled stellar winds in the \textsc{IllustrisTNG} simulations \citep{Pillepich2018} are launched with a yet higher speed, $\approx 800$ \kms for a $z=0$ Milky Way-mass halo, which leads to an overall more efficient stellar feedback than in the \textsc{Illustris} model.

 In contrast, at numerical resolutions much higher than in our work, e.g., in simulations of dwarf galaxies, feedback from SNe can be modelled as a direct thermal dump, without relying on an intricate subgrid prescription \citep[e.g.][]{Gutcke2021,2022MNRAS.513.1372G,2022arXiv220810528H}. Moreover, if the resolution $m_{\rm gas} \lesssim 10 \, \rm M_\odot$, the SN feedback is always thermal-kinetic. A detonating SN from a single star produces a blast wave, which consists of a hot, low-density bubble with temperatures exceeding $10^7$ K, and a dense, colder shell made up of the initial SN ejecta and the ISM gas swept up by the blast \citep[e.g.][]{2011piim.book.....D}. If the energy-conserving phase of the blast evolution is well resolved, then regardless of how the SN energy has been initially injected (thermally or kinetically), the ratio of the thermal to kinetic energy within the blast wave will approach $0.39$ during this phase \citep[e.g.][]{2015ApJ...802...99K}. At some point later in time, radiative cooling in the shell will become dominant and the blast will enter its momentum-conserving phase when the total radial momentum reached by the blast is\footnote{Assuming an SN energy of $10^{51}$ erg, an average ISM density of $n_{\rm H} = 1 \, \rm cm^{-3}$, and solar metallicity.} $\approx 3\times 10^5 \, \rm M_\odot$ \kms{} \citep[e.g.][]{2015ApJ...802...99K}, which scales with the SN energy (in units of $10^{51}$ erg) as $f_{\rm E}^{13/14}$ \citep[e.g.][]{1988ApJ...334..252C}. We can use equation (\ref{eq:number_of_kicks}) to estimate the total momentum injected into the gas by our model for different values of $f_{\rm kin}$ and $\Delta v_{\rm kick}$. We find
 \begin{align}
    \langle p_{\rm tot} \rangle &= 2\, \langle N_{\rm kick,tot}\rangle \, m_{\rm gas} \, \Delta v_{\rm kick} \nonumber \\
    &= 2.4 \times 10^{5} \, \mathrm{M_\odot \, km \, s^{-1}} \, f_{\rm E}\,  \left(\frac{f_{\rm kin}}{0.1}\right) \left(\frac{m_{\rm *}}{\mathrm{10^2 \, \rm M_\odot}}\right)\,\left(\frac{\Delta v_{\rm kick}}{50 \, \rm km \, s^{-1}}\right)^{-1} \, ,
 \end{align}
 where the value $m_{*} = 10^2 \, \rm M_\odot$ is comparable to the mass of a simple stellar population expected to result in a single SN. For our fiducial values of $f_{\rm kin}$ and $\Delta v_{\rm kick}$ this gives a value for $\langle p_{\rm tot} \rangle$ that is close to the theoretical expectation, though we did not try to match it. Thus, physically, the low-velocity kicks in the kinetic channel, which disrupt molecular clouds and drive turbulence, can be thought of as SN blasts that entered the momentum-conserving phase, whereas the role of the thermal channel with powerful energy injections, besides driving galactic-scale outflows, is to generate the hot ISM phase ($T \gtrsim 10^6$ K) that is expected from clustered SN feedback, but which cannot arise naturally at our relatively low resolution.

\subsection{The effect of correcting for relative gas-star motion}

The potential significance of the relative gas-star motion was pointed out already in the earliest hydrodynamical simulations including kinetic feedback from SNe \citep{NavarroWhite1993}. As explained by \citet{NavarroWhite1993} and more recently by \citet{hopkinsfeedback2018}, including the correction for the relative motion has pros and cons. Namely, accounting for the relative star-gas motion enables exact energy conservation in the SN feedback but may result in very large kick velocities if the gas is rapidly and coherently approaching the star particle. In this case, the kick velocity will be such that the direction of motion of the converging gas flow is reversed. The implementation with the absence of the star-gas motion correction does not suffer from this issue but possesses a different potential problem: since it is unable to conserve energy, kicking gas particles that are rapidly receding from the star particles leads to an excessive amount of kinetic energy being injected. 

\citet{hopkinsfeedback2018} studied the importance of accounting for the gas-star motion in the framework of their model for `mechanical' SN feedback, which releases momentum and thermal energy whose values are taken from high-resolution simulations of isolated SNe. They ran zoom-in simulations of a Milky Way-mass galaxy and a dwarf galaxy at several gas-mass resolutions using the mesh-free, Lagrangian code \textsc{gizmo} \citep{2015MNRAS.450...53H} in its finite-mass mode. \citet{hopkinsfeedback2018} showed that for a resolution of $m_{\rm gas} = 4.5 \times 10^{5} \, \rm M_{\odot}$ in a Milky Way-mass galaxy and $m_{\rm gas} = 2 \times 10^{3} \, \rm M_{\odot}$ in a dwarf galaxy, their implementations with and without the relative-motion correction lead to differences in galaxy properties that are small, and which become even smaller with increasing resolution (based on additional tests not shown in their paper). They attributed this outcome to the fact that the events with coherently (and rapidly) inflowing or outflowing gas around star particles are quite rare, and that if they do occur, then the differences will tend to average out in time and space. In Appendix \ref{sec:rel_motion_multi_kick} we show that in our simulations the differences caused by the relative motion correction are also small for our fiducial model. However, for purely kinetic feedback and a low desired kick velocity the correction is significant, though not as important as preventing particles from being kicked multiple times per time-step.

\section{Conclusions}
\label{sec: conclusions}

We presented a new stochastic isotropic thermal-kinetic model for SN feedback that is suitable for large cosmological simulations of galaxy formation, including those that (partly) resolve a cold ISM. Releasing SN energy in two different forms accomplishes two different goals: strong galactic winds and the hot ISM phase are generated by powerful but rare injections of thermal energy, while small but frequent kinetic injections help drive turbulence in the neutral ISM. These two SN feedback channels can be thought of as representing, respectively, superbubbles resulting from clustered SNe and the momentum injected by isolated SNe. Our model for SN feedback manifestly conserves energy, linear and angular momentum, and is statistically isotropic.

Our model builds on the earlier works of \citetalias{DallaVecchiaSchaye2008} and \citetalias{2012MNRAS.426..140D} and is fully specified by four free parameters: (i) the amount of energy per single SN in units of $10^{51}$ erg, $f_{\rm E}$; (ii) the fraction of SN energy injected in kinetic form, $f_{\rm kin}$; (iii) the temperature increase $\Delta T$, parametrizing the amount of energy deposited in one thermal injection event; and (iv) the desired kick velocity, $\Delta v_{\rm kick}$, defining the energy of one kick event. Our main findings are as follows:

\begin{itemize}
    \item The purely thermal model ($f_{\rm kin}=0$) with a high heating temperature ($\Delta T=10^{7.5}$ K) and the purely kinetic model ($f_{\rm kin}=1$) with low-energy kicks ($\Delta v_{\rm kick} = 50$ \kms) result in galaxy properties that differ in many respects. The kinetic model yields a greater amount of gas in the ISM (Fig. \ref{fig:morphology_M12_main}), whose surface density rises more steeply towards the galactic centre (Fig. \ref{fig:surface_density_M12_main}), and has higher velocity dispersion in both the H\textsc{i} and H$_2$ gas (Fig. \ref{fig:velocity_dispersion}). In contrast, the thermal model generates a hot phase of the ISM (Fig. \ref{fig:morphology_M12_main}) and is able to sustain strong galactic winds (Fig. \ref{fig:Mass_loading}). These differences imply that in the thermal model, star formation is regulated mainly by the ejection of gas from the disc, while in the kinetic model with low-velocity kicks, it is regulated mainly through the increase of the ISM velocity dispersion. 
    
    \item In the models including kinetic feedback, accounting for the gas motion around stars when kicking the gas neighbours, which is necessary in order to conserve energy, leads to a distribution of the actual kick velocities (Fig. \ref{fig:distribution_v_kick}). The width of the distribution decreases for higher desired kick velocities, $\Delta v_{\rm kick}$, with the distribution becoming more peaked around $\Delta v_{\rm kick}$.
    
    \item Shortly after the low-energy kicks ($\Delta v_{\rm kick} = 50$ \kms), the turbulent velocity dispersion in the kicked neutral gas increases by a factor of a few, which in turn leads to a drop in the mass fraction of the neutral gas that is star-forming (Fig. \ref{fig:velocity_dispersion_vs_time}).  However, the drop in the local fraction of star-forming gas does not necessary lead to a decrease in the galaxy's total SFR. In fact, the models with higher $f_{\rm kin}$ have higher total SFRs at late times ($t > 0.5$ Gyr, Fig. \ref{fig:SFH_main}) because the galaxy is able to retain more gas within the ISM. A larger $f_{\rm kin}$ results in a higher gas fraction because it leads to a higher velocity dispersion, which prolongs the time-scale on which gas is converted into stars, and because it implies less clustered SNe with fewer high thermal energy injections, which reduces the amount of outflowing gas.
        
    \item Neither the purely thermal nor the purely kinetic model can fully match the observed KS star-formation relation in the Milky Way-mass galaxy (Fig. \ref{fig:KS_law_main}). In order to obtain a relation with the (asymptotically) correct slope, a small (but non-zero) fraction of kinetic energy ($f_{\rm kin}\approx 0.1$) is required. Otherwise, if $f_{\rm kin}=0$, the relation becomes too steep, whereas if $f_{\rm kin}=1$, the relation is cut off below a too-high gas surface density and undershoots the observed data.
    
    \item Irrespective of $f_{\rm kin}$, the spatially resolved H\textsc{i} velocity dispersion is an increasing function of the SFR surface density (Fig. \ref{fig:dispersion_SFR}) and all values of $f_{\rm kin}$ yield reasonable agreement with the observational data from \citet{2017MNRAS.470.4573Z} and \citet{2022ApJ...928...58L}, with the models with $f_{\rm kin}=0.1$ and $0.3$ being closest to the average values reported by those observations.
    
    \item In the purely kinetic models, the H\textsc{i} turbulent velocity dispersion (wind mass loading) decreases (increases) with increasing $\Delta v_{\rm kick}$ for $10^2 \lesssim \Delta v_{\rm kick}\lesssim 10^3$ \kms{} and converges for $\Delta v_{\rm kick}\lesssim 10^2$ \kms{} (Figs. \ref{fig:vary_vkick_compilation_of_plots} and \ref{fig:dispersion_SFR_kick}). The galaxy SFRs are also mostly converged with decreasing $\Delta v_{\rm kick}$ for $\Delta v_{\rm kick}\lesssim 10^2$ \kms{}. This indicates that as long as $\Delta v_{\rm kick} \lesssim 10^2$ \kms{}, the galaxy properties are largely insensitive to the exact value of $\Delta v_{\rm kick}$. 
    
    \item For our fiducial model ($f_{\rm kin} = 0.1$, $\Delta v_{\rm kick} = 50$ \kms), the SFRs and wind mass loading factors show excellent convergence with the numerical resolution over several orders of magnitude in gas particle mass, which holds for both the Milky Way-mass galaxy and the dwarf galaxy (Fig. \ref{fig:SFR_and_mass_loading_resolution}).

\end{itemize}

We conclude that the thermal channel with a high heating temperature and the kinetic channel with low-energy kicks naturally complement one another. Together they enable simulation predictions that are remarkably insensitive to the numerical resolution and that reproduce key galaxy observables like the spatially resolved star formation rates and H\textsc{i} velocity dispersion. In future work, we will show how both feedback channels perform in a cosmological simulation of galaxy formation including a cold ISM.

\section*{Acknowledgements}

We thank Camila Correa, Alexander Richings, and the anonymous referee for their useful comments. This work used the DiRAC@Durham facility managed by the Institute for Computational Cosmology on behalf of the STFC DiRAC HPC Facility (www.dirac.ac.uk). The equipment was funded by BEIS capital funding via STFC capital grants ST/K00042X/1, ST/P002293/1, ST/R002371/1 and ST/S002502/1, Durham University and STFC operations grant ST/R000832/1. DiRAC is part of the National e-Infrastructure. EC is supported by the funding from the European Union's Horizon 2020 research and innovation programme under the Marie Skłodowska-Curie grant agreement No 860744 (BiD4BESt). ABL acknowledges support from the European Research Council (ERC) under the European Union’s Horizon 2020 research and innovation program under the grant agreement 101026328, and UNIMIB’s Fondo di Ateneo Quota Competitiva (project 2020-ATESP-0133). The research in this paper made use of the \textsc{swift} open-source simulation code (\url{http://www.swiftsim.com}, \citealt{2018ascl.soft05020S}) version 0.9.0. The data analysis was carried out with the use of \textsc{swiftsimio} \citep{Borrow2020simio,2021arXiv210605281B}, \textsc{numpy} \citep{2020Natur.585..357H}, and \textsc{matplotlib} \citep{2007CSE.....9...90H}.

\section*{Data Availability}

The \textsc{swift} simulation code is publicly available at \href{http://www.swiftsim.com}{http://www.swiftsim.com}. The data underlying this article will be shared on reasonable request to the corresponding author.



\bibliographystyle{mnras}
\bibliography{example} 

\begin{thebibliography}{}
\makeatletter
\relax
\def\mn@urlcharsother{\let\do\@makeother \do\$\do\&\do\#\do\^\do\_\do\%\do\~}
\def\mn@doi{\begingroup\mn@urlcharsother \@ifnextchar [ {\mn@doi@}
  {\mn@doi@[]}}
\def\mn@doi@[#1]#2{\def\@tempa{#1}\ifx\@tempa\@empty \href
  {http://dx.doi.org/#2} {doi:#2}\else \href {http://dx.doi.org/#2} {#1}\fi
  \endgroup}
\def\mn@eprint#1#2{\mn@eprint@#1:#2::\@nil}
\def\mn@eprint@arXiv#1{\href {http://arxiv.org/abs/#1} {{\tt arXiv:#1}}}
\def\mn@eprint@dblp#1{\href {http://dblp.uni-trier.de/rec/bibtex/#1.xml}
  {dblp:#1}}
\def\mn@eprint@#1:#2:#3:#4\@nil{\def\@tempa {#1}\def\@tempb {#2}\def\@tempc
  {#3}\ifx \@tempc \@empty \let \@tempc \@tempb \let \@tempb \@tempa \fi \ifx
  \@tempb \@empty \def\@tempb {arXiv}\fi \@ifundefined
  {mn@eprint@\@tempb}{\@tempb:\@tempc}{\expandafter \expandafter \csname
  mn@eprint@\@tempb\endcsname \expandafter{\@tempc}}}

\bibitem[\protect\citeauthoryear{{Asplund}, {Grevesse}, {Sauval}  \&
  {Scott}}{{Asplund} et~al.}{2009}]{2009ARA&A..47..481A}
{Asplund} M.,  {Grevesse} N.,  {Sauval} A.~J.,   {Scott} P.,  2009, \mn@doi
  [\araa] {10.1146/annurev.astro.46.060407.145222}, \href
  {https://ui.adsabs.harvard.edu/abs/2009ARA&A..47..481A} {47, 481}

\bibitem[\protect\citeauthoryear{{Bigiel}, {Leroy}, {Walter}, {Brinks}, {de
  Blok}, {Madore}  \& {Thornley}}{{Bigiel} et~al.}{2008}]{Bigiel2008AJ}
{Bigiel} F.,  {Leroy} A.,  {Walter} F.,  {Brinks} E.,  {de Blok} W.~J.~G.,
  {Madore} B.,   {Thornley} M.~D.,  2008, \mn@doi [\aj]
  {10.1088/0004-6256/136/6/2846}, \href
  {https://ui.adsabs.harvard.edu/abs/2008AJ....136.2846B} {136, 2846}

\bibitem[\protect\citeauthoryear{{Bigiel}, {Leroy}, {Walter}, {Blitz},
  {Brinks}, {de Blok}  \& {Madore}}{{Bigiel} et~al.}{2010}]{Bigiel2010AJ}
{Bigiel} F.,  {Leroy} A.,  {Walter} F.,  {Blitz} L.,  {Brinks} E.,  {de Blok}
  W.~J.~G.,   {Madore} B.,  2010, \mn@doi [\aj] {10.1088/0004-6256/140/5/1194},
  \href {https://ui.adsabs.harvard.edu/abs/2010AJ....140.1194B} {140, 1194}

\bibitem[\protect\citeauthoryear{Borrow \& Borrisov}{Borrow \&
  Borrisov}{2020}]{Borrow2020simio}
Borrow J.,  Borrisov A.,  2020, \mn@doi [Journal of Open Source Software]
  {10.21105/joss.02430}, 5, 2430

\bibitem[\protect\citeauthoryear{{Borrow} \& {Kelly}}{{Borrow} \&
  {Kelly}}{2021}]{2021arXiv210605281B}
{Borrow} J.,  {Kelly} A.~J.,  2021, arXiv e-prints, \href
  {https://ui.adsabs.harvard.edu/abs/2021arXiv210605281B} {p. arXiv:2106.05281}

\bibitem[\protect\citeauthoryear{{Borrow}, {Schaller}, {Bower}  \&
  {Schaye}}{{Borrow} et~al.}{2022}]{2022MNRAS.511.2367B}
{Borrow} J.,  {Schaller} M.,  {Bower} R.~G.,   {Schaye} J.,  2022, \mn@doi
  [\mnras] {10.1093/mnras/stab3166}, \href
  {https://ui.adsabs.harvard.edu/abs/2022MNRAS.511.2367B} {511, 2367}

\bibitem[\protect\citeauthoryear{{Chabrier}}{{Chabrier}}{2003}]{Chabrier2003}
{Chabrier} G.,  2003, \mn@doi [\pasp] {10.1086/376392}, \href
  {https://ui.adsabs.harvard.edu/abs/2003PASP..115..763C} {115, 763}

\bibitem[\protect\citeauthoryear{{Chaikin}, {Schaye}, {Schaller}, {Bah{\'e}},
  {Nobels}  \& {Ploeckinger}}{{Chaikin} et~al.}{2022}]{Chaikin2022}
{Chaikin} E.,  {Schaye} J.,  {Schaller} M.,  {Bah{\'e}} Y.~M.,  {Nobels} F.
  S.~J.,   {Ploeckinger} S.,  2022, \mn@doi [\mnras] {10.1093/mnras/stac1132},
  \href {https://ui.adsabs.harvard.edu/abs/2022MNRAS.514..249C} {514, 249}

\bibitem[\protect\citeauthoryear{{Cioffi}, {McKee}  \& {Bertschinger}}{{Cioffi}
  et~al.}{1988}]{1988ApJ...334..252C}
{Cioffi} D.~F.,  {McKee} C.~F.,   {Bertschinger} E.,  1988, \mn@doi [\apj]
  {10.1086/166834}, \href
  {https://ui.adsabs.harvard.edu/abs/1988ApJ...334..252C} {334, 252}

\bibitem[\protect\citeauthoryear{{Crain} et~al.,}{{Crain}
  et~al.}{2015}]{Crain2015}
{Crain} R.~A.,  et~al., 2015, \mn@doi [\mnras] {10.1093/mnras/stv725}, \href
  {https://ui.adsabs.harvard.edu/abs/2015MNRAS.450.1937C} {450, 1937}

\bibitem[\protect\citeauthoryear{{Dalla Vecchia} \& {Schaye}}{{Dalla Vecchia}
  \& {Schaye}}{2008}]{DallaVecchiaSchaye2008}
{Dalla Vecchia} C.,  {Schaye} J.,  2008, \mn@doi [\mnras]
  {10.1111/j.1365-2966.2008.13322.x}, \href
  {https://ui.adsabs.harvard.edu/abs/2008MNRAS.387.1431D} {387, 1431}

\bibitem[\protect\citeauthoryear{{Dalla Vecchia} \& {Schaye}}{{Dalla Vecchia}
  \& {Schaye}}{2012}]{2012MNRAS.426..140D}
{Dalla Vecchia} C.,  {Schaye} J.,  2012, \mn@doi [\mnras]
  {10.1111/j.1365-2966.2012.21704.x}, \href
  {https://ui.adsabs.harvard.edu/abs/2012MNRAS.426..140D} {426, 140}

\bibitem[\protect\citeauthoryear{{Dav{\'e}}, {Angl{\'e}s-Alc{\'a}zar},
  {Narayanan}, {Li}, {Rafieferantsoa}  \& {Appleby}}{{Dav{\'e}}
  et~al.}{2019}]{2019MNRAS.486.2827D}
{Dav{\'e}} R.,  {Angl{\'e}s-Alc{\'a}zar} D.,  {Narayanan} D.,  {Li} Q.,
  {Rafieferantsoa} M.~H.,   {Appleby} S.,  2019, \mn@doi [\mnras]
  {10.1093/mnras/stz937}, \href
  {https://ui.adsabs.harvard.edu/abs/2019MNRAS.486.2827D} {486, 2827}

\bibitem[\protect\citeauthoryear{{Draine}}{{Draine}}{2011}]{2011piim.book.....D}
{Draine} B.~T.,  2011, {Physics of the Interstellar and Intergalactic Medium}

\bibitem[\protect\citeauthoryear{{Dubois}, {Volonteri}, {Silk}, {Devriendt},
  {Slyz}  \& {Teyssier}}{{Dubois} et~al.}{2015}]{Dubois2015}
{Dubois} Y.,  {Volonteri} M.,  {Silk} J.,  {Devriendt} J.,  {Slyz} A.,
  {Teyssier} R.,  2015, \mn@doi [\mnras] {10.1093/mnras/stv1416}, \href
  {https://ui.adsabs.harvard.edu/abs/2015MNRAS.452.1502D} {452, 1502}

\bibitem[\protect\citeauthoryear{{Dubois} et~al.,}{{Dubois}
  et~al.}{2021}]{2021A&A...651A.109D}
{Dubois} Y.,  et~al., 2021, \mn@doi [\aap] {10.1051/0004-6361/202039429}, \href
  {https://ui.adsabs.harvard.edu/abs/2021A&A...651A.109D} {651, A109}

\bibitem[\protect\citeauthoryear{{Durier} \& {Dalla Vecchia}}{{Durier} \&
  {Dalla Vecchia}}{2012}]{2012MNRAS.419..465D}
{Durier} F.,  {Dalla Vecchia} C.,  2012, \mn@doi [\mnras]
  {10.1111/j.1365-2966.2011.19712.x}, \href
  {https://ui.adsabs.harvard.edu/abs/2012MNRAS.419..465D} {419, 465}

\bibitem[\protect\citeauthoryear{{Eldridge}, {Stanway}, {Xiao}, {McClelland },
  {Taylor}, {Ng}, {Greis}  \& {Bray}}{{Eldridge} et~al.}{2017}]{BPASS2017}
{Eldridge} J.~J.,  {Stanway} E.~R.,  {Xiao} L.,  {McClelland } L.~A.~S.,
  {Taylor} G.,  {Ng} M.,  {Greis} S.~M.~L.,   {Bray} J.~C.,  2017, \mn@doi
  [\pasa] {10.1017/pasa.2017.51}, \href
  {https://ui.adsabs.harvard.edu/abs/2017PASA...34...58E} {34, e058}

\bibitem[\protect\citeauthoryear{{Faucher-Gigu{\`e}re}}{{Faucher-Gigu{\`e}re}}{2020}]{2020MNRAS.493.1614F}
{Faucher-Gigu{\`e}re} C.-A.,  2020, \mn@doi [\mnras] {10.1093/mnras/staa302},
  \href {https://ui.adsabs.harvard.edu/abs/2020MNRAS.493.1614F} {493, 1614}

\bibitem[\protect\citeauthoryear{{Feldmann} et~al.,}{{Feldmann}
  et~al.}{2023}]{2023MNRAS.522.3831F}
{Feldmann} R.,  et~al., 2023, \mn@doi [\mnras] {10.1093/mnras/stad1205}, \href
  {https://ui.adsabs.harvard.edu/abs/2023MNRAS.522.3831F} {522, 3831}

\bibitem[\protect\citeauthoryear{{Ferland} et~al.,}{{Ferland}
  et~al.}{2017}]{Cloudy17}
{Ferland} G.~J.,  et~al., 2017, \rmxaa, \href
  {https://ui.adsabs.harvard.edu/abs/2017RMxAA..53..385F} {53, 385}

\bibitem[\protect\citeauthoryear{{Gentry}, {Madau}  \& {Krumholz}}{{Gentry}
  et~al.}{2020}]{Gentry2020}
{Gentry} E.~S.,  {Madau} P.,   {Krumholz} M.~R.,  2020, \mn@doi [\mnras]
  {10.1093/mnras/stz3440}, \href
  {https://ui.adsabs.harvard.edu/abs/2020MNRAS.492.1243G} {492, 1243}

\bibitem[\protect\citeauthoryear{{Gerritsen}}{{Gerritsen}}{1997}]{1997PhDT........19G}
{Gerritsen} J.~P.~E.,  1997, PhD thesis, -

\bibitem[\protect\citeauthoryear{{Gutcke}, {Pakmor}, {Naab}  \&
  {Springel}}{{Gutcke} et~al.}{2021}]{Gutcke2021}
{Gutcke} T.~A.,  {Pakmor} R.,  {Naab} T.,   {Springel} V.,  2021, \mn@doi
  [\mnras] {10.1093/mnras/staa3875}, \href
  {https://ui.adsabs.harvard.edu/abs/2021MNRAS.501.5597G} {501, 5597}

\bibitem[\protect\citeauthoryear{{Gutcke}, {Pakmor}, {Naab}  \&
  {Springel}}{{Gutcke} et~al.}{2022}]{2022MNRAS.513.1372G}
{Gutcke} T.~A.,  {Pakmor} R.,  {Naab} T.,   {Springel} V.,  2022, \mn@doi
  [\mnras] {10.1093/mnras/stac867}, \href
  {https://ui.adsabs.harvard.edu/abs/2022MNRAS.513.1372G} {513, 1372}

\bibitem[\protect\citeauthoryear{{Harris} et~al.,}{{Harris}
  et~al.}{2020}]{2020Natur.585..357H}
{Harris} C.~R.,  et~al., 2020, \mn@doi [\nat] {10.1038/s41586-020-2649-2},
  \href {https://ui.adsabs.harvard.edu/abs/2020Natur.585..357H} {585, 357}

\bibitem[\protect\citeauthoryear{{Hayward} \& {Hopkins}}{{Hayward} \&
  {Hopkins}}{2017}]{2017MNRAS.465.1682H}
{Hayward} C.~C.,  {Hopkins} P.~F.,  2017, \mn@doi [\mnras]
  {10.1093/mnras/stw2888}, \href
  {https://ui.adsabs.harvard.edu/abs/2017MNRAS.465.1682H} {465, 1682}

\bibitem[\protect\citeauthoryear{{Hernquist}}{{Hernquist}}{1990}]{hernquist1990}
{Hernquist} L.,  1990, \mn@doi [\apj] {10.1086/168845}, \href
  {https://ui.adsabs.harvard.edu/abs/1990ApJ...356..359H} {356, 359}

\bibitem[\protect\citeauthoryear{{Hopkins}}{{Hopkins}}{2015}]{2015MNRAS.450...53H}
{Hopkins} P.~F.,  2015, \mn@doi [\mnras] {10.1093/mnras/stv195}, \href
  {https://ui.adsabs.harvard.edu/abs/2015MNRAS.450...53H} {450, 53}

\bibitem[\protect\citeauthoryear{{Hopkins}, {Quataert}  \& {Murray}}{{Hopkins}
  et~al.}{2012}]{2012MNRAS.421.3522H}
{Hopkins} P.~F.,  {Quataert} E.,   {Murray} N.,  2012, \mn@doi [\mnras]
  {10.1111/j.1365-2966.2012.20593.x}, \href
  {https://ui.adsabs.harvard.edu/abs/2012MNRAS.421.3522H} {421, 3522}

\bibitem[\protect\citeauthoryear{{Hopkins}, {Kere{\v{s}}}, {O{\~n}orbe},
  {Faucher-Gigu{\`e}re}, {Quataert}, {Murray}  \& {Bullock}}{{Hopkins}
  et~al.}{2014}]{2014MNRAS.445..581H}
{Hopkins} P.~F.,  {Kere{\v{s}}} D.,  {O{\~n}orbe} J.,  {Faucher-Gigu{\`e}re}
  C.-A.,  {Quataert} E.,  {Murray} N.,   {Bullock} J.~S.,  2014, \mn@doi
  [\mnras] {10.1093/mnras/stu1738}, \href
  {https://ui.adsabs.harvard.edu/abs/2014MNRAS.445..581H} {445, 581}

\bibitem[\protect\citeauthoryear{{Hopkins} et~al.,}{{Hopkins}
  et~al.}{2018a}]{hopkinsfeedback2018}
{Hopkins} P.~F.,  et~al., 2018a, \mn@doi [\mnras] {10.1093/mnras/sty674}, \href
  {https://ui.adsabs.harvard.edu/abs/2018MNRAS.477.1578H} {477, 1578}

\bibitem[\protect\citeauthoryear{{Hopkins} et~al.,}{{Hopkins}
  et~al.}{2018b}]{Hopkins2018Fire2}
{Hopkins} P.~F.,  et~al., 2018b, \mn@doi [\mnras] {10.1093/mnras/sty1690},
  \href {https://ui.adsabs.harvard.edu/abs/2018MNRAS.480..800H} {480, 800}

\bibitem[\protect\citeauthoryear{{Hu} et~al.,}{{Hu}
  et~al.}{2022}]{2022arXiv220810528H}
{Hu} C.-Y.,  et~al., 2022, arXiv e-prints, \href
  {https://ui.adsabs.harvard.edu/abs/2022arXiv220810528H} {p. arXiv:2208.10528}

\bibitem[\protect\citeauthoryear{{Hunter}}{{Hunter}}{2007}]{2007CSE.....9...90H}
{Hunter} J.~D.,  2007, \mn@doi [Computing in Science and Engineering]
  {10.1109/MCSE.2007.55}, \href
  {https://ui.adsabs.harvard.edu/abs/2007CSE.....9...90H} {9, 90}

\bibitem[\protect\citeauthoryear{{Joung} \& {Mac Low}}{{Joung} \& {Mac
  Low}}{2006}]{2006ApJ...653.1266J}
{Joung} M.~K.~R.,  {Mac Low} M.-M.,  2006, \mn@doi [\apj] {10.1086/508795},
  \href {https://ui.adsabs.harvard.edu/abs/2006ApJ...653.1266J} {653, 1266}

\bibitem[\protect\citeauthoryear{{Katz}}{{Katz}}{1992}]{Katz1992}
{Katz} N.,  1992, \mn@doi [\apj] {10.1086/171366}, \href
  {https://ui.adsabs.harvard.edu/abs/1992ApJ...391..502K} {391, 502}

\bibitem[\protect\citeauthoryear{{Katz}, {Weinberg}  \& {Hernquist}}{{Katz}
  et~al.}{1996}]{Katz1996}
{Katz} N.,  {Weinberg} D.~H.,   {Hernquist} L.,  1996, \mn@doi [\apjs]
  {10.1086/192305}, \href
  {https://ui.adsabs.harvard.edu/abs/1996ApJS..105...19K} {105, 19}

\bibitem[\protect\citeauthoryear{{Kennicutt}}{{Kennicutt}}{1998}]{Kennicutt1998ApJ}
{Kennicutt} Robert~C. J.,  1998, \mn@doi [\apj] {10.1086/305588}, \href
  {https://ui.adsabs.harvard.edu/abs/1998ApJ...498..541K} {498, 541}

\bibitem[\protect\citeauthoryear{{Kennicutt} Robert~C. et~al.,}{{Kennicutt}
  et~al.}{2007}]{2007ApJ...671..333K}
{Kennicutt} Robert~C. J.,  et~al., 2007, \mn@doi [\apj] {10.1086/522300}, \href
  {https://ui.adsabs.harvard.edu/abs/2007ApJ...671..333K} {671, 333}

\bibitem[\protect\citeauthoryear{{Kim} \& {Ostriker}}{{Kim} \&
  {Ostriker}}{2015}]{2015ApJ...802...99K}
{Kim} C.-G.,  {Ostriker} E.~C.,  2015, \mn@doi [\apj]
  {10.1088/0004-637X/802/2/99}, \href
  {https://ui.adsabs.harvard.edu/abs/2015ApJ...802...99K} {802, 99}

\bibitem[\protect\citeauthoryear{{Law} et~al.,}{{Law}
  et~al.}{2022}]{2022ApJ...928...58L}
{Law} D.~R.,  et~al., 2022, \mn@doi [\apj] {10.3847/1538-4357/ac5620}, \href
  {https://ui.adsabs.harvard.edu/abs/2022ApJ...928...58L} {928, 58}

\bibitem[\protect\citeauthoryear{{Lu}, {Blanc}  \& {Benson}}{{Lu}
  et~al.}{2015}]{2015ApJ...808..129L}
{Lu} Y.,  {Blanc} G.~A.,   {Benson} A.,  2015, \mn@doi [\apj]
  {10.1088/0004-637X/808/2/129}, \href
  {https://ui.adsabs.harvard.edu/abs/2015ApJ...808..129L} {808, 129}

\bibitem[\protect\citeauthoryear{{McCarthy}, {Schaye}, {Bird}  \& {Le
  Brun}}{{McCarthy} et~al.}{2017}]{2017MNRAS.465.2936M}
{McCarthy} I.~G.,  {Schaye} J.,  {Bird} S.,   {Le Brun} A. M.~C.,  2017,
  \mn@doi [\mnras] {10.1093/mnras/stw2792}, \href
  {https://ui.adsabs.harvard.edu/abs/2017MNRAS.465.2936M} {465, 2936}

\bibitem[\protect\citeauthoryear{{Mitchell}, {Schaye}, {Bower}  \&
  {Crain}}{{Mitchell} et~al.}{2020}]{2020MNRAS.494.3971M}
{Mitchell} P.~D.,  {Schaye} J.,  {Bower} R.~G.,   {Crain} R.~A.,  2020, \mn@doi
  [\mnras] {10.1093/mnras/staa938}, \href
  {https://ui.adsabs.harvard.edu/abs/2020MNRAS.494.3971M} {494, 3971}

\bibitem[\protect\citeauthoryear{{Mitra}, {Dav{\'e}}  \& {Finlator}}{{Mitra}
  et~al.}{2015}]{2015MNRAS.452.1184M}
{Mitra} S.,  {Dav{\'e}} R.,   {Finlator} K.,  2015, \mn@doi [\mnras]
  {10.1093/mnras/stv1387}, \href
  {https://ui.adsabs.harvard.edu/abs/2015MNRAS.452.1184M} {452, 1184}

\bibitem[\protect\citeauthoryear{{Muratov}, {Kere{\v{s}}},
  {Faucher-Gigu{\`e}re}, {Hopkins}, {Quataert}  \& {Murray}}{{Muratov}
  et~al.}{2015}]{2015MNRAS.454.2691M}
{Muratov} A.~L.,  {Kere{\v{s}}} D.,  {Faucher-Gigu{\`e}re} C.-A.,  {Hopkins}
  P.~F.,  {Quataert} E.,   {Murray} N.,  2015, \mn@doi [\mnras]
  {10.1093/mnras/stv2126}, \href
  {https://ui.adsabs.harvard.edu/abs/2015MNRAS.454.2691M} {454, 2691}

\bibitem[\protect\citeauthoryear{{Naab} \& {Ostriker}}{{Naab} \&
  {Ostriker}}{2017}]{2017ARA&A..55...59N}
{Naab} T.,  {Ostriker} J.~P.,  2017, \mn@doi [\araa]
  {10.1146/annurev-astro-081913-040019}, \href
  {https://ui.adsabs.harvard.edu/abs/2017ARA&A..55...59N} {55, 59}

\bibitem[\protect\citeauthoryear{{Navarro} \& {White}}{{Navarro} \&
  {White}}{1993}]{NavarroWhite1993}
{Navarro} J.~F.,  {White} S.~D.~M.,  1993, \mn@doi [\mnras]
  {10.1093/mnras/265.2.271}, \href
  {https://ui.adsabs.harvard.edu/abs/1993MNRAS.265..271N} {265, 271}

\bibitem[\protect\citeauthoryear{{Navarro}, {Frenk}  \& {White}}{{Navarro}
  et~al.}{1996}]{1996ApJ...462..563N}
{Navarro} J.~F.,  {Frenk} C.~S.,   {White} S. D.~M.,  1996, \mn@doi [\apj]
  {10.1086/177173}, \href
  {https://ui.adsabs.harvard.edu/abs/1996ApJ...462..563N} {462, 563}

\bibitem[\protect\citeauthoryear{{Oppenheimer} \& {Dav{\'e}}}{{Oppenheimer} \&
  {Dav{\'e}}}{2006}]{OppenheimerDave2006}
{Oppenheimer} B.~D.,  {Dav{\'e}} R.,  2006, \mn@doi [\mnras]
  {10.1111/j.1365-2966.2006.10989.x}, \href
  {https://ui.adsabs.harvard.edu/abs/2006MNRAS.373.1265O} {373, 1265}

\bibitem[\protect\citeauthoryear{{Ostriker} \& {Shetty}}{{Ostriker} \&
  {Shetty}}{2011}]{2011ApJ...731...41O}
{Ostriker} E.~C.,  {Shetty} R.,  2011, \mn@doi [\apj]
  {10.1088/0004-637X/731/1/41}, \href
  {https://ui.adsabs.harvard.edu/abs/2011ApJ...731...41O} {731, 41}

\bibitem[\protect\citeauthoryear{{Pillepich} et~al.,}{{Pillepich}
  et~al.}{2018}]{Pillepich2018}
{Pillepich} A.,  et~al., 2018, \mn@doi [\mnras] {10.1093/mnras/stx2656}, \href
  {https://ui.adsabs.harvard.edu/abs/2018MNRAS.473.4077P} {473, 4077}

\bibitem[\protect\citeauthoryear{{Ploeckinger} \& {Schaye}}{{Ploeckinger} \&
  {Schaye}}{2020}]{2020MNRAS.497.4857P}
{Ploeckinger} S.,  {Schaye} J.,  2020, \mn@doi [\mnras]
  {10.1093/mnras/staa2172}, \href
  {https://ui.adsabs.harvard.edu/abs/2020MNRAS.497.4857P} {497, 4857}

\bibitem[\protect\citeauthoryear{{Portinari}, {Chiosi}  \&
  {Bressan}}{{Portinari} et~al.}{1998}]{Portinari1998}
{Portinari} L.,  {Chiosi} C.,   {Bressan} A.,  1998, \aap, \href
  {https://ui.adsabs.harvard.edu/abs/1998A&A...334..505P} {334, 505}

\bibitem[\protect\citeauthoryear{{Price}}{{Price}}{2012}]{price2012}
{Price} D.~J.,  2012, \mn@doi [Journal of Computational Physics]
  {10.1016/j.jcp.2010.12.011}, \href
  {https://ui.adsabs.harvard.edu/abs/2012JCoPh.231..759P} {231, 759}

\bibitem[\protect\citeauthoryear{{Rela{\~n}o}, {Beckman}, {Zurita}, {Rozas}  \&
  {Giammanco}}{{Rela{\~n}o} et~al.}{2005}]{2005A&A...431..235R}
{Rela{\~n}o} M.,  {Beckman} J.~E.,  {Zurita} A.,  {Rozas} M.,   {Giammanco} C.,
   2005, \mn@doi [\aap] {10.1051/0004-6361:20040483}, \href
  {https://ui.adsabs.harvard.edu/abs/2005A&A...431..235R} {431, 235}

\bibitem[\protect\citeauthoryear{{Rosdahl}, {Schaye}, {Dubois}, {Kimm}  \&
  {Teyssier}}{{Rosdahl} et~al.}{2017}]{RosdahlSchaye2017}
{Rosdahl} J.,  {Schaye} J.,  {Dubois} Y.,  {Kimm} T.,   {Teyssier} R.,  2017,
  \mn@doi [\mnras] {10.1093/mnras/stw3034}, \href
  {https://ui.adsabs.harvard.edu/abs/2017MNRAS.466...11R} {466, 11}

\bibitem[\protect\citeauthoryear{{Schaller}, {Gonnet}, {Chalk}  \&
  {Draper}}{{Schaller} et~al.}{2016}]{2016pasc.conf....2S}
{Schaller} M.,  {Gonnet} P.,  {Chalk} A. B.~G.,   {Draper} P.~W.,  2016, in
  Proceedings of the Platform for Advanced Scientific Computing Conference.
  p.~2 (\mn@eprint {arXiv} {1606.02738}), \mn@doi{10.1145/2929908.2929916}

\bibitem[\protect\citeauthoryear{{Schaller}, {Gonnet}, {Draper}, {Chalk},
  {Bower}, {Willis}  \& {Hausammann}}{{Schaller}
  et~al.}{2018}]{2018ascl.soft05020S}
{Schaller} M.,  {Gonnet} P.,  {Draper} P.~W.,  {Chalk} A. B.~G.,  {Bower}
  R.~G.,  {Willis} J.,   {Hausammann} L.,  2018, {SWIFT: SPH With
  Inter-dependent Fine-grained Tasking} (\mn@eprint {ascl} {1805.020})

\bibitem[\protect\citeauthoryear{{Schaye} \& {Dalla Vecchia}}{{Schaye} \&
  {Dalla Vecchia}}{2008}]{2008MNRAS.383.1210S}
{Schaye} J.,  {Dalla Vecchia} C.,  2008, \mn@doi [\mnras]
  {10.1111/j.1365-2966.2007.12639.x}, \href
  {https://ui.adsabs.harvard.edu/abs/2008MNRAS.383.1210S} {383, 1210}

\bibitem[\protect\citeauthoryear{{Schaye} et~al.,}{{Schaye}
  et~al.}{2010}]{2010MNRAS.402.1536S}
{Schaye} J.,  et~al., 2010, \mn@doi [\mnras]
  {10.1111/j.1365-2966.2009.16029.x}, \href
  {https://ui.adsabs.harvard.edu/abs/2010MNRAS.402.1536S} {402, 1536}

\bibitem[\protect\citeauthoryear{{Schaye} et~al.,}{{Schaye}
  et~al.}{2015}]{2015MNRAS.446..521S}
{Schaye} J.,  et~al., 2015, \mn@doi [\mnras] {10.1093/mnras/stu2058}, \href
  {https://ui.adsabs.harvard.edu/abs/2015MNRAS.446..521S} {446, 521}

\bibitem[\protect\citeauthoryear{{Schmidt}}{{Schmidt}}{1959}]{1959ApJ...129..243S}
{Schmidt} M.,  1959, \mn@doi [\apj] {10.1086/146614}, \href
  {https://ui.adsabs.harvard.edu/abs/1959ApJ...129..243S} {129, 243}

\bibitem[\protect\citeauthoryear{{Schroetter}, {Bouch{\'e}}, {P{\'e}roux},
  {Murphy}, {Contini}  \& {Finley}}{{Schroetter}
  et~al.}{2015}]{2015ApJ...804...83S}
{Schroetter} I.,  {Bouch{\'e}} N.,  {P{\'e}roux} C.,  {Murphy} M.~T.,
  {Contini} T.,   {Finley} H.,  2015, \mn@doi [\apj]
  {10.1088/0004-637X/804/2/83}, \href
  {https://ui.adsabs.harvard.edu/abs/2015ApJ...804...83S} {804, 83}

\bibitem[\protect\citeauthoryear{{Schroetter} et~al.,}{{Schroetter}
  et~al.}{2019}]{2019MNRAS.490.4368S}
{Schroetter} I.,  et~al., 2019, \mn@doi [\mnras] {10.1093/mnras/stz2822}, \href
  {https://ui.adsabs.harvard.edu/abs/2019MNRAS.490.4368S} {490, 4368}

\bibitem[\protect\citeauthoryear{{Smith}, {Sijacki}  \& {Shen}}{{Smith}
  et~al.}{2018}]{Smith2018}
{Smith} M.~C.,  {Sijacki} D.,   {Shen} S.,  2018, \mn@doi [\mnras]
  {10.1093/mnras/sty994}, \href
  {https://ui.adsabs.harvard.edu/abs/2018MNRAS.478..302S} {478, 302}

\bibitem[\protect\citeauthoryear{{Smith}, {Sijacki}  \& {Shen}}{{Smith}
  et~al.}{2019}]{2019MNRAS.485.3317S}
{Smith} M.~C.,  {Sijacki} D.,   {Shen} S.,  2019, \mn@doi [\mnras]
  {10.1093/mnras/stz599}, \href
  {https://ui.adsabs.harvard.edu/abs/2019MNRAS.485.3317S} {485, 3317}

\bibitem[\protect\citeauthoryear{{Somerville} \& {Dav{\'e}}}{{Somerville} \&
  {Dav{\'e}}}{2015}]{2015ARA&A..53...51S}
{Somerville} R.~S.,  {Dav{\'e}} R.,  2015, \mn@doi [\araa]
  {10.1146/annurev-astro-082812-140951}, \href
  {https://ui.adsabs.harvard.edu/abs/2015ARA&A..53...51S} {53, 51}

\bibitem[\protect\citeauthoryear{{Springel} \& {Hernquist}}{{Springel} \&
  {Hernquist}}{2003}]{SpringelHernquist2003}
{Springel} V.,  {Hernquist} L.,  2003, \mn@doi [\mnras]
  {10.1046/j.1365-8711.2003.06206.x}, \href
  {https://ui.adsabs.harvard.edu/abs/2003MNRAS.339..289S} {339, 289}

\bibitem[\protect\citeauthoryear{{Springel}, {Di Matteo}  \&
  {Hernquist}}{{Springel} et~al.}{2005}]{springel2005}
{Springel} V.,  {Di Matteo} T.,   {Hernquist} L.,  2005, \mn@doi [\mnras]
  {10.1111/j.1365-2966.2005.09238.x}, \href
  {https://ui.adsabs.harvard.edu/abs/2005MNRAS.361..776S} {361, 776}

\bibitem[\protect\citeauthoryear{{Stanway} \& {Eldridge}}{{Stanway} \&
  {Eldridge}}{2018}]{BPASS2018}
{Stanway} E.~R.,  {Eldridge} J.~J.,  2018, \mn@doi [\mnras]
  {10.1093/mnras/sty1353}, \href
  {https://ui.adsabs.harvard.edu/abs/2018MNRAS.479...75S} {479, 75}

\bibitem[\protect\citeauthoryear{{Stinson}, {Seth}, {Katz}, {Wadsley},
  {Governato}  \& {Quinn}}{{Stinson} et~al.}{2006}]{Stinson2006}
{Stinson} G.,  {Seth} A.,  {Katz} N.,  {Wadsley} J.,  {Governato} F.,   {Quinn}
  T.,  2006, \mn@doi [\mnras] {10.1111/j.1365-2966.2006.11097.x}, \href
  {https://ui.adsabs.harvard.edu/abs/2006MNRAS.373.1074S} {373, 1074}

\bibitem[\protect\citeauthoryear{{Veilleux}, {Cecil}  \&
  {Bland-Hawthorn}}{{Veilleux} et~al.}{2005}]{2005ARA&A..43..769V}
{Veilleux} S.,  {Cecil} G.,   {Bland-Hawthorn} J.,  2005, \mn@doi [\araa]
  {10.1146/annurev.astro.43.072103.150610}, \href
  {https://ui.adsabs.harvard.edu/abs/2005ARA&A..43..769V} {43, 769}

\bibitem[\protect\citeauthoryear{{Vogelsberger}, {Genel}, {Sijacki}, {Torrey},
  {Springel}  \& {Hernquist}}{{Vogelsberger} et~al.}{2013}]{Vogelsberger2013}
{Vogelsberger} M.,  {Genel} S.,  {Sijacki} D.,  {Torrey} P.,  {Springel} V.,
  {Hernquist} L.,  2013, \mn@doi [\mnras] {10.1093/mnras/stt1789}, \href
  {https://ui.adsabs.harvard.edu/abs/2013MNRAS.436.3031V} {436, 3031}

\bibitem[\protect\citeauthoryear{{Vogelsberger} et~al.,}{{Vogelsberger}
  et~al.}{2014}]{2014MNRAS.444.1518V}
{Vogelsberger} M.,  et~al., 2014, \mn@doi [\mnras] {10.1093/mnras/stu1536},
  \href {https://ui.adsabs.harvard.edu/abs/2014MNRAS.444.1518V} {444, 1518}

\bibitem[\protect\citeauthoryear{{Vogelsberger}, {Marinacci}, {Torrey}  \&
  {Puchwein}}{{Vogelsberger} et~al.}{2020}]{2020NatRP...2...42V}
{Vogelsberger} M.,  {Marinacci} F.,  {Torrey} P.,   {Puchwein} E.,  2020,
  \mn@doi [Nature Reviews Physics] {10.1038/s42254-019-0127-2}, \href
  {https://ui.adsabs.harvard.edu/abs/2020NatRP...2...42V} {2, 42}

\bibitem[\protect\citeauthoryear{{Wiersma}, {Schaye}, {Theuns}, {Dalla Vecchia}
   \& {Tornatore}}{{Wiersma} et~al.}{2009}]{2009MNRAS.399..574W}
{Wiersma} R. P.~C.,  {Schaye} J.,  {Theuns} T.,  {Dalla Vecchia} C.,
  {Tornatore} L.,  2009, \mn@doi [\mnras] {10.1111/j.1365-2966.2009.15331.x},
  \href {https://ui.adsabs.harvard.edu/abs/2009MNRAS.399..574W} {399, 574}

\bibitem[\protect\citeauthoryear{{Zhou} et~al.,}{{Zhou}
  et~al.}{2017}]{2017MNRAS.470.4573Z}
{Zhou} L.,  et~al., 2017, \mn@doi [\mnras] {10.1093/mnras/stx1504}, \href
  {https://ui.adsabs.harvard.edu/abs/2017MNRAS.470.4573Z} {470, 4573}

\makeatother
\end{thebibliography}




\appendix

\section{Gas-star relative motion and multiple kicks in SN kinetic feedback}
\label{sec:rel_motion_multi_kick}

\begin{figure}
    \centering
    \includegraphics[width=0.46\textwidth]{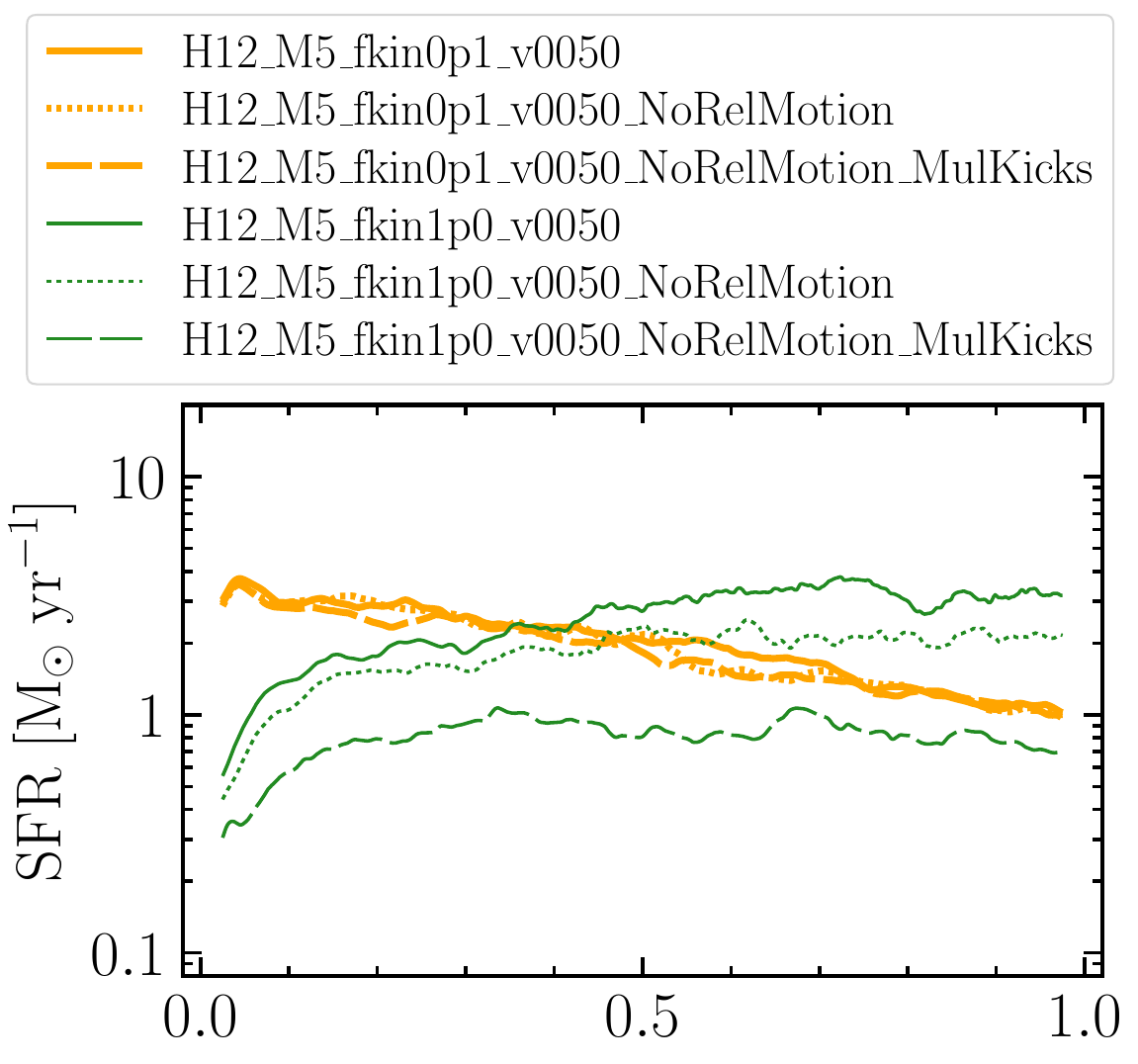} \\
    \includegraphics[width=0.46\textwidth]{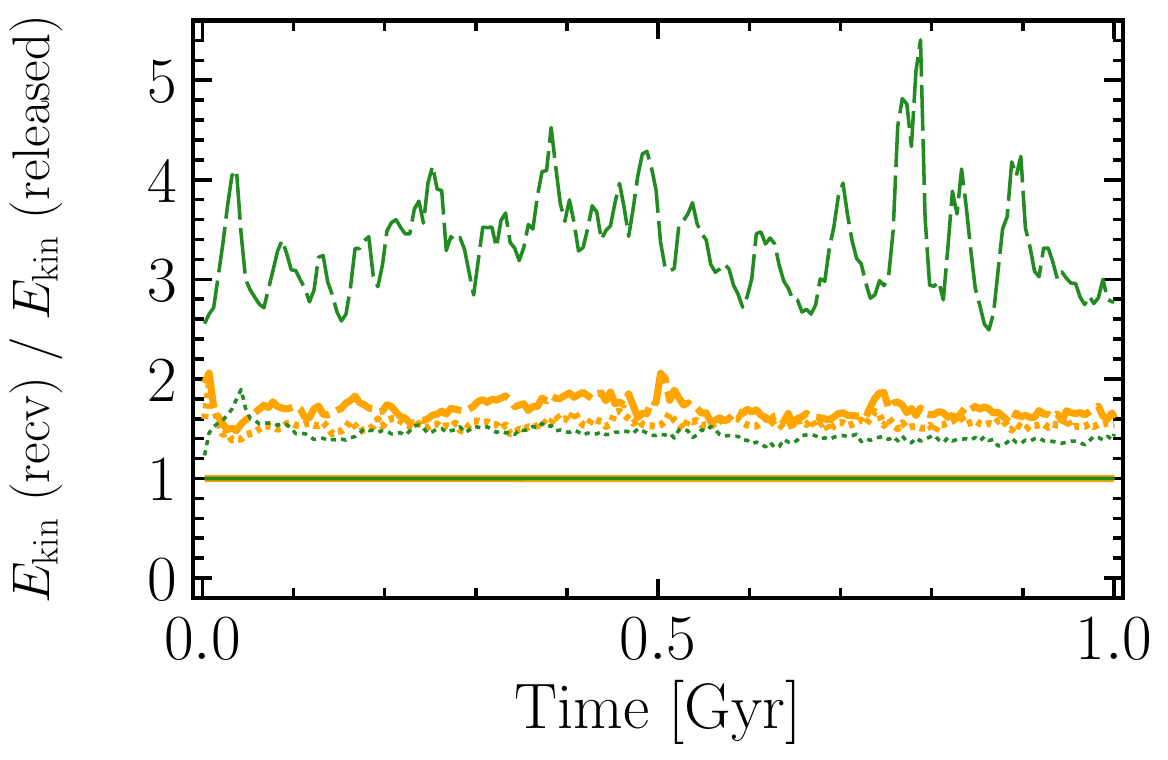}
    \caption{Star formation rate versus time (\textit{top panel}) and ratio between the total SN kinetic energies released by stars and received by gas (\textit{bottom panel}), for the H12 galaxy with M5 resolution with $f_{\rm kin}=0.1$ (orange) and $f_{\rm kin}=1$ (green). The desired kick velocity is $\Delta v_{\rm kick} = 50$ \kms. For each value of $f_{\rm kin}$, we show the model with the fiducial SN kinetic feedback (solid) and its variations where we do not account for the star-gas relative motion (short-dashed) and where we additionally allow gas particles to be kicked more than once in a single time-step (long-dashed), which results in different amounts of energies released and received. The greater the (absolute) excess in the received kinetic energy, the larger the drop in the galaxy SFR.}
    \label{fig:SFH_vary_kinetic_model}
\end{figure}

In this appendix, we quantify the importance of accounting for relative star-gas motion and the effect of limiting the number of kicks per gas particle per time-step to one in our model for SN feedback.

The top panel of Fig. \ref{fig:SFH_vary_kinetic_model} shows the star formation histories of the H12 galaxy with M5 resolution for the SN feedback models with $f_{\rm kin}=0.1$ (orange) and $f_{\rm kin}=1$ (green). The desired kick velocity is set to $\Delta v_{\rm kick} = 50$ \kms{} in both cases. For each $f_{\rm kin}$, the solid curves show our fiducial model, the short-dashed curves show the model where we switch off the correction due to star-gas relative motion in the SN kinetic feedback, while the long-dashed curves describe the runs where we additionally allow gas particles to be kicked more than once in a single time-step. 

When $f_{\rm kin}=0.1$, the star formation histories for all three models are nearly identical. For $f_{\rm kin}=1$, neglecting the relative star-gas motion results in an SFR that is lower by a factor of $\approx 1.4$, while not limiting the number of kicks per gas particle per time-step (along with neglecting the relative motion) leads to a drop in SFR by another factor of $\approx 3$.

The bottom panel shows the ratio between the SN total kinetic energies released by stars and received by gas, which can be different if the relative gas-star motion is not accounted for and/or gas particles receive multiple kicks in a single time-step. When the gas-star relative motion is neglected, the gas receives $\approx 50$ per cent more kinetic energy than it should, which is true for both $f_{\rm kin}=0.1$ and $1$. Allowing gas neighbours to be kicked multiplies times in a single time-step has a large impact if $f_{\rm kin}=1$. In this case, the energy received by the gas can be more than 4 times greater compared to the energy released by the stars. Conversely, when $f_{\rm kin}=0.1$, star particles kick their gas neighbours rarely enough that the number of kick collisions remains small and allowing multiple kicks in the simulation makes negligible difference. On average, we find that the greater the absolute excess in the received kinetic energy, the larger the drop in the SFR.

We note that galaxy properties other than the SFRs may be affected too depending on whether relative star-gas motion is accounted for and/or whether kick collisions are prevented. Exploring these other properties is, however, beyond the scope of this work.

\section{Impact of the star formation criterion}
\label{sec:sf_crit}

\begin{figure}
    \centering
    \includegraphics[width=0.46\textwidth]{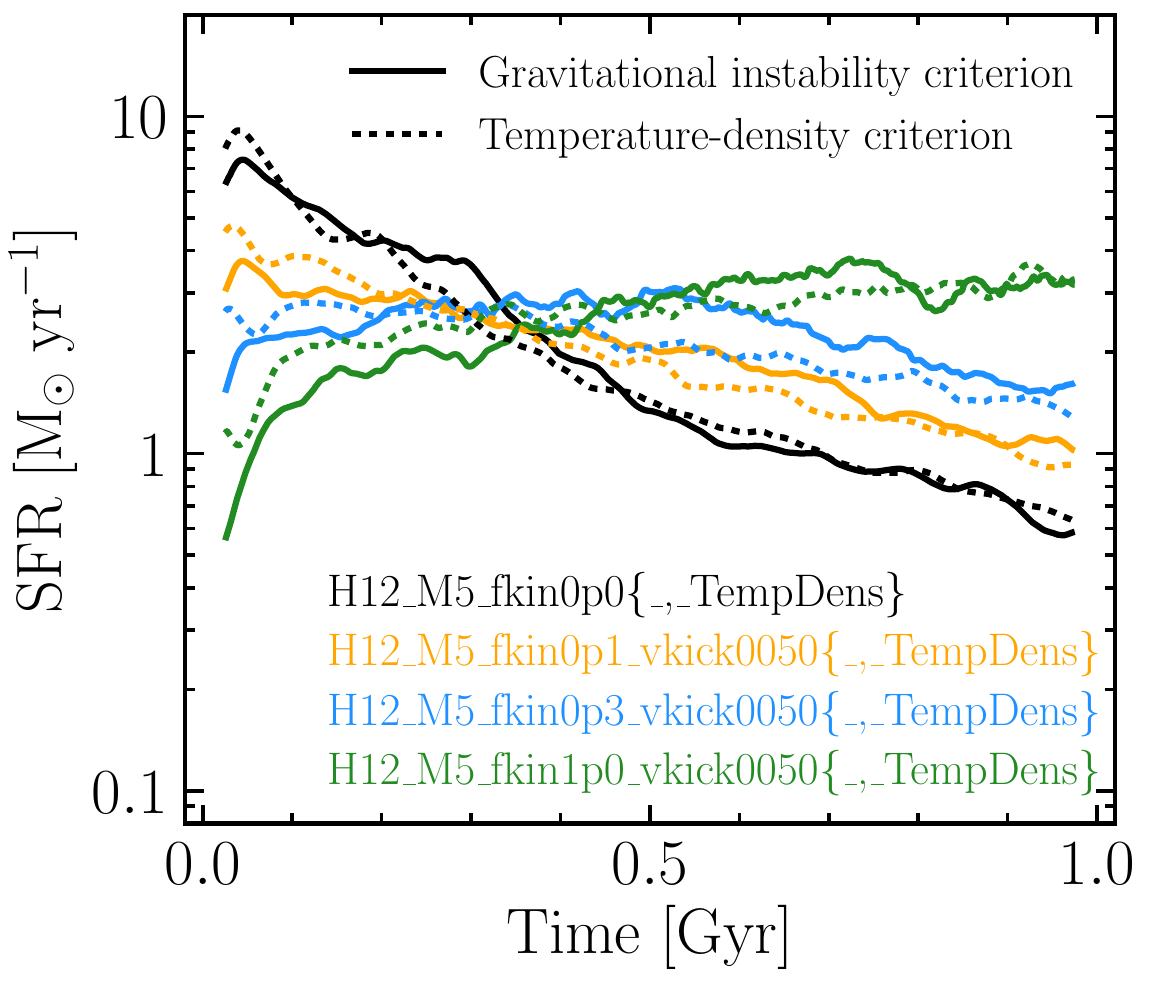}
    \caption{Star formation rate versus time in the Milky Way-mass galaxy with M5 resolution for different values of $f_{\rm kin}$ (colours) and two versions of the star formation criterion: the gravitational instability criterion (solid) and the temperature-density criterion (dashed). The desired kick velocity is $\Delta v_{\rm kick} = 50$ \kms{} in the runs including SN kinetic feedback. At a fixed $f_{\rm kin}$, replacing one star formation criterion by the other has only a marginal impact on the galaxy star formation rate.}
    \label{fig:SFH_vary_sf_crit}
\end{figure}

Fig. \ref{fig:SFH_vary_sf_crit} shows the star formation rate versus time in the Milky Way-mass galaxy with M5 resolution for different fractions of energy released in kinetic form, $f_{\rm kin}$, and for two criteria for star formation. The parameter $f_{\rm kin}$ equals $0$ (black), $0.1$ (orange), $0.3$ (blue), or $1$ (green), while the star formation criterion is either the gravitational instability criterion (solid curves) or the temperature-density criterion (dashed curves). The desired kick velocity is set to $\Delta v_{\rm kick} = 50$ \kms{} in the runs with $f_{\rm kin}>0$.

By comparing the curves at a fixed $f_{\rm kin}$, we find that as we replace one star formation criterion by the other, the galaxy SFR remains largely unaffected. In other words, changing $f_{\rm kin}$ has a similar impact on the total SFR regardless of which star formation criterion is employed. This implies that our results are not driven by the choice of the star formation criterion. 

In section $\S$\ref{sec:vel_disp_and_sfr} we showed that the increase in the turbulent velocity dispersion due to the injection of kinetic energy (temporarily) stops the gas from satisfying the gravitational instability criterion for star formation. This raises the question of why the results are so similar if we instead use the temperature-density criterion for star formation.  We believe that there are two main reasons why the gravitational instability and temperature-density criteria yield such similar star formation histories:
 \begin{itemize}
     \item According to the temperature-density criterion, the gas has to have temperatures below $T=10^3$ K (or hydrogen number densities greater than $10^2$ cm$^{-3}$) in order to be star-forming, whereas a kick with $\Delta v_{\rm kick} = 50$ \kms{} corresponds (energy-wise) to a temperature increase of $\Delta T\sim 10^5$ K. Therefore, if the kinetic feedback-induced turbulence is (partly) thermalised, the temperature of the gas particles can easily exceed the threshold value of $T=10^3$ K and this gas will cease forming stars. The net effect is thus roughly similar to how turbulence modulates the SFR in the case of the gravitational instability criterion.
     
     \item Kicking gas particles in dense, star-forming gas clumps has on average a disruptive effect on the clumps. As a result of the kicks, the density in these clumps should decrease, leading to a reduction in the SFR through the \citet{1959ApJ...129..243S} law. The regulation of the SFR in this way is not affected by the change from one star formation criterion to the other.
 \end{itemize}

\bsp	
\label{lastpage}
\end{document}